\documentclass[12pt]{article}

\usepackage[english]{babel}
\usepackage[T1]{fontenc}
\usepackage[utf8]{inputenc}
\usepackage{amssymb,amsmath,array,hhline,wasysym}
\usepackage{latexsym}
\usepackage{graphicx}
\usepackage[final]{pdfpages}
\usepackage{verbatim}

\title{Signature and spectral flow \\ of $J$-unitary $\SM^1$-Fredholm operators}

\author{Hermann Schulz-Baldes
\\
\\
{\small Department Mathematik, Universit\"at Erlangen-N\"urnberg, Germany}
}

\date{}

\newtheorem{theo}{Theorem}
\newtheorem{defini}{Definition}
\newtheorem{proposi}{Proposition}
\newtheorem{lemma}{Lemma}
\newtheorem{coro}{Corollary}

\newcommand{\BM}{{\mathbb B}}
\newcommand{\CM}{{\mathbb C}}

\newcommand{\NM}{{\mathbb N}}
\newcommand{\RM}{{\mathbb R}}
\newcommand{\SM}{{\mathbb S}}

\newcommand{\ZM}{{\mathbb Z}}
\newcommand{\LM}{{\mathbb L}}

\newcommand{\DM}{{\mathbb D}}
\newcommand{\GM}{{\mathbb G}{\mathbb U}}
\newcommand{\UM}{{\mathbb U}}

\newcommand{\FM}{{\mathbb F}}

\newcommand{\Ee}{{\cal E}}
\newcommand{\Ff}{{\cal F}}

\newcommand{\Oo}{{\cal O}}
\newcommand{\Tr}{\mbox{\rm Tr}}
\newcommand{\Tt}{{\cal T}}

\newcommand{\Kk}{{\cal K}}
\newcommand{\Hh}{{\cal H}}

\newcommand{\one}{{\bf 1}}

\newcommand{\Ker}{{\mbox{\rm Ker}}}
\newcommand{\Ran}{{\mbox{\rm Ran}}}
\newcommand{\Ind}{{\mbox{\rm Ind}}}
\newcommand{\Sig}{{\mbox{\rm Sig}}}

\newcommand{\SF}{{\mbox{\rm SF}}}
\newcommand{\BMa}{{\mbox{\rm BM}}}

\newcommand{\IN}{{\mbox{\rm IN}}}
\newcommand{\sech}{{\mbox{\rm sech}}}
\newcommand{\ess}{{\mbox{\rm\tiny ess}}}

\begin{document}

\maketitle

\begin{abstract}
Bijective operators conserving the indefinite scalar product on a Krein space $(\Kk,J)$ are called $J$-unitary. Such an operator $T$ is defined to be $\SM^1$-Fredholm if $T-z\,\one$ is Fredholm for all $z$ on the unit circle $\SM^1$, and essentially $\SM^1$-gapped if there is only discrete spectrum on $\SM^1$. For paths in the $\SM^1$-Fredholm operators an intersection index similar to the Conley-Zehnder index is introduced. The strict subclass of essentially $\SM^1$-gapped operators has a countable number of components which can be distinguished by a homotopy invariant given by the signature of $J$ restricted to the eigenspace of all eigenvalues on $\SM^1$. These concepts are illustrated by several examples.

\vspace{.1cm}

\noindent MSCN: 47B50, 53D12, 58J30
\end{abstract}

\section{Introduction and short overview}
\label{sec-intro}

A separable Krein space $\Kk$ is a separable complex Hilbert space furnished with a so-called fundamental symmetry which is a bounded operator $J\in\BM(\Kk)$ satisfying $J^2=\one $ and $J=J^*$. In this work $\Kk=\Hh\oplus\Hh$ with a Hilbert space $\Hh$ and in this grading 
$$
J\;=\;
\left(
\begin{array}{cc}
\one & 0 \\
0 & -\one
\end{array}
\right)
\;.
$$
The interest is mainly on the case of infinite $n=\dim(\Hh)$. The fundamental symmetry can also be seen as a non-degenerate quadratic form of signature $(n,n)$, and it induces a sesquilinear form on $\Kk$ by $\phi,\psi\in\Kk\mapsto \phi^*J\psi$, also called an indefinite scalar product in this context. Here $\phi^*$ denotes the linear functional associated to $\phi$ by the Riesz theorem, that is, the bra in the Dirac notation. An invertible, bounded operator $T\in\BM(\Kk)$ is called $J$-unitary if it conserves the indefinite scalar product, or equivalently, if it satisfies $T^*JT=J$ where $T^*$ denotes the usual Hilbert space adjoint in $\Kk$. A lot of basic facts about Krein spaces and $J$-unitary operators can be found in the monographs \cite{Bog,AI}, for finite dimension $n$ also in \cite{YS,GLR}. The spectrum $\sigma(T)$ of a $J$-unitary $T$ has the $\SM^1$-reflection property $\overline{\sigma(T)}=\sigma(T)^{-1}$. Furthermore, the discrete spectrum on $\SM^1$ (composed by normal eigenvalues, namely isolated eigenvalues of finite multiplicity) enjoys a remarkable stability under perturbations as discovered by Krein \cite{Kre} and Gelfand and Lidskii \cite{GL} in the finite dimensional case. Even eigenvalue collisions stay on $\SM^1$ whenever $J$ is definite on the associated eigenspace. These results readily extend to the discrete spectrum in the infinite dimensional situation, and hold under certain further conditions also for essential spectrum (see Theorem 2.5.23 in \cite{AI}). For the convenience of the reader and because it is relevant in the sequel, the main facts of stability theory for the discrete spectrum on $\SM^1$ are reviewed in Section~\ref{sec-Krein}.

\vspace{.2cm}

This work is about Fredholm properties of $J$-unitary operators and homotopy invariants for associated operator classes. A $J$-unitary $T$ is said to be $\SM^1$-Fredholm if $T-z\,\one$ is a Fredholm operator for all $z\in\SM^1$. It is said to be essentially $\SM^1$-gapped if there is only discrete spectrum on $\SM^1$. The associated operator classes are denoted by $\FM(\Kk,J)$ and $\GM(\Kk,J)$. Both are open subsets of the $J$-unitary operators. The essentially $\SM^1$-gapped $J$-unitaries form a strict subset of the $\SM^1$-Fredholm operators, as is shown by an explicit example in Section~\ref{sec-exampleperturbation}. This example also shows that the property to be essentially $\SM^1$-gapped is not stable under compact (actually even finite dimensional) perturbations, while the $\SM^1$-Fredholm property clearly is compactly stable. 

\vspace{.2cm}

The first main topic of the paper is the signature $\Sig(T)$ of an essentially $\SM^1$-gapped $J$-unitary $T$. It is defined as the difference of the number of positive and negative eigenvalues of $J$ restricted to the generalized eigenspace of all eigenvalues on $\SM^1$ (see Section~\ref{sec-FredholmSigDif}). The signature is a homotopy invariant (Theorem~\ref{theo-sigdiffhomo}, based on Krein stability theory) which is trivial in finite dimension, but in infinite dimension it permits to split the set $\GM(\Kk,J)$ of essentially $\SM^1$-gapped $J$-unitaries into disjoint open components, similar as the Fredholm index does for the Fredholm operators. It is actually simple to produce examples with non-trivial signature, see Section~\ref{sec-FredholmSigDif}, so that all components are non-empty. Furthermore, non-trivial signatures appear in applications. Section~\ref{sec-nontrivialex} consideres the $J$-unitary transfer operator of a half-space discrete magnetic Sch\"odinger operator (Harper model). Its signature is then equal to the number of edge modes weighted by the sign of the group velocity, and this latter number is known to be equal to a Chern number of the planar Harper model. It is reasonable to expect that there is also a theory of  essentially $\RM$-gapped $J$-self-adjoint operators and associated signature invariants.

\vspace{.2cm}

The second main topic concerns the homotopy theory of paths in $\FM(\Kk,J)$ of $\SM^1$-Fredholm operators. This uses the construction (Theorem~\ref{theo-TtoU0}) of a unitary operator $V(T)$ on $\Kk$ which can be associated to every $J$-unitary operator $T$ because its graph is Lagrangian. It is explicitly given by
$$ 
T
\;=\;
\left(
\begin{array}{cc}
a & b \\
c & d
\end{array}
\right)
\;\;\;\;\;\;
\mapsto
\;\;\;\;\;\;
V(T)
\;=\;
\left(
\begin{array}{cc}
(a^*)^{-1} & bd^{-1} \\
-\,d^{-1}c & d^{-1} 
\end{array}
\right)
\;.
$$
One then has the equivalence of $T\phi=\phi$, $\phi\in\Kk$, with $V(T)\phi=\phi$ (Theorem~\ref{theo-TtoU}). Furthermore, it turns out (Theorem~\ref{theo-S1Fredholm}) that for an $\SM^1$-Fredholm operator $T$, the point $1$ is not in the essential spectrum of $V(T)$. Hence for paths $t\mapsto T_t$ in the $\SM^1$-Fredholm operators one can associate a spectral flow of $t\mapsto V(T_t)$ through $1$. Due to the above simple connection between the eigenvalue equations, this spectral flow is an intersection index counting the weighted number of periodic solutions $\phi_t$ of $T_t\phi_t=\phi_t$ along the path (see Section~\ref{sec-S1FredCZ}). Actually, this index can also be understood as a Bott-Maslov index \cite{Bot,Mas}, albeit in an infinite dimensional space. More precisely, the unitary $V(T_t)$ describes a Lagrangian subspace given by a certain adequate graph of $T_t$, called the twisted graph. Therefore the above spectral flow can be considered as an infinite dimensional and modified version of the Conley-Zehnder index \cite{CZ}. Because of this connection, this paper also contains in Sections~\ref{sec-basicintersect} and \ref{sec-Index} a streamlined exposition of the Bott-Maslov index in infinite dimension. It is based on the use of unitaries to describe Lagrangian subspaces and the associated spectral flow similar as above. This approach does not use the Souriau map as prior works \cite{Fur}, or several charts as the approach in \cite{GPP}.

\vspace{.2cm}

The Bott-Maslov index and the above intersection number (of Conley-Zehnder type) is of little use if one has no information about the orientation of intersections (or equivalently the orientation of the crossings of the spectral flow). These orientations have to be studied for every given concrete path. A standard example is a linear Hamiltonian system (here in infinite dimension) for which there is some monotonicity (Theorem~\ref{theo-osci}) that is actually at the basis of Sturm-Liouville oscillation theory \cite{Bot,Lid,SB,SB2}. Another example of a monotonous path counts the bound states of Schr\"odinger equations (see Section~\ref{sec-application}). Furthermore, associated to every essentially $\SM^1$-gapped $J$-unitary $T$ there is a path $t\in[0,2\pi)\mapsto e^{-\imath t}T$ in $\FM(\Kk,J)$. Its intersection number is equal to the signature $\Sig(T)$ (Theorem~\ref{theo-sigSFeq}). This particular path is not monotonous, but the orientation of every crossing is given by the inertia of the corresponding eigenvalue of $T$  on $\SM^1$.

\vspace{.2cm}

The author's initial motivation to study $\SM^1$-Fredholm operators roots in applications to solid state physics systems. Just as for the Harper model alluded to above, the half-space transfer operators of periodic systems at energies in a gap of the full-space operator are $\SM^1$-Fredholm operators. Each eigenvalue on the unit circle then corresponds to an edge mode (a peculiarity is that a continuum of eigenvalues on $\SM^1$ corresponds to a flat band of edge modes). These edge modes determine the boundary physics, and systems for which it is non-trivial are called topological insulators. Therefore the signature (and its variants for classes of operators with further symmetries) of associated half-space transfer operators allows to distinguish between different topological insulators and its homotopy invariance also shows their structural stability. This application will be explored in a subsequent publication \cite{SV}.

\vspace{.2cm}

Finally a few words on the structure and style of the present paper. It is basically self-contained with detailed proofs except for a few functional analytical facts resembled in the appendices. Of course, this makes the paper lengthy. However, a large part of the text is anyway necessary in order to introduce notations. Moreover, the author hopes that some readers may appreciate the concise treatment of some already known results and that experts can easily localize the new results. Section~\ref{sec-subspaceKrein} concerns the basic geometry of Krein spaces and introduces the Bott-Maslov index. Section~\ref{sec-Junitary} recalls some important properties of $J$-unitaries including Krein stability analysis, and provides a detailed analysis of the unitary $V(T)$ introduced above. Then Sections~\ref{sec-S1Fred} and \ref{sec-S1gapped} study the $\SM^1$-Fredholm operators and essentially $\SM^1$-gapped operators respectively, as well as the intersection number and signature. Finally Section~\ref{sec-example} provides examples to illustrate the general concepts of the paper.

\vspace{.2cm} 

\noindent {\bf Notations:} Vectors in Hilbert space are denoted by $v,w\in\Hh$ and $\phi,\psi\in\Kk$; frames in $\Kk$ (see Appendix~\ref{app-angles} for a definition) are denoted by $\Phi$, $\Psi$; subspaces of $\Kk$ (or $\Hh$ in  Appendix~\ref{app-angles}) are described by $\Ee$, $\Ff$; bounded operators on $\Hh$ are described by small letters $a,b,c,d\in\BM(\Hh)$, while bounded operators on $\Kk$ rather by capital letters $T,S,K\in\BM(\Kk)$; unitary operators are denoted by $u\in\UM(\Hh)$ and $U,V\in\UM(\Kk)$; the star $*$ is used for the passage to dual space in $\Kk$ and $\Hh$ (notably, $\phi^*\psi$ denotes the scalar product of $\phi,\psi\in\Kk$)  as well as the Hilbert space adjoint operator; norms of vectors and the operator norm are written as $\|\,.\,\|$; finally the essential spectrum $\sigma_\ess(T)$ is the spectrum of $T$ without the discrete spectrum (thus in this work the essential spectrum is {\bf not} given by those $z$ for which $T-z\one$ is Fredholm).

\section{Geometry of subspaces of a Krein space}
\label{sec-subspaceKrein}

\subsection{Isotropy and orthogonality in Krein space}
\label{sec-isotropy}

This section recalls a few definitions ({\it e.g.} \cite{Bog,AI}), then provides some formulas for projections in terms of frames used later on.

\begin{defini}
\label{def-subspaces} Let $\Ee$ be a subspace of a Krein space $ (\Kk,J)$.

\vspace{.1cm}

\noindent {\rm (i)} $\Ee$ is called positive definite for $ J$ if and
only if $\phi^*J \phi>0$ for all non-zero $\phi\in\Ee$. Similarly, $\Ee$ 

is called negative definite for $ J$ if
$\phi^*J \phi<0$ for all non-zero $\phi\in\Ee$. Moreover, $\Ee$ is called 

definite if it is either
positive or negative definite.

\vspace{.1cm}

\noindent {\rm (ii)} $\Ee$ is called degenerate for $ J$ if and only if there
is a non-zero vector $\phi\in\Ee$ such that $\phi^* J \psi=0$ 

for all  $\psi\in\Ee$. If $\Ee$ is not degenerate, it is called non-degenerate.

\vspace{.1cm}

\noindent {\rm (iii)}  $\Ee$ is called isotropic for $ J$ if and only if
$\phi^* J \psi=0$ for all  $\phi,\psi\in\Ee$.

\vspace{.1cm}

\noindent {\rm (iv)} A maximally isotropic subspace is called Lagrangian.

\vspace{.1cm}

\noindent {\rm (v)}  $\Ee$ is called coisotropic for $ J$ if and only if
$\phi^* J \psi=0$ for all  $\psi\in\Ee$ implies that $\phi\in\Ee$.



  
\end{defini}

Let us point out that the maximality condition implies that Lagrangian subspaces are always closed.  Note also that definite subspaces are clearly non-degenerate, but the inverse is not true.  

\begin{defini}
\label{def-subspacesum} Let $\Ee$ and $\Ee'$ be two subspaces of a Krein space $ (\Kk,J)$.

\vspace{.1cm}

\noindent {\rm (i)} $\Ee$ and $\Ee'$ are called $ J$-orthogonal if and
only if $\phi^*J \psi=0$ for all $\phi\in\Ee$ and $\psi\in\Ee'$. We then 

also write $\Ee\widehat{\perp}\,\Ee'$.

\vspace{.1cm}

\noindent {\rm (ii)} Given $\Ee$, its $ J$-orthogonal complement
$\Ee^{\widehat{\perp}}$ is given by all vectors $\phi\in \Kk$
satisfying $\phi^*J \psi=0$ 

for all $\psi\in\Ee$.

\vspace{.1cm}

\noindent {\rm (iii)} If $\Ee$ and $\Ee'$ are $ J$-orthogonal and have
trivial intersection, their sum is denoted by $\Ee\widehat{+}\,\Ee'$.

\end{defini}

Just as in euclidean geometry, $J$-orthogonal complements are closed. The following lemma collects a few direct connections with the notions above.

\begin{lemma}
\label{lem-connections} Let $\Ee$ be a subspace of $ \Kk$.

\vspace{.1cm}

\noindent {\rm (i)} $(\Ee^{\widehat{\perp}})^{\widehat{\perp}}=\overline{\Ee}$

\vspace{.1cm}

\noindent {\rm (ii)} $\Ee$ is isotropic if and only if $\Ee\subset \Ee^{\widehat{\perp}}$

\vspace{.1cm}

\noindent {\rm (iii)} $\Ee$ is coisotropic if and only if $\Ee^{\widehat{\perp}}\subset\Ee$

\vspace{.1cm}

\noindent {\rm (iv)} $\Ee$ is Lagrangian if and only if $\Ee= \Ee^{\widehat{\perp}}$

\vspace{.1cm}

\noindent {\rm (v)} $\Ee$ is non-degenerate if and only if $\Ee\cap\Ee^{\widehat{\perp}}=\{0\}$

\end{lemma}

Other than in euclidean geometry, $\Ee^{\widehat{\perp}}$ and $\Ee$ do not span $ \Kk$, in general. But the following result shows that the dimensions add up as usual (a statement that is not very interesting in infinite dimension). 

\begin{proposi}
\label{prop-dimensionadd}
Let $\Ee$ and $\Ff$ be subspaces of  $ \Kk$. Then:

\vspace{.1cm}

\noindent {\rm (i)} $\Ee\subset\Ff\;\;\;\Longleftrightarrow\;\;\;\Ff^{\widehat{\perp}}\subset \Ee^{\widehat{\perp}}$

\vspace{.1cm}

\noindent {\rm (ii)} 
$\Ee^{\widehat{\perp}}+ \Ff^{\widehat{\perp}}=(\Ee\cap \Ff)^{\widehat{\perp}}$ and
$(\Ee+ \Ff)^{\widehat{\perp}}=\Ee^{\widehat{\perp}}\cap \Ff^{\widehat{\perp}}$

\vspace{.1cm}

\noindent {\rm (iii)} $\Ker( J|_\Ee)=\Ee\cap\Ee^{\widehat{\perp}}$ 

\vspace{.1cm}

\noindent {\rm (iv)} $\dim(\Ee)+\dim(\Ee^{\widehat{\perp}})= 2n$

\vspace{.1cm}

\noindent {\rm (v)} Lagrangian subspaces of $ \Kk$ have dimension $n$.

\end{proposi}

\noindent {\bf Proof.} (i) to (iii) are obvious. (iv) For the usual orthogonal complement $\Ee^\perp$ to $\Ee$ (w.r.t. to the euclidean scalar product on $ \Kk$), one has $\dim(\Ee)+\dim(\Ee^\perp)= 2n$. It is hence sufficient to provide an isomorphism between $\Ee^\perp$ and $\Ee^{\widehat{\perp}}$. Indeed, one has $ J\Ee^\perp=\Ee^{\widehat{\perp}}$. (v) follows from $\Ee^{\widehat{\perp}}=\Ee$. 
\hfill $\Box$

\vspace{.2cm}

Two closed subspaces are said to form a Fredholm pair if both their intersection and the codimension of their sum are finite dimensional. A frame for a subspace is a partial isometry from some Hilbert space onto it. More details on these Hilbert space concepts are recollected in Appendix~\ref{app-angles}.

\begin{proposi}
\label{proposi-FredpairIsotropiProj}
Let $\Ee$ and $\Ff$ be a Fredholm pair of Lagrangian subspaces with trivial intersection and let $\Phi$ and $\Psi$ be frames for them respectively. Suppose that $\Hh=\Ee+\Ff$. Then the oblique projection $P$ with range $\Ee$ and kernel $\Ff$ is given by
$$
P
\;=\;
\Phi\,\bigr(\Psi^* J\,\Phi\bigl)^{-1}\,\Psi^*\, J
\;.
$$
\end{proposi}

\noindent {\bf Proof.} This follows directly from Proposition~\ref{prop-projectionconstruct} because $J\Psi$ is a frame for $\Ff^\perp$.
\hfill $\Box$

\vspace{.2cm}

As already pointed out, $\Ee^{\widehat{\perp}}$ and $\Ee$ do not span $ \Kk$ in general.
If, however, $\Ee$ is non-degenerate, then this holds as shown in the following result.

\begin{proposi}
\label{prop-subspaceJorthcomp}
Let $\Ee$ be a closed non-degenerate subspace of $\Kk$ with frame $\Phi$. Let us suppose that  $0\not\in\sigma_\ess(\Phi^*J\Phi)$. Then 
\begin{equation}
\label{eq-perpdecomp}
 \Kk\;=\;
\Ee\,\widehat{+}\,\Ee^{\widehat{\perp}}
\;,
\end{equation}
and the oblique projection $P$ with range $\Ee$ and kernel $\Ee^{\widehat{\perp}}$ is given by
\begin{equation}
\label{eq-projformula}
P
\;=\;
 \Phi\,( \Phi^* J \, \Phi)^{-1}\, \Phi^* J
\;.
\end{equation}
It satsifies $P^*= JPJ$.
\end{proposi}

\noindent {\bf Proof.} For \eqref{eq-perpdecomp} one has to show that every $\phi\in \Kk$ can be written as $\phi= \Phi v+ \psi$ with unique vectors $u$ and $\psi\in \Ee^{\widehat{\perp}}$. For that purpose, let us multiply this equation from the left by $ \Phi^* J$. As $ \Phi^* J \psi=0$, this gives $\Phi^* J \,\phi= \Phi^* J\,  \Phi\,v$. Now the non-degeneracy of $\Ee$ shows that $0$ is not an eigenvalue of the self-adjoint operator $\Phi^*J\Phi$. As there is no essential spectrum by hypothesis, it follows that $ \Phi^*J  \Phi$ is invertible. Thus
$$
v\;=\;\bigl( \Phi^* J\,  \Phi\bigr)^{-1}
 \Phi^* J \,\phi\;.
$$
Then $\psi=\phi- \Phi\,v$ is by construction in $\Ee^{\widehat{\perp}}$. As there was no freedom in this construction, uniqueness of $v$ and $\psi$ is guarenteed. Furthermore, the projection $P$ is given by $P \phi= \Phi v$ leading to \eqref{eq-projformula}. From this the last formula is now readily deduced. 
\hfill $\Box$

\vspace{.2cm}

\subsection{The inertia and signature of subspaces}
\label{sec-signature}

First let us recall that for any self-adjoint operator $H=H^*$ (identified with a quadratic form), the inertia $\nu(H)=(\nu_+,\nu_-,\nu_0)$ is defined by the dimensions of the spectral projection of $H$ on $(0,\infty)$, $(-\infty,0)$ and $\{0\}$ respectively. The inertia is sometimes also called signature, but here the signature of $H$ is $\Sig(H)=\nu_+-\nu_-$. Another number used in this context is the Witt index $\min\{\nu_+,\nu_-\}+\nu_0$ (this is equal to the dimension of the maximally isotropic subspaces). If $\nu_0$ vanishes, it is often suppressed in the inertia. If $H$ acts on an infinite dimensional Hilbert space, all entries of the inertia can be infinite, but, of course, one is most interested in cases where some of them are finite. The main result about the inertia used below is Sylvester's law stating that $\nu(C^*HC)=\nu(H)$ for any invertible operator $C$. Furthermore, one has $\nu(H^{-1})=\nu(H)$ for any invertible $H$.

\begin{defini}
\label{def-subspacesignature} 
Let $\Ee$ be a subspace of a Krein space $ (\Kk,J)$. If $\Phi$ is a frame for $\Ee$, then let us set $ J_\Ee= \Phi^* J  \Phi$. The inertia $\nu(\Ee)=(\nu_+(\Ee),\nu_-(\Ee),\nu_0(\Ee))$ and signature $\Sig(\Ee)$ of $\Ee$ are the inertia and signature of $ J_\Ee$, that is, $\nu(\Ee)=\nu( J_\Ee)$ and $\Sig(\Ee)=\Sig(J_\Ee)$.  
\end{defini}

Sylvester's law of inertia implies that the definition of $\nu(\Ee)$ and $\Sig(\Ee)$ is independent of the choice of the frame $ \Phi$.  Let us note a few obvious connections between inertia and the other notions introduced above.

\begin{lemma}
\label{lem-signaturenotion} Let $\Ee$ be a subspace of $ \Kk$.

\vspace{.1cm}

\noindent {\rm (i)} $\Ee$ isotropic $\Longleftrightarrow$ $\nu_+(\Ee)=\nu_-(\Ee)=0$

\vspace{.1cm}

\noindent {\rm (ii)} $\Ee$ non-degenerate $\Longleftrightarrow$ $\nu_0(\Ee)=0$

\vspace{.1cm}

\noindent {\rm (iii)} $\Ee$ positive definite $\Longleftrightarrow$ $\nu_0(\Ee)=\nu_-(\Ee)=0$

\vspace{.1cm}

\noindent {\rm (iv)} $\Ee$ negative definite $\Longleftrightarrow$ $\nu_0(\Ee)=\nu_+(\Ee)=0$

\vspace{.1cm}

\noindent {\rm (v)} $\nu_+(\Ee)+\nu_-(\Ee)+\nu_0(\Ee)=\dim(\Ee)$

\end{lemma}

The following result provides an alternative way to calculate the inertia and the signature of a non-degenerate subspace.

\begin{proposi}
\label{prop-signaturecalc}
Let $\Ee$ be a closed non-degenerate subspace of $\Kk$ with frame $\Phi$. Let us suppose that  $0\not\in\sigma_\ess(\Phi^*J\Phi)$. Further let $P$ be the oblique projection given by {\rm \eqref{eq-projformula}}. Then $\Ee^{\widehat{\perp}}$ is also non-degenerate and
$$
 \Kk\;=\;\Ee\,\widehat{+}\,\Ee^{\widehat{\perp}}\;,
\qquad
\nu(\Ee)\,+\,\nu(\Ee^{\widehat{\perp}})\;=\;(n,n,0)\;.
$$
Furthermore
$$
\nu_+(\Ee)
\;=\;
\nu_+(P^* J P)\;,
\qquad
\nu_-(\Ee)
\;=\;
\nu_-(P^* J P)
\;.
$$
\end{proposi}

\noindent {\bf Proof.} Let $ \Phi$ and $\Psi$ be frames for $\Ee$ and $\Ee^{\widehat{\perp}}$. Then $(\Phi,\Psi)$ is invertible in $ \Kk$ and
$$
( \Phi,\Psi)^*\, J\,( \Phi,\Psi)
\;=\;
\begin{pmatrix}
 \Phi^* J  \Phi & 0 \\
0 & \Psi^* J  \Psi
\end{pmatrix}
\;.
$$
From this the first claims can be deduced. As to the last one, let us first of all note
$$
P^* J P
\;=\;
 J  \Phi( \Phi^* J  \Phi)^{-1} \Phi^* J
\;.
$$
Therefore
$$
\nu(P^* J P)
\;=\;
\nu( \Phi( \Phi^* J  \Phi)^{-1} \Phi^*)
\;.
$$
Because the kernel of $ \Phi^*$ has dimension $ 2n-\dim(\Ee)$, it now follows that
$$
\nu(P^* J P)
\;=\;
\nu(( \Phi^* J  \Phi)^{-1})
\,-\,\bigl(0,0, 2n-\dim(\Ee)\bigr)
\;.
$$
From this the result follows.
\hfill $\Box$

\subsection{Lagrangian subspaces}
\label{sec-basicintersect}

This section further analyzes the Grassmanian $\LM(\Kk,J)$ of $J$-Lagrangian subspaces. It will be useful to describe Lagrangian subspaces by Lagrangian frames. Because Lagrangian subspaces are half-dimensional and closed, it is suggestive to choose these Lagrangian frames as linear maps $\Phi:\Hh\to\Kk=\Hh\oplus\Hh$ with $\Phi^*\Phi=\one$ and such that $\Ran(\Phi)$ is a Lagrangian subspace. This latter fact is equivalent to $\Phi^*J\Phi=0$. The set of Lagrangian frames is a $\UM(\Hh)$-cover of $\LM(\Kk,J)$ because $\Phi$ and $\Phi u$ span the same subspace for every $u\in\UM(\Hh)$.  Moreover, $\Phi\Phi^*$ is an orthogonal projection in $\Kk$ with range given by the range of $\Phi$. The complementary orthogonal projection $ J\Phi( J\Phi)^*$ projects on the $J$-Lagrangian subspace with frame $J\Phi$ and one has
\begin{equation}
\label{eq-LagProjAdd}
\one\;=\;
\Phi\Phi^*\,+\, J\Phi( J\Phi)^*
\;.
\end{equation}
Clearly, also the set of all orthogonal projections with range given by a $J$-Lagrangian subspace is in bijection with $\LM(\Kk,J)$. 

\begin{theo}
\label{theo-diffeo} 
Let $\Psi$ be a fixed Lagrangian frame. For any other Lagrangian frame $\Phi$, introduce bounded operators $x$ and $y$ on $\Hh$ by
\begin{equation}
\label{eq-coordinaterep}
\Phi
\;=\;
\Psi\,x
\;+\;
J\,\Psi\,y
\;,
\end{equation}
and define the stereographic projection of $\Phi$ along $\Psi$ by
\begin{equation}
\label{eq-coordinateUdef}
\pi_\Psi(\Phi)
\;=\;
(x+y)(x-y)^{-1}
\;=\;
(x+y)(x-y)^*
\;.
\end{equation}
Then $\pi_\Psi$ is well-defined and unitary, namely $\pi_\Psi(\Phi)\in\UM(\Hh)$. Moreover, one has  $\pi_\Psi(\Phi)=\pi_\Psi(\Phi u)$ for any $u\in\UM(\Hh)$ so that $\pi_\Psi$ factors to a map on $\LM(\Kk,J)$ also denoted by $\pi_\Psi$. It establishes a bijection $\pi_\Psi:\LM(\Kk,J)\to\UM(\Hh)$ with inverse given by
\begin{equation}
\label{eq-Piinv}
\pi_\Psi^{-1}(u)
\;=\;
\Psi\,
\frac{1}{2}(u+\one)
\;+\;
J\,\Psi\,
\frac{1}{2}(u-\one)
\;,
\end{equation}
where the r.h.s.'s of this equation gives one representative in $\LM(\Kk,J)$. 
\end{theo}

\noindent {\bf Proof.}   Equation \eqref{eq-coordinaterep} can be rewritten as
$$
\Phi
\;=\;
(\Psi,J\,\Psi)
\begin{pmatrix}
x \\ y
\end{pmatrix}
\;.
$$
Now one directly checks that $(\Psi,J\,\Psi)$ is unitary. Thus 
$$
\begin{pmatrix}
x \\ y
\end{pmatrix}
\;=\;
(\Psi,J\,\Psi)^*
\;\Phi
$$
is a frame, that is $x^*x+y^*y=\one$. Furthermore, one has
$$
0\;=\;
\Phi^*J\Phi
\;=\;
\begin{pmatrix}
x \\ y
\end{pmatrix}^*
(\Psi,J\,\Psi)^*J(\Psi,J\,\Psi)
\begin{pmatrix}
x \\ y
\end{pmatrix}
\;=\;
\begin{pmatrix}
x \\ y
\end{pmatrix}^*
\begin{pmatrix}
0 & \one \\ \one & 0
\end{pmatrix}
\begin{pmatrix}
x \\ y
\end{pmatrix}
\;=\;
x^*y+y^*x
\;,
$$
so that
$$
(x\pm y)^*(x\pm y)\;=\;\one\;.
$$
This shows  that $\pi_\Psi$ is well-defined and unitary. That the inverse is indeed given by \eqref{eq-Piinv} can be readily checked.
\hfill $\Box$

\vspace{.2cm}

\noindent {\bf Remarks} A standard choice for a reference $J$-Lagrangian subspace is $\Psi=2^{-\frac{1}{2}}\binom{\one}{\one}$. In this case, the stereographic projection is simply denoted by $\pi=\pi_\Psi$ and takes the following particularly simple form:
\begin{equation}
\label{eq-stprodef}
\pi(\Phi)\;=\;ab^{-1}\;,
\qquad
\Phi\;=\;\binom{a}{b}
\;.
\end{equation}
This also shows that $\pi_\Psi$ is indeed a generalization of the standard stereographic projection in $\RM^2$. If $\Phi:\Hh\to\Kk$ spans a Lagrangian subspace $\Ee$, but only $\Phi c$ is a frame for some invertible $c\in\BM(\Hh)$, then one can still define $x$ and $y$ by \eqref{eq-coordinaterep} and the first formula in \eqref{eq-coordinateUdef} remains valid, namely $\pi(\Ee)=(x+y)(x-y)^{-1}$. However, the factors $x\pm y$ are not unitary any more.  Furthermore, $\pi$ allows to calculate $\pi_\Psi$ via
\begin{equation}
\label{eq-stprojeq}
\pi_\Psi(\Phi)
\;=\;
(\sqrt{2}\beta)^*\pi(\Psi)^*\pi(\Phi)(\sqrt{2}\beta)
\;,
\qquad
\Psi\;=\;\binom{\alpha}{\beta}
\;.
\end{equation}
As $\sqrt{2}\beta$ is unitary, this shows that the spectra of $\pi_\Psi(\Phi)$ and $\pi(\Psi)^*\pi(\Phi)$ coincide. Let us also note that for any $u\in\UM(\Hh)$ 
$$
\pi_{\Psi u}(\Phi)\;=\;u^*\pi_{\Psi}(\Phi)u
\;.
$$
In particular, the spectrum of  $\pi_{\Psi}(\Phi)$ does not depend on $\Psi$, but only the Lagrangian subspace spanned by it. These spectral properties are of importance in view of the following result.
\hfill $\diamond$

\vspace{.2cm}

The dimension of the intersection of two Lagrangian subspaces can be conveniently read off from the spectral theory of the associated stereographic projection, as shows the next proposition.

\begin{proposi}
\label{prop-Wronski} 
Let $\Ee$ and $\Ff$ be two $J$-Lagrangian subspaces of $\Kk$ with Lagrangian frames ${\Phi}$ and $\Psi$ respectively.  Then
$$
\dim\bigl(\Ee\cap\Ff\bigr)
\;=\;
\dim\bigl(\mbox{\rm Ker}({\Psi}^*{ J}\,{\Phi})\,\bigr)
\;=\;
\dim\bigl(\,
\mbox{\rm Ker}(\pi_\Psi(\Phi)-\one)\,\bigr)
\;=\;
\mbox{\rm codim}(\Ee+\Ff)
\;.
$$
\end{proposi}

\noindent {\bf Proof.} Let us begin with the inequality $\leq$ of the first equality. Let $p\in\NM\cup\{\infty\}$ be the dimension of $\Ee\cap\Ff$. Then there are two partial isometries $v,w:\ell^2(\{1,\ldots,p\})\to\Hh$ such that $\Phi v=\Psi w$. Then $\Psi^* J\Phi w=\Psi^* J\Psi v=0$ so that the kernel of $\Psi^* J\Phi$ is at least of dimension $p$. Inversely, given an isometry $w:\ell^2(\{1,\ldots,p\})\to\Hh$ such that $\Psi^* J\Phi w=0$, one deduces that $( J\Psi)^*\Phi w=0$. As the spans of $\Psi$ and $ J\Psi$ are orthogonal and span all $\Kk$ by \eqref{eq-LagProjAdd}, it follows that the span of $\Psi w$ lies in the span of $\Phi$. This shows the other inequality and hence proves the first equality. Next  let us first note that the dimension of the kernel of $\Psi^* J\Phi$ does not depend on the choice of the representative. Using the representative given in \eqref{eq-Piinv}, one finds
\begin{equation}
\label{eq-PJP}
{\Psi}^*{ J}\,{\Phi}\;=\;
\frac{1}{2}\,(\pi_\Psi(\Phi)-\one)\;,
\end{equation}
which implies the second equality. Finally the third equality follows from
$$
\mbox{\rm codim}(\Ee+\Ff)
\;=\;
\mbox{\rm dim}\bigl((\Ee+\Ff)^\perp\bigr)
\;=\;
\mbox{\rm dim}\bigl(\Ee^\perp\cap(\Ff)^\perp\bigr)
\;=\;
\mbox{\rm dim}\bigl( J\Ee\cap J\Ff\bigr)
\;,
$$
and the fact that $ J$ is an isomorphism.
\hfill $\Box$

\begin{theo}
\label{theo-Fredholm} 
Let $\Ee$ and $\Ff$ be two $J$-Lagrangian subspaces with associated Lagrangian frames ${\Phi}$ and ${\Psi}$. Then the following are equivalent:

\vspace{.1cm}

\noindent {\rm (i)} $\Ee$ and $\Ff$ form a Fredholm pair

\vspace{.1cm}

\noindent {\rm (ii)} $\Psi^* J\Phi$ is a Fredholm operator on $\Hh$

\vspace{.1cm}

\noindent {\rm (iii)} $\pi_\Psi(\Phi)-\one$ is a Fredholm operator on $\Hh$

\vspace{.1cm}

\noindent {\rm (iv)} $1$ is not in the essential spectrum of $\pi_\Psi(\Phi)$

\vspace{.1cm}

\noindent The index $\Ind(\Ee,\Ff)$ associated to the Fredholm pair of Lagrangian subspaces vanishes.
\end{theo}

\noindent {\bf Proof.}  The equivalence of (i) and (ii) follows from Theorem~\ref{theo-Fredholmpair} in Appendix~\ref{app-angles}. The equivalence of (ii) and (iii) follows from the identity \eqref{eq-PJP}. The equivalence of (iii) and (iv) holds for any unitary operator $\pi_\Psi(\Phi)$, by the same argument showing that a selfadjoint operator is Fredholm if and only if $0$ is not in the essential spectrum. The last claim follows immediately from the definition of the index (see Appendix~\ref{app-angles}) and Proposition~\ref{prop-Wronski}.
\hfill $\Box$

\vspace{.2cm}

The theorem shows that the spectrum of the unitary $\pi_\Psi(\Phi)$ allows to determine a distance between the two subspaces. If its eigenvalues are phases close to $0$, then one is near an intersection between the subspaces. The link of the angle spectrum $\sigma(\Ee,\Ff)$ between the subspaces (see Appendix~\ref{app-angles} for a definition) to the spectrum of $\pi_\Psi(\Phi)$ is discussed in the following result.

\begin{proposi}
\label{prop-anglecompare} 
Let ${\Phi}$ and ${\Psi}$ be Lagrangian frames. Then
$$
e^{\imath\varphi}\in\sigma(\pi_\Psi(\Phi))\;\; \mbox{\rm with }\varphi\in(-\pi,\pi]
\qquad
\Longleftrightarrow
\qquad
\frac{|\varphi|}{2}\in \sigma(\Ee,\Ff)
\;.
$$
\end{proposi}

\noindent {\bf Proof.} Using \eqref{eq-Piinv}, one finds
$$
\Psi^*\Phi\;=\;
\frac{1}{2}(\pi_\Psi(\Phi)+\one)
\;,
$$
so than
$$
\Phi^*\Psi\Psi^*\Phi\;=\;
\frac{1}{4}(\pi_\Psi(\Phi)+\one)^*(\pi_\Psi(\Phi)+\one)
\;.
$$
Now by spectral calculus of  $\pi_\Psi(\Phi)$ the claim follows from
$\frac{1}{4}|e^{\imath\varphi}+1|^2=\cos^2(\frac{\varphi}{2})$.
\hfill $\Box$

\vspace{.2cm}

Note that the spectrum of $\pi_\Psi(\Phi)$ contains more information than the angle spectrum, namely a supplementary sign for each eigenvalue. In other words, a Lagrangian subspace $\Ff$ comes with a second Lagrangian frame $J\Ff$ orthogonal to $\Ff$, and within the associated quarter plane splitting, one can define angles together with an orientation.  

\subsection{The Bott-Maslov index in infinite dimensions}
\label{sec-Index}

Given a fixed Lagrangian subspace $\Ff$ in $(\Kk,J)$ and associated frame $\Psi$ for $\Ff$, let us introduce the Fredholm Lagrangian Grassmanian w.r.t. $\Ff$ by
$$
\FM\LM(\Kk,J,\Ff)
\;=\;
\left\{
\Ee\in\LM(\Kk,J)\,\left|\,(\Ee,\Ff)\;\;\mbox{\rm form Fredholm pair}\,\right.\right\}
\;.
$$
One is now interested in (continuous) paths $\gamma=(\gamma_t)_{t\in[t_0,t_1)}$ in $\FM\LM(\Kk,J,\Ff)$ and in counting the number of points $t$ with non-trivial intersections $\gamma_t\cap\Ff$, however, with an orientation as weight. This weighted sum of intersections is then the Bott-Maslov index which will now be defined in detail as a spectral flow. As in the finite dimensional case \cite{Arn}, the singular cycle of Lagrangian subspaces with non-trivial intersections is the stratified space defined by
$$
\SM(\Ff)\;=\;
\bigcup_{l\geq 1}\;
\SM_l(\Ff)
\;,
\qquad
\;\;\;\;\SM_l(\Ff)
\;=\;
\left\{
\Ee\in\FM\LM(\Kk,J,\Ff)
\;\left|
\;
\dim\bigl(\Ee\cap\Ff\bigr)=l\,
\right.\right\}
\;.
$$
Note that for $\Ee\in\FM\LM(\Kk,J,\Ff)$ it never happens that $\dim(\Ee\cap\Ff)=\infty$.  Under the stereographic projection $\pi_\Psi$ one gets from Theorem~\ref{theo-Fredholm}  
$$
\pi_\Psi\bigl(\SM(\Ff)\bigr)
\;=\;
\UM_\ess(\Hh)
\;,
$$
where $\UM_\ess(\Hh)$ are those $u\in \UM(\Hh)$ satisfying $1\not\in\sigma_\ess(u)$, and furthermore from Proposition~\ref{prop-Wronski}
\begin{equation}
\label{eq-PiSing}
\pi_\Psi\bigl(\SM_l(\Ff)\bigr)
\;=\;
\left\{\;u\in \UM(\Hh)
\;\left|\;
1\not\in\sigma_\ess(u)\;
\mbox{\rm and }
\dim(\ker(u-\one))=l\,
\right.\right\}
\;.
\end{equation}
Let now $\gamma=(\gamma_t)_{t\in[t_0,t_1)}$ be a (continuous) 
path as above for which, for sake of simplicity, the number of intersections  $\{t\in
[t_0,t_1)\;|\;\gamma_t\in\SM(\Ff)  \}$ is finite and does not contain the initial point $t_0$. Associated to $\gamma$ is the path
$$
u_t\;=\;\pi_\Psi(\gamma_t)\in\UM_\ess(\Hh)
\;,
$$
for which a spectral flow $\SF((u_t)_{t\in[t_0,t_1)})$ through $1$ is defined in an obvious manner, as recalled in Appendix~\ref{sec-SF}. This spectral flow now defines the Bott-Maslov intersection number or index of the path $\gamma$ w.r.t. the singular cycle $\SM(\Ff)$: 
\begin{equation}
\label{eq-intersec}
\BMa(\gamma,\Ff)
\;=\;
\SF\bigl((u_t)_{t\in[t_0,t_1)}\bigr)
\;.
\end{equation}
Let us collect without detailed proof a few basic properties of the index.

\begin{proposi}
\label{prop-index} For $T\in\UM(\Kk,J)$ and $\gamma$ as above, set $T\,\gamma=(T\,\gamma_t)_{t\in[t_0,t_1)}$. 

\vspace{.1cm}

\noindent {\rm (i)} 
Let the path $\gamma+\gamma'$ denote the concatenation with a second path $\gamma'=(\gamma_t)_{t\in[t_1,t_2)}$ in $\FM\LM(\Kk,J,\Ff)$. 

Then
$$
\BMa(\gamma+\gamma',\Ff)
\;=\;
\BMa(\gamma,\Ff) + \BMa(\gamma',\Ff)
\;.
$$

\vspace{.1cm}

\noindent {\rm (ii)}  Given a second path $\gamma'=(\gamma'_t)_{[t_0,t_1)}$ in $\FM\LM(\Kk',J',\Ff')$ where $\Ff'$ is a Lagrangian subspace in a 

Krein space $(\Kk',J')$, one has, with $\widehat{\oplus}$ denoting the symplectic direct sum, 
$$
\BMa(\gamma\,\widehat{\oplus}\,\gamma',{\Ff}\,\widehat{\oplus}\,\Ff')
\;=\;
\BMa(\gamma,\Ff)\, +\, \BMa(\gamma',\Ff')
\;.
$$

\vspace{.1cm}

\noindent {\rm (iii)} One has
$\BMa(T\,\gamma,T\,{\Ff}) =\BMa(\gamma,\Ff)$.

\vspace{.1cm}

\noindent {\rm (iv)} 
For a closed path $\gamma$, $\BMa(\gamma,{\Ff})$ is independent of $\Ff$ as long as it remains in $\FM\LM(\Kk,J,\Ff)$.

\end{proposi}

\section{Basic analysis of $ J$-unitary operators}
\label{sec-Junitary}

Let $\UM(\Kk,J)$ denote the set of $J$-unitary operators, namely all invertible $T\in\BM(\Kk)$  satisfying $T^*JT=J$. 

\subsection{M\"obius action of $J$-unitaries}
\label{sec-basicscat}

\begin{proposi}
\label{prop-basicpropUKL}
$\UM(\Kk,J)$ is a $*$-invariant group. One has 
\begin{eqnarray}
\label{eq-ULL}
\UM(\Kk,J)
& = &
\left\{
\left.\,\left(
\begin{array}{cc}
a & b \\  c & d
\end{array}
\right)\;
\right|
\;
a^*a-c^*c=\one\;,\;
d^*d-b^*b=\one\;,\;
a^*b=c^*d\;
\right\}
\\
& = &
\left\{
\left.\,\left(
\begin{array}{cc}
a & b \\  c & d
\end{array}
\right)\;
\right|
\;
aa^*-bb^*=\one\;,\;
dd^*-cc^*=\one\;,\;
ac^*=bd^*\;
\right\}
\;,
\nonumber
\end{eqnarray}
and in this representation $a$ and $d$ are invertible and satisfy $\|a^{-1}\|\leq 1$, $\|d^{-1}\|\leq 1$. Also $\|a^{-1}b\|<1$, $\|d^{-1}c\|<1$, $\|bd^{-1}\|<1$ and $\|ca^{-1}\|<1$.
\end{proposi}

\noindent {\bf Proof.}  The group property is obvious. Inverting $T^*JT=T$ shows $T^{-1}J(T^*)^{-1}=J$ so that $J=TJT^*$. The relations in \eqref{eq-ULL} are equivalent to $T^*JT=T$ and $TJT^*=T$. The fact that $a$ is invertible follows from $aa^*\geq \one$. Furthermore $aa^*-bb^*={\bf 1}$ implies that $a^{-1}b(a^{-1}b)^* =\one - a^{-1}(a^{-1})^*<\one$, so that $\|a^{-1} b\|<1$. The same argument leads to the other inequalities. 
\hfill $\Box$ 

\vspace{.2cm}

The following result is well-known \cite{KS}.

\begin{theo}
\label{theo-MoebDyn}
The group $\UM(\Kk,J)$ acts on the Siegel disc $\DM(\Hh)=\{u\in\BM(\Hh)\,|\,\|u\|<1\}$ and also on $\UM(\Hh)$ by M\"obius transformation denoted by a dot and defined by:
$$
\left(
\begin{array}{cc}
a & b \\
c & d
\end{array}
\right)
\cdot u
\;=\;
(au+b)(cu+d)^{-1}
\;,
\qquad
u\in\BM(\Hh)
\;.
$$
\end{theo}

\noindent {\bf Proof.} One first has to show that for $u\in\BM(\Hh)$ with $\|u\|<1$ and $T\in\UM(\Kk,J)$ the inverse in the M\"obius transformation $T\cdot u$ is well-defined. By Proposition~\ref{prop-basicpropUKL},  $(cu+d)=d(\one+d^{-1}cu)$ is indeed invertible. Then the identities of Proposition~\ref{prop-basicpropUKL} imply
\begin{equation}
\label{eq-crux}
(cu+d)^*\,(cu+d)\;-\;
(au+b)^*\,(au+b)
\;=\;
\one\;-\;u^*\,u
\;.
\end{equation}
Now multiplying \eqref{eq-crux} from the left by $((cu+d)^*)^{-1}$ and the right by $(cu+d)^{-1}$ and using $\one-u^*u> 0$ shows $(T\cdot u)^*(T\cdot u)< \one$ so that $T\cdot u\in\DM(\Hh)$. By the same argument, if $u\in\UM(\Hh)$, then $T\cdot u\in\UM(\Hh)$. A short algebraic calculation also shows that $(TT')\cdot u=T\cdot(T'\cdot u)$.
\hfill $\Box$

\vspace{.2cm}

The action of $\UM(\Kk,J)$ on $\UM(\Hh)$ is the stereographic projection of the natural geometric action of $\UM(\Kk,J)$ on the Lagrangian Grassmannian $\LM(\Kk,J)$. This natural action sends a subspace $\Ee\in\LM(\Kk,J)$ to $T\Ee\in\LM(\Kk,J)$. Let $\Phi$ be a frame for $\Ee$, then $T\Phi$ is not a again a frame, but $T\cdot\Phi=T\Phi\bigl((T\Phi)^*(T\Phi)\bigr)^{-\frac{1}{2}}$ is a frame for $T\Ee$. Thus one indeed has using the above M\"obius action:
\begin{equation}
\label{eq-Moeb}
\pi\bigl(T\cdot\Phi\bigr)
\;=\;
 T\cdot\pi\bigl(\Phi\bigr)
\;.
\end{equation}
The following calculation will turn out to be useful later on.

\begin{proposi} 
\label{prop-derivcalc} 
Let $t\mapsto T_t=\binom{a_t\;b_t}{c_t\;d_t}$ be a differentiable path in $\UM(\Kk,J)$ and $u\in\UM(\Hh)$. Then
$$
(T_t\cdot u)^*\partial_t(T_t\cdot u)
\;=\;
\bigl((c_tu+d_t)^{-1}\bigr)^*
\begin{pmatrix}
u \\ \one
\end{pmatrix}^*
\bigl(T_t^*J\partial_t T_t\bigr)
\begin{pmatrix}
u \\ \one
\end{pmatrix}
(c_tu+d_t)^{-1}
\;.
$$
\end{proposi}

\noindent {\bf Proof.} For sake of notational simplicity, let us suppress the index $t$. Using $(T\cdot u)^*=(T\cdot u)^{-1}$ and the laws of operator differentiation, one finds
\begin{eqnarray*}
(T\cdot u)^*\partial_t(T\cdot u)
& = &
(T_t\cdot u)^*\bigl(\partial (au+b)\bigr)(cu+d)^{-1}\,-\,\bigl(\partial (cu+d)\bigr)(cu+d)^{-1}
\\
& = &
\bigl((cu+d)^{-1}\bigr)^*
\bigl[
(au+b)^*\partial (au+b)\,-\,
(cu+d)^*\partial (cu+d)\bigr](cu+d)^{-1}
\;.
\end{eqnarray*}
This directly leads to the identity.
\hfill $\Box$

\vspace{.2cm}

\subsection{Basic spectral properties and Riesz projections of $J$-unitaries}
\label{sec-Riesz}

Most results in the section go at least back to \cite{Lan}, and can also be found in \cite{Bog}. Standard notations for the spectrum as well as the continuous, discrete, point  and residual spectrum are used.

\begin{proposi}
\label{prop-specana}
Let $T$ be a $J$-unitary.

\vspace{.1cm}

\noindent {\rm (i)} $\sigma( T)=\overline{\sigma( T)}^{-1}$

\vspace{.1cm}

\noindent {\rm (ii)} $\sigma_c( T)=\overline{\sigma_c( T)}^{-1}$ 

\vspace{.1cm}

\noindent {\rm (iii)} $\sigma_d( T)=\overline{\sigma_d( T)}^{-1}$ 

\vspace{.1cm}

\noindent {\rm (iv)}  $z\in\sigma_r( T)$ implies $\overline{z}^{-1}\in\sigma_p(T)$

\vspace{.1cm}

\noindent {\rm (v)} $z\in\sigma_p( T)$ implies $\overline{z}^{-1}\in\sigma_p(T)\cup\sigma_r(T)$

\vspace{.1cm}

\noindent {\rm (vi)} $\sigma_r( T)\cap\SM^1=\emptyset$ 
\end{proposi}

\noindent {\bf Proof.}  The main identity used in the proof is
$$
T-z\one
\;=\;
J^*(( T^*)^{-1}-z\one) J
\;=\;
z\,J( T^*)^{-1}(T-\overline{z}^{-1}\one )^*J
\;.
$$
(i) As $J$ and $T^*$ are invertible (with bounded inverses), invertibility of $ T-z\one$ is equivalent to invertibility of $T-\overline{z}^{-1}\one$.  (ii) Recall that the continuous spectrum consists of those points $z$ for which $T-z\one$ is bijective and has dense range. Again these properties are conserved in the above identity. (iii) Due to (i) only remains to show that the isolated points $z,\overline{z}^{-1}\in\sigma(T)$ both have finite same multiplicity. But this follows from Proposition~\ref{prop-RieszProperties} proved independently below, when combined with Proposition~\ref{prop-RieszProj}(v). (iv) Let $z\in\sigma_p(T)$ with eigenvector $\phi$. Then by (i) one has $\overline{z}^{-1}\in\sigma(T)$ and wants to exclude $\overline{z}^{-1}\in\sigma_c(T)$, namely that $T-\overline{z}^{-1}\one$ is bijective with dense range. Let us suppose the contrary. Then there exists a $\psi\in\Kk$ with
$$
0\;\not =\;
\phi^*J(T-\overline{z}^{-1}\one)\psi
\;=\;
\phi^*((T^{-1})^*-\overline{z}^{-1}\one)J \psi
\;=\;
\overline{z}^{-1}\;
((z\one-T)\phi)^*
(T^{-1})^*
J\psi
\;.
$$
But this is a contradiction to $T\phi=z\phi$. (v) Let $z\in\sigma_r(T)$. Then there exists some $J\psi\in\Kk$ such that for all $\phi\in\Kk$
$$
0\;=\;(J\psi)^*(T-z\one)\phi
\;=\;
z\,((T-\overline{z}^{-1}\one)\psi)^*(T^*)^{-1}J\phi
\;.
$$
But this shows that $T\psi=\overline{z}^{-1}\psi$. (vi) is a corollary of (iv) and (v).
\hfill $\Box$

\vspace{.2cm}

Let $T$ be a $J$-unitary and  $\Delta\subset\sigma(T)$ be a separated spectral subset, namely a closed subset which has trivial intersection with the closure of $\sigma(T)\setminus\Delta$. Then $P_\Delta$ denotes the Riesz projection of $T$ on $\Delta$ and furthermore $\Ee_\Delta=\Ran(P_\Delta)$ and $\Ff_\Delta=\Ker(P_\Delta)$. If $\Delta=\{\lambda\}$, let us also simply write $P_\lambda$, $\Ee_\lambda$ and $\Ff_\lambda$. The definition of $P_\Delta$ and a few basic properties which do not pend on the Krein space structure are recalled in Appendix~\ref{app-Riesz}. Next follows a list of properties of Riesz projections of $J$-unitaries.

\begin{proposi}
\label{prop-RieszProperties}
Let $ T$ be a $J$-unitary and $\Delta$ a separated spectral subset. Then
$$
(P_\Delta)^*
\;=\;
J^*\,P_{(\overline{\Delta})^{-1}}\, J
\;,
\qquad
\Ff_\Delta^\perp\;=\;J\,\Ee_{(\overline{\Delta})^{-1}}
\;.
$$ 
\end{proposi}

\noindent {\bf Proof.}
First of all, let us note that indeed $(\overline{\Delta})^{-1}$ is in the spectrum of $ T$ by Proposition~\ref{prop-specana}, and thus by the spectral mapping theorem one also knows that $\overline{\Delta}$ is in the spectrum of $ T^{-1}$.  Let us take the adjoint of the definition \eqref{eq-RieszProj} of the Riesz projection:
$$
(P_\Delta)^*
\;=\;
\oint_{\overline{\Gamma}} \frac{dz}{2\pi\imath}\;
(z- T^*)^{-1}
\;,
$$
where $\overline{\Gamma}$ is the complex conjugate of $\Gamma$, hence encircling $\overline{\Delta}$ instead of $\Delta$. It is also positively oriented even though the complex conjugated of the path $\Gamma$ would have inverse orientation, but the imaginary factor compensates this. Thus $P_\Delta(T)^*=P_{\overline{\Delta}}(T^*)$ if one adds the initial operator as an argument to the Riesz projection. Next let us use $ T^*= J^*T^{-1} J$:
$$
(P_\Delta)^*
\;=\;
J^*\;\oint_{\overline{\Gamma}} \frac{dz}{2\pi\imath}\;
(z- T^{-1})^{-1}
\; J
\;.
$$
Now Proposition~\ref{prop-RieszProj}(ii) concludes the proof of the first identity. As to the second,
$$
\Ff_\Delta^\perp
\;=\;
\Ker(P_\Delta)^\perp
\;=\;
\Ran(P_\Delta^*)
\;=\;
\Ran(J\,P_{(\overline{\Delta})^{-1}}\, J)
\;=\;
J\,\Ee_{(\overline{\Delta})^{-1}}
\;,
$$
so that the proof is complete.
\hfill $\Box$

\vspace{.2cm}

When combined with Proposition~\ref{prop-RieszProj}(v) one obtains the following.

\begin{coro}
\label{coro-dimension}
Let $\Delta$ be a disjoint separated spectral subset of a $J$-unitary satisfying $\overline{\Delta}\cap\Delta^{-1}=\emptyset$. Then $\dim(\Ee_\Delta)=\dim(\Ee_{\overline{\Delta}^{-1}})$.
\end{coro}

\begin{proposi}
\label{prop-eigenspacesJortho}
Let $\Delta$ and $\Delta'$ be disjoint separated spectral subsets of a $J$-unitary satisfying $\overline{\Delta'}\cap\Delta^{-1}=\emptyset$. Then $\Ee_\Delta$ and $\Ee_{\Delta'}$ are $J$-orthogonal. Similarly, $\Ff_\Delta^\perp$ and $\Ff_{\Delta'}^\perp$ are $J$-orthogonal.
\end{proposi}

\noindent {\bf Proof.} By Propositions~\ref{prop-RieszProperties} and \ref{prop-RieszProj}, one has
$$
P_{\Delta}^*J\,P_{\Delta'}
\;=\;
J\,P_{\overline{\Delta}^{-1}}P_{\Delta'}
\;=\;0
\;.
$$
But this is implies that $\Ee_\Delta$ and $\Ee_{\Delta'}$ are $J$-orthogonal. The other claim is proved similarly.
\hfill $\Box$

\begin{proposi}
\label{prop-decomp}
Let $T$ be a $J$-unitary and suppose that its spectrum is decomposed $\sigma(T)=\bigcup_{l=1}^L\Delta_l$ into disjoint separated spectral subsets $\Delta_l$ satisfying $\overline{\Delta_l}=(\Delta_l)^{-1}$. Then, with pairwise $J$-orthogonal subspaces, 
\begin{equation}
\label{eq-C2Ldecomp}
\Kk
\;=\;
\Ee_{\Delta_1}\,\widehat{+}\ldots\widehat{+}\, 
\Ee_{\Delta_L}
\;.
\end{equation}
Moreover, each summand is non-degenerate. 
\end{proposi}

\noindent {\bf Proof.} 
The fact that $\Kk=\Ee_{\Delta_1}+\ldots +\Ee_{\Delta_L}$ follows from Proposition~\ref{prop-RieszProj}(iv). Moreover, Proposition~\ref{prop-eigenspacesJortho} implies that the subspaces are pairwise $J$-orthogonal. To prove the last claim, let us suppose that $\phi\in\Ee_{\Delta_l}$ is in the kernel of $J|_{\Ee_{\Delta_l}}$. Then $\phi$ is $J$-orthogonal to $\Ee_{\Delta_l}$ and hence by the above $J$-orthogonal to all $\Kk$. But as $J$ is non-degenerate, this implies that $\phi=0$.
\hfill $\Box$

\vspace{.2cm}

\subsection{Signatures of generalized eigenspaces of normal eigenvalues}
\label{sec-eigenspaceGunitary}

In this section, let us fix $T\in\UM(\Kk,J)$ which has $K$ normal eigenvalue pairs off the unit circle and $k$ normal eigenvalues on the unit circle.  The eigenvalues off the circle come in pairs and are denoted by $(\lambda_l,(\overline{\lambda_l})^{-1})$ with $|\lambda_l|<1$ and $l=1,\ldots,K$, and further let the eigenvalues on the circle be $\lambda'_1,\ldots,\lambda'_k$. Let us denote the span of all associated generalized eigenspaces by $\Ee_=$, a notation which then coincides with the choice made for essentially $\SM^1$-gapped operators in Section~\ref{sec-S1gapped}. Now Proposition~\ref{prop-decomp} can be applied and it yields
\begin{equation}
\label{eq-E=decomp}
\Ee_=
\;=\;
\Ee_{\lambda'_1}\,
\widehat{+}\ldots\widehat{+}\, 
\Ee_{\lambda'_k}
\,\widehat{+}\,
\bigl( \Ee_{\lambda_1}\,+\,\Ee_{(\overline{\lambda_1})^{-1}}\bigr)
\,\widehat{+}\ldots\widehat{+}\,
\bigl( \Ee_{\lambda_K}\,+\,\Ee_{(\overline{\lambda_K})^{-1}}\bigr)
\;,
\end{equation}
with pairwise $J$-orthogonal non-degenerate summands which all have trivial intersection. As to the dimensions, one has
$$
\dim(\Ee_=)
\;=\;
\sum_{l=1}^k\dim(\Ee_{\lambda'_l})
\;+\;
2\,\sum_{l=1}^K\dim(\Ee_{\lambda_l})
\;.
$$
Let us set $n_==\dim(\Ee_=)$. Let $\Phi_=$ be a frame for $\Ee_=$ and introduce the $n_=\times n_=$ matrices $J_==\Phi_=^*J\Phi_=$ and $T_==\Phi_=^* T\Phi_=$. Then $T_=$ is $J_=$-unitary.  For sake of concreteness it can be useful to choose an adequate basis change $N$ corresponding to \eqref{eq-E=decomp} which brings $J_=$ into a particularly simple normal form. It is important to note that $N$ is in general not unitary.

\begin{proposi}
\label{prop-decomp2}
There exists an invertible $ n_=\times  n_=$ matrix $N$ such that both $N^{-1} T_= N$ and $ N^* J_= N$ are block diagonal with blocks of the size of the summands in {\rm \eqref{eq-E=decomp}}. Moreover, $ N$ can be chosen such that the first blocks of $ N^* J_= N$ corresponding to eigenvalues on the unit circle are diagonal with diagonal entries equal to $1$ or $-1$ and that the blocks of $ N^* J_= N$ corresponding to eigenvalue pairs off the unit circle are $\binom{0\,-\imath}{\imath\;\;\,0}$. 
\end{proposi}

\noindent {\bf Proof.} Let $\Psi=(\psi_1,\ldots,\psi_{n_=})$ be composed by basis vectors of the summands in \eqref{eq-E=decomp}, respecting the order of \eqref{eq-E=decomp}. Then define $N'$ by $\Phi_=N'=\Psi$. Then clearly $( N')^{-1} T_= N'$ is block-diagonal, but because these summands are $ J$-orthogonal, it follows that also $( N')^* J_= N'$ is block-diagonal. Moreover, $( N')^* J_= N'$ is self-adjoint and has no vanishing eigenvalue. In a second step, one can diagonalize the blocks corresponding to eigenvalues on the unit circle and then multiply with adequate invertible diagonal matrices in order to produce eigenvalues $1$ and $-1$. Finally, by Proposition~\ref{prop-eigenspacesJortho} (applied to $\Delta=\Delta'=\{\lambda_l\}$), the blocks corresponding to eigenvalues off the unit circle are of the form
$$
\begin{pmatrix}
0 & A \\
A^* & 0
\end{pmatrix}
\;,
$$
where, moreover, $A$ is invertible by Proposition~\ref{prop-decomp}, as otherwise the block would be degenerate. Now the basis change
$$
\begin{pmatrix}
\one & 0 \\
0 & -\imath\,A^{-1}
\end{pmatrix}^*\,
\begin{pmatrix}
0 & A \\
A^* & 0
\end{pmatrix}
\,
\begin{pmatrix}
\one & 0 \\
0 & -\imath\,A^{-1}
\end{pmatrix}
\;=\;
\begin{pmatrix}
0 & -\imath \,\one\\
\imath\,\one & 0
\end{pmatrix}
$$
can be done within this block. Combining these blockwise operations one obtains $ N$ from $ N'$.
\hfill $\Box$



\begin{defini}
\label{def-signatures}
Let $\lambda$ be a normal eigenvalue of $T\in\UM(\Kk,J)$ with generalized eigen\-space $\Ee_\lambda$. Its inertia $(\nu_+(\lambda),\nu_-(\lambda))$ is defined as follows:

\vspace{.1cm}

\noindent {\rm (i)} If $|\lambda|=1$, then the inertia is the inertia of $\Ee_\lambda$, namely that of the form $ J_{\Ee_\lambda}$ on $\Ee_\lambda$.

\vspace{.1cm}

\noindent {\rm (ii)} If $|\lambda|>1$, then the inertia is equal to $(\dim(\Ee_\lambda),0)$. 

\vspace{.1cm}

\noindent {\rm (iii)} If $|\lambda|<1$, then the inertia is equal to $(0,\dim(\Ee_\lambda))$. 

\vspace{.1cm}
 
\noindent Eigenvalues for which $\nu_+(\lambda)=0$ or $\nu_-(\lambda)=0$ are called negative and positive definite respectively. An eigenvalue which is not definite is called indefinite or of mixed inertia.
\end{defini}

In the russian literature \cite{Kre,GL,YS}, the terms {\it of first kind}, {\it of second kind} and {\it of mixed kind} are used instead of positive definitive, negative definite and indefinite. Just as for quadratic forms, the inertia is also sometimes called signature and in the present context Krein signature. The inertia of the eigenvalues on the unit circle can be obtained by counting the entries $1$ and $-1$ of the corresponding block in Proposition~\ref{prop-decomp2} and there is indeed no vanishing eigenvalue (which is where $\nu_0(\lambda)=0$ is suppressed in the notation). For eigenvalues off the unit circle the definition of the inertia is rather a convention, which is, however, consistent with Proposition~\ref{prop-decomp2} because the inertia of blocks corresponding to $\Ee_\lambda+ \Ee_{\overline{\lambda}^{-1}}$ with $|\lambda|\not =1$ is $(\dim(\Ee_\lambda),\dim(\Ee_\lambda))$.

\vspace{.2cm}

The following result shows that for eigenvalues on the unit circle the definiteness can be checked by only looking at eigenvectors (and hence not the generalized eigenvectors, often also called root vectors).

\begin{proposi}
\label{prop-definitenesscheck}
Let $\lambda$ be a unit eigenvalue of $T\in\UM(\Kk,J)$. Then
$$
\nu_\pm(\lambda)\;=\;0
\qquad
\Longleftrightarrow
\qquad
\mp\;\phi^*J \phi \;>\; 0
\;\;\mbox{ for all eigenvectors }\phi \mbox{ of }\lambda\;.
$$
\end{proposi}

\noindent {\bf Proof.} The implication ``$\Longrightarrow$'' is clear. For the converse, let us show that the condition on the r.h.s. actually implies that $\Ee_\lambda$ only consists of eigenvectors so that again the definiteness follows. Hence let us suppose that there is a non-trivial Jordan block, namely that there are vectors $\phi ,\psi \in\Ee_\lambda$ such that $ T \phi =\lambda \phi $ and $ T \psi =\lambda \psi +\phi $. Then
$$
 \phi^*J \psi \
\;=\;
\phi^* T^* J T \psi 
\;=\;
(\lambda \phi )^* J (\lambda \psi +\phi) 
\;=\;
|\lambda|^2 \, \phi^*J \psi  +\overline{\lambda} \,\phi^*J \phi 
\;.
$$
Hence
$$
\overline{\lambda} \, \phi^*J \phi 
\;=\;
(1-|\lambda|^2) \,\phi^*J \psi 
\;=\;
0
\;.
$$
But this shows $\phi^* J\phi =0$, which is a contradiction to the hypothesis.
\hfill $\Box$

\vspace{.2cm}

The previous proof actually also shows the following result.

\begin{coro}
\label{coro-signaturesJordan}
Let $\lambda$ be a unit eigenvalue of $T\in\UM(\Kk,J)$. 

\vspace{.1cm}

\noindent {\rm (i)} If there is a non-diagonal Jordan block for $\lambda$, then $\lambda$ is indefinite and there exists an eigenvector 

$\phi $ such that $\phi^* J \phi =0$.

\vspace{.1cm}

\noindent {\rm (ii)} If $\lambda$ is definite, then all Jordan blocks are diagonal.
\end{coro}

The following is a conservation law of the inertia of all eigenvalues in $\Ee_=$.

\begin{proposi}
\label{prop-signaturessum}
The inertia of $T\in\UM(\Kk,J)$ as above satisfy
\begin{equation}
\label{eq-signaturesum}
\sum_{\lambda\in\sigma(T_=)}\;\nu_\pm(\lambda)
\;=\;
\nu_\pm(\Ee_=)\;.
\end{equation}
\end{proposi}

\noindent {\bf Proof.}  The inertia of the quadratic form $ J_=$ is $(\nu_+(\Ee_=),\nu_-(\Ee_=))$.  By Sylvester's law it is also equal to the sum of the inertia of the summands it in the decomposition \eqref{eq-E=decomp}. Therefore Definition~\ref{def-signatures} completes the proof.
\hfill $\Box$

\begin{proposi}
\label{prop-stabilitysignature}
The inertia $\nu(\lambda)$ of normal eigenvalues of $T\in\UM(\Kk,J)$ depends continuously on the $ T$.
\end{proposi}

\noindent {\bf Proof.} First of all, let us recall \cite[IV.3.5]{Kat} that the normal eigenvalues of $ T$ depend continuously on $ T$. Hence the result can be restated as follows. For any continuous path $t\in\RM\mapsto T(t)$ with $ T(0)= T$, the eigenvalues $\lambda(t)$ can be labelled (at eigenvalue crossings) such that the inertia $\nu(\lambda(t))$ is constant in the sense that the sum of all inertia of colliding eigenvalues is constant. Clearly it is sufficient to prove this continuity at one point, say $t=0$. As the eigenvalues of $ T$ off the unit circle stay off the unit circle for $t$ sufficiently small, their inertia are preserved as well. Therefore one only has to consider eigenvalues of $T$ on the unit circle. Hence let $\lambda$ be such an eigenvalue and $P(t)$ be the finite dimensional Riesz projection of $ T(t)$ on all eigenvalues of $ T(t)$ converging to $\lambda$ as $t\to 0$. By \cite[Theorem 3.1]{GK}, the dimension of $\Ran(P(t))$ is constant in $t$. Let $\Phi(t)$ be a frame for $P(t)$. Then the quadratic form $\Phi(0)^*J\Phi(t)$ has precisely  $\nu_+(\lambda)$ positive eigenvalues and $\nu_-(\lambda)$ negative eigenvalues, and $0$ is not an eigenvalue because $\Ee_\lambda$ is non-degenerate. By continuity of the eigenvalues of finite dimensional matrices, these properties remain conserved for small $t$. But $\nu(\Phi(t)^* J \Phi(t))$ is precisely the sum of the inertias of all eigenvalues converging to $\lambda$ as $t\to 0$.
\hfill $\Box$

\subsection{Krein stability analysis}
\label{sec-Krein}

\begin{proposi}
\label{proposi-Kreinstability0}
Let  $\lambda$ be a simple normal unit eigenvalue of $T\in\UM(\Kk,J)$. Then there is a neighborhood of $ T$ such that the eigenvalue close to $\lambda$ stays on the unit circle. 
\end{proposi}

\noindent {\bf Proof.} This follows from the symmetry of the
spectrum of $ T$ stating that, if $\lambda$ is an eigenvalue, so is
$\overline{\lambda}^{-1}$. Thus if $\lambda$ would leave the unit
circle, it would have to lead to two eigenvalues.
\hfill $\Box$

\vspace{.2cm}

Proposition~\ref{proposi-Kreinstability0} is of elementary nature. However, it does not
allow to say anything about  structural stability in presence of
degenerate eigenvalues appearing. These issues were first addressed by Krein, and the
following result is basically due to him  \cite{Kre} and Gelfand and Lidskii \cite{GL}.

\begin{theo}
\label{theo-Kreinstability1}
Let  $\lambda$ be an normal unit eigenvalue of $T\in\UM(\Kk,J)$. Suppose that $\lambda$ is definite. Then there is a neighborhood of $ T$ in which all operators have the property that all eigenvalues close to $\lambda$ lay on the unit circle and have diagonal Jordan blocks.
\end{theo}

\noindent {\bf First proof.} 
Let us suppose the contrary. Then there exists a sequence of $J$-unitary operators $( T_n)_{n\geq 1}$ converging to $ T$ with corresponding eigenvalues $(\lambda_n)_{n\geq 1}$ converging to $\lambda\in\SM^1$ such that for each $n$ either (i) $|\lambda_n|\not = 1$ or (ii) $|\lambda_n|=1$ with a non-diagonal Jordan block. Let $\phi _n\in \Kk$ be corresponding normalized eigenvectors, namely $ T_n\phi _n=\lambda_n \phi _n$, such that $\phi _n^* J \phi _n=0$ for all $n\geq 1$. Indeed, this holds in case (i) for all eigenvectors due to Proposition~\ref{prop-eigenspacesJortho} (applied with two equal eigenvalues) and in case (ii) it can be assured due to Corollary~\ref{coro-signaturesJordan}(i). By weak compactness of the set of unit vectors, there now exists a subsequence also denoted by $\phi _n$ that converges weakly to a vector $\phi $. Then, for any $\psi\in\Kk$,
$$
\psi^*T\phi
\;=\;
\lim_{n\to\infty}\,\psi^*T\phi_n
\;=\;
\lim_{n\to\infty}\, \psi^*(T-T_n)\phi_n\;+\;\psi^*T_n\phi_n
\;=\;
\lambda\;\psi^*\phi
\;.
$$
Thus $ T \phi =\lambda \phi $ and similarly $\phi^* J \phi =0$ so that $\lambda$ would not be definite.
\hfill $\Box$

\vspace{.2cm}

\noindent {\bf Second proof.} Let $t\mapsto  T(t)$ be the path of $ J$-unitaries with $ T(0)= T$. By Proposition~\ref{prop-stabilitysignature} the inertia has to stay definite for $t$ sufficiently small. Hence it cannot leave the unit circle because this would lead to an indefinite inertia of the eigenvalue group by Proposition~\ref{prop-decomp2}.
\hfill $\Box$

\vspace{.2cm}

Let us point out that both non-trivial Jordan blocks with eigenvalues on $\SM^1$ and diagonal Jordan blocks with mixed inertia are unstable in the sense that in any neighborhood of them are operators with eigenvalues off the unit circle. This can be shown by adapting the arguments in finite dimension from \cite{YS}.

\subsection{The twisted graph of a $ J$-unitary}
\label{sec-twistedgraph}

This section is an improved and generalized version of what is given in \cite{SB2}.  Let us associate to an operator $ T$ on $\Kk=\Hh\oplus\Hh$ a new operator $\widehat{ T}=\one\widehat{\oplus} T$ on $\widehat{\Kk}=\Kk\oplus\Kk$ where the $\widehat{\oplus}$ denotes the symplectic checker board sum given by
\begin{equation}
\label{eq-expanding}
\left(
\begin{array}{cc}
a & b \\
c & d
\end{array}
\right)
\,\widehat{\oplus}\,
\left(
\begin{array}{cc}
a' & b' \\
c' & d'
\end{array}
\right)
\;=\;
\left(
\begin{array}{cccc}
a & 0 & b & 0 \\
0 & a' & 0 & b' \\
c & 0 & d & 0 \\
0 & c' & 0 & d'
\end{array}
\right)
\;.
\end{equation}
If $ T\in\UM(\Kk,J)$, then $\widehat{ T}=\one\widehat{\oplus}{ T}$ is in the group $\UM(\widehat{\Kk}, \widehat{J})$ of operators conserving the form $\widehat{ J}={ J}\,\widehat{\oplus}\, J$.  The stereographic projection from $\LM(\widehat{\Kk}, \widehat{J})$ to $\UM( \Kk)$ defined as in \eqref{eq-stprodef} is denoted by $\widehat{\pi}$. Furthermore, one can also define the symplectic sum of frames by 
\begin{equation}
\label{eq-Phisum}
\left(
\begin{array}{cc}
a \\
b
\end{array}
\right)
\,\widehat{\oplus}\,
\left(
\begin{array}{cc}
a' \\
b'
\end{array}
\right)
\;=\;
\left(
\begin{array}{cc}
a & 0  \\
0 & a' \\
b & 0  \\
0 &  b'
\end{array}
\right)
\;.\,
\qquad
a,b,a',b'\in\BM(\Hh)\;.
\end{equation}
If $\Phi$ and $\Psi$ are $ J$-Lagrangian, then $\Phi\,\widehat{\oplus}\,\Psi$ is $\widehat{ J}$-Lagrangian. Let us use a particular frame which is not of this form:
\begin{equation}
\label{eq-Phiexpand}
\widehat{\Psi}_0
\;=\;
\frac{1}{\sqrt{2}}\;
\left(
\begin{array}{cc}
0 & \one \\
\one & 0 \\
\one & 0  \\ 
0 & \one 
\end{array}
\right)
\;.
\end{equation}
One has $\widehat{\Psi}_0^*\,\widehat{ J}\,\widehat{\Psi}_0=0$ so that $\widehat{\Psi}_0$ defines a $\widehat{ J}$-Lagrangian subspace. Moreover, 
\begin{equation}
\label{eq-Psi0rep}
\widehat{\pi}(\widehat{\Psi}_0)
\;=\;
\begin{pmatrix} 0 & \one \\
\one & 0 \end{pmatrix}
\;,
\qquad
\widehat{\Psi}_0
\;=\;
\frac{1}{\sqrt{2}}\;
\left(
\begin{array}{c}
\widehat{\pi}(\widehat{\Psi}_0) \\ \one
\end{array}
\right)
\;.
\end{equation}
Note that in our notations the hat always designates objects in the doubled Krein space $\widehat{\Kk}$, but, moreover, $\widehat{ T}$ is a particular operator in $\UM(\widehat{\Kk}, \widehat{J})$ associated to a given $ T\in\UM(\Kk,J)$. Now as $\widehat{\Psi}_0$ is $\widehat{J}$-Lagrangian, so is $\widehat{ T}\,\widehat{\Psi}_0$ and one can therefore associate a unitary $\widehat{\pi}(\widehat{ T}\,\widehat{\Psi}_0)$ to it. This unitary will turn out to be particularly useful for the spectral analysis of $ T\in\UM(\Kk,J)$ (see Theorem~\ref{theo-TtoU} below). 

\begin{theo}
\label{theo-TtoU0} To a given $ T\in\UM(\Kk,J)$ let us associate a unitary $V( T)$ by
\begin{equation}
\label{eq-Uhatdef}
 V(T)
\;=\;
\widehat{\pi}(\widehat{\Psi}_0)^*
\,\widehat{\pi}(\widehat{ T}\,\widehat{\Psi}_0)
\;\in\;
\UM( \Kk)
\;.
\end{equation} 
If $T=\binom{a\;b}{c\;d}$, then
$$
 V(T)
\;=\;
\left(
\begin{array}{cc}
a-bd^{-1}c & bd^{-1} \\
-\,d^{-1}c & d^{-1} 
\end{array}
\right)
\;=\;
\left(
\begin{array}{cc}
(a^*)^{-1} & bd^{-1} \\
-\,d^{-1}c & d^{-1} 
\end{array}
\right)
\;.
$$
Furthermore $V(T)=\widehat{\pi}_{\widehat{\Psi}_0}(\widehat{ T}\,\widehat{\Psi}_0)$. The map $ T\in\UM(\Kk,J)\mapsto  V(T)\in\UM( \Kk)$
is a continuous dense embedding  with image
\begin{equation}
\label{eq-Uhatcoeff}
\left\{
\left.\,\left(
\begin{array}{cc}
\alpha & \beta \\ \gamma & \delta
\end{array}
\right)\;\in\;\UM(\Kk)\;
\right|
\;
\alpha,\delta\in\mbox{\rm invertible}
\;
\right\}
\;.
\end{equation}
\end{theo}

\noindent {\bf Proof.}  By \eqref{eq-ULL} one has $d^*d\geq \one$ so that $d$ is invertible (similarly $a$ is invertible). Now by \eqref{eq-Moeb},
$$
\widehat{ T}
\cdot
\pi(\Psi_0)
\;=\;
\left(
\begin{array}{cccc}
\one & 0 & 0 & 0 \\
0 & a & 0 & b \\
0 & 0 & \one & 0 \\
0 & c & 0 & d
\end{array}
\right)
\cdot
\left(
\begin{array}{cc}
0 & \one \\
\one & 0 
\end{array}
\right)
\;,
$$
so that writing out the M\"obius transformation, one gets
\begin{eqnarray*}
\widehat{ T}
\cdot
\pi(\Psi_0)
& = & 
\left(
\begin{array}{cc}
0 & \one \\
a & b
\end{array}
\right)
\left(
\begin{array}{cc}
\one & 0  \\
c & d 
\end{array}
\right)^{-1}
\;=\;
\left(
\begin{array}{cc}
0 & \one \\
a & b 
\end{array}
\right)
\left(
\begin{array}{cc}
d^{-1} & 0  \\
0 & d^{-1} 
\end{array}
\right)
\left(
\begin{array}{cc}
d & 0  \\
-\,c & \one
\end{array}
\right)
\;.
\end{eqnarray*}
From this the first formula for $ V(T)$ follows, and the second results from the relations in $\UM(\Kk,J)$. Clearly its upper left and lower right entries, the matrices denoted by $\alpha$ and $\delta$ in \eqref{eq-Uhatcoeff}, are invertible. It can readily be checked that $V(T)$ is equal to $\widehat{\pi}_{\widehat{\Psi}_0}(\widehat{ T}\,\widehat{\Psi}_0)$ and not just unitarily equivalent to it as in \eqref{eq-stprojeq}. 

\vspace{.2cm}

Finally let us show that the map $ T\in\UM(\Kk,J)\mapsto  V(T)\in\UM( \Kk)$ is surjective onto the set \eqref{eq-Uhatcoeff}. Indeed, given an element of this set, it is natural to set $d=\delta^{-1}$, $b=-\beta\delta^{-1}$, $c=\delta^{-1}\gamma$ and $a=\alpha-\beta\delta^{-1}\gamma$. With some care one then checks that the equations in \eqref{eq-ULL} indeed hold.
\hfill $\Box$

\vspace{.2cm}

\begin{proposi}
\label{prop-TtoUrelations} Given $ T\in\UM(\Kk,J)$, one has
$$
 V( T)^{-1}
\;=\;
 V( T)^*
\;=\;
 V( T^{-1})
\;=\;
 J\, V( T^*)\, J
\;.
$$ 
\end{proposi}

\noindent {\bf Proof.} Using $ T^{-1}= J T^* J$, one finds
$$
 V( T^*)
\;=\;
\begin{pmatrix}
a^{-1} & c^*(d^*)^{-1} \\ -(d^*)^{-1}b^* & (d^*)^{-1} 
\end{pmatrix}
\;=\;
\begin{pmatrix}
a^{-1} & a^{-1}b \\ -ca^{-1} & (d^*)^{-1} 
\end{pmatrix}
\;,
$$
and
\begin{equation}
\label{eq-VTinv}
 V( T^{-1})
\;=\;
\begin{pmatrix}
a^{-1} & -a^{-1}b \\ ca^{-1} & (d^*)^{-1} 
\end{pmatrix}
\;=\;
\begin{pmatrix}
a^{-1} & -c^*(d^*)^{-1} \\ (d^*)^{-1}b^* & (d^*)^{-1} 
\end{pmatrix}
\;.
\end{equation}
This shows the claim.
\hfill $\Box$

\vspace{.2cm}

The following result justifies the above construction and, in particular, the choice of $\widehat{\Psi}_0$.

\begin{theo}
\label{theo-TtoU}
Let $ T$ and $ V(T)$ be as in {\rm Theorem \ref{theo-TtoU0}}. 
Then
$$
\mbox{\rm geometric multiplicity of }1\;\mbox{\rm as eigenvalue of } T
\;=\;
\mbox{\rm multiplicity of }1\;\mbox{\rm as eigenvalue of } V(T)\;,
$$
and the eigenvectors coincide.
\end{theo}

\noindent {\bf Proof.} One then has for all vectors $v,v',w,w'\in\Hh$
$$
 V(T)\,\binom{w}{w'}
\;=\;
\binom{v'}{v}
\qquad
\Longleftrightarrow
\qquad
 T\binom{w}{v}
\;=\;
\binom{v'}{w'}
\;,
$$
as can readily be seen by writing everything out:
$$
\left(
\begin{array}{cc}
a-bd^{-1}c & bd^{-1}  \\
 -d^{-1}c & d^{-1} 
\end{array}
\right)
\binom{w}{w'}
\;=\;
\binom{v'}{v}
\qquad
\Longleftrightarrow
\qquad
\begin{pmatrix} a & b \\ c & d \end{pmatrix}
\binom{w}{v}
\;=\;
\binom{v'}{w'}
\;.
$$
In particular, studying an eigenvalue $\lambda$ of $ T$ one has
\begin{equation}
\label{eq-eigenvalcorresp}
 V(T)\;\binom{w}{\lambda\,v}
\;=\;
\binom{\lambda\,w}{v}
\qquad
\Longleftrightarrow
\qquad
 T\binom{w}{v}
\;=\;
\lambda\,\binom{w}{v}
\;,
\end{equation}
or similarly eigenvalues $\lambda$ of $V$:
\begin{equation}
\label{eq-eigenvalcorresp2}
 V(T)\;\binom{w}{w'}
\;=\;
\lambda\,\binom{w}{w'}
\qquad
\Longleftrightarrow
\qquad
 T\binom{w}{\lambda w'}
\;=\;
\binom{\lambda w}{w'}
\;,
\end{equation}
which means that $T$ conserses the length of $\binom{\lambda w}{w'}$. Both equations are particularly interesting in the case $\lambda=1$:
\begin{equation}
\label{eq-eigenvectorconnection}
 V(T)\,\binom{w}{v}
\;=\;
\binom{w}{v}
\qquad
\Longleftrightarrow
\qquad
 T\binom{w}{v}
\;=\;
\binom{w}{v}
\;.
\end{equation}
This equivalence proves the theorem.
\hfill $\Box$

\vspace{.2cm}

\noindent {\bf Remark} 
An alternative proof of Theorem~\ref{theo-TtoU} is obtained by studying the Bott-Maslov intersection of $\widehat{ T}\,\widehat{\Psi}_0$ with $\widehat{\Psi}_0$. Suppose that these frames have a non-trivial intersection. This means that there are vectors $v,w,v',w'\in\Hh$ such that such $\widehat{\Psi}_0\binom{v}{w}=\widehat{ T}\,\widehat{\Psi}_0\binom{v'}{w'}$. The first and third line of this vector equality imply $w=w'$ and $v=v'$, the other two that $ T\binom{v}{w}=\binom{v}{w}$. This shows 
$$
\mbox{\rm geometric multiplicity of }1\;\mbox{\rm as eigenvalue of } T
\;=\;
\dim\bigl(\,\widehat{ T}\,\widehat{\Psi}_0\, \Kk\,\cap\,
\widehat{\Psi}_0\, \Kk\bigr)
\;.
$$
But now Proposition~\ref{prop-Wronski} can be applied to calculate the r.h.s., and this leads again to a proof of Theorem~\ref{theo-TtoU}.
\hfill $\diamond$

\vspace{.2cm}

\noindent {\bf Remark}
If $\binom{w}{v}$ is an eigenvector of $T\in\UM(\Kk,J)$ with eigenvalue off the unit circle, then $\|w\|=\|v\|$. Indeed, by \eqref{eq-eigenvalcorresp} and the fact that $ V(T)$ is unitary and therefore isometric follows that 
$$
\|w\|^2+|\lambda|^2\|v\|^2
\;=\;
|\lambda|^2\|w\|^2+\|v\|^2
\;.
$$
As $|\lambda|\not = 1$ the claim follows.
This fact is compatible with the fact that the eigenspace $\Ee_\lambda$ with $\lambda$ a normal eigenvalue off the unit circle is $J$-orthogonal to itself (see Proposition~\ref{prop-eigenspacesJortho}). However, in the above argument it is not required that $\lambda$ is normal. 
\hfill $\diamond$

\vspace{.2cm}

Theorem~\ref{theo-TtoU} as well as the connection between eigenvectors can easily be adapted to study other eigenvalues on the unit circle. Indeed, if $ T v=z v$ for $z\in\SM^1$, then also $(\overline{z}\, T) v= v$. But the operator $\overline{z}\, T$ is also $ J$-unitary so that one can apply the above again to construct an associated unitary. This shows the following.

\begin{proposi}
\label{prop-TtoU2}
Let $ T=\binom{a\;b}{c\;d}$ be a $ J$-unitary and set, for $z\in\SM^1$,
\begin{equation}
\label{eq-VZdef} 
V(\overline{z}\,T)
\;=\;
\widehat{\pi}(\widehat{\Psi}_0)^*
\,\widehat{\pi}(\widehat{\overline{z}\, T}\,\widehat{\Psi}_0)
\;=\;
\left(
\begin{array}{cc}
\overline{z}\,(a^*)^{-1} & bd^{-1} \\
-\,d^{-1}c & z\,d^{-1} 
\end{array}
\right)
\;.
\end{equation}
Then
$$
\mbox{\rm geometric multiplicity of }z\;\mbox{\rm as eigenvalue of } T
\;=\;
\mbox{\rm multiplicity of }1\;\mbox{\rm as eigenvalue of } V(\overline{z}\,T)\;.
$$
\end{proposi}

Therefore, the unitaries $V(\overline{z}\,T)$ are a tool to study eigenvalues of $T$ which lie on the unit circle. Let us focus again on $z=1$. Theorem~\ref{theo-TtoU} concerns the kernel of $V(T)-\one$. It is natural to analyze how much more spectrum $V(T)$ has close to $1$, or, what is equivalent, how much spectrum the self-adjoint operator $\Re e(V(T))=\frac{1}{2}(V(T)+V(T)^*)$ has close to $1$. For this purpose let us calculate $\Re e(V(T))$.

\begin{proposi}
\label{prop-anglesagain}
Let $ T$ be a $ J$-unitary and $V( T)$ as above. Then 
\begin{equation}
\label{eq-realanglecalc}
\Re e( V(T))
\;=\;
(\one+ T)
\bigl(\one+ T^* T\bigr)^{-1}
(\one+ T)^*
\;-\;\one
\;.
\end{equation}
\end{proposi}

\noindent {\bf Proof.} Let us begin by calculating
$$
\Re e( V(T))
\;=\;
\frac{1}{2}\,( V(T)+ V(T)^*)
\;=\;
\frac{1}{2}\,( V(T)+\one)( V(T)+\one)^*\;-\;\one
\;.
$$
Now $ V(T)=\widehat{\pi}(\widehat{\Psi}_0)^*\,\widehat{\pi}(\widehat{ T}\,\widehat{\Psi}_0)$. The frame $\widehat{\Psi}_0$ defined as in \eqref{eq-Phiexpand}  satisfies $\widehat{\Psi}_0^*\widehat{\Psi}_0=\one$. On the other hand, $\widehat{ T}\,\widehat{\Psi}_0$ is not a frame, but
$$
\widehat{\Phi}\;=\;\widehat{ T}\,\widehat{\Psi}_0
\;
\bigl(\,(\widehat{ T}\,\widehat{\Psi}_0)^*\widehat{ T}\,\widehat{\Psi}_0\,\bigr)^{-\frac{1}{2}}
\;.
$$
is a frame with range equal to the range of  $\widehat{ T}\,\widehat{\Psi}_0$. Furthermore, using the representation formula \eqref{eq-Piinv} for Lagrangian frames by their unitaries as well as \eqref{eq-Psi0rep}, one shows that there is a unitary $ U \in\UM(\Kk)$ such that
$$
 U\,\widehat{\Phi}^*\widehat{\Psi}_0
\;=\;
\frac{1}{2}
\;
\begin{pmatrix}
\widehat{\pi}(\widehat{ T}\,\widehat{\Psi}_0)  \\
\one
\end{pmatrix}^*
\begin{pmatrix}
\widehat{\pi}(\widehat{\Psi}_0)  \\
\one
\end{pmatrix}
\;=\;
\frac{1}{2}\;( V(T)^*+\one)
$$
Replacing this in the above shows that
$$
\Re e( V(T))
\;= \;
2\;
\widehat{\Psi}_0^*\widehat{\Phi}\,\widehat{\Phi}^*\widehat{\Psi}_0
\;-\;\one
\; = \;
2\;
\widehat{\Psi}_0^*
\widehat{ T}\,\widehat{\Psi}_0
\;
\bigl(\,(\widehat{ T}\,\widehat{\Psi}_0)^*\widehat{ T}\,\widehat{\Psi}_0\,\bigr)^{-1}
\,(\widehat{ T}\,\widehat{\Psi}_0)^*
\widehat{\Psi}_0
\;-\;\one
\;.
$$
Hence remains to calculate the appearing products:
$$
(\widehat{ T}\,\widehat{\Psi}_0)^*\widehat{ T}\,\widehat{\Psi}_0
\;=\;
\frac{1}{2}\;
\begin{pmatrix}
0 & \one \\
a & b \\
\one & 0 \\
c & d
\end{pmatrix}^*
\begin{pmatrix}
0 & \one \\
a & b \\
\one & 0 \\
c & d
\end{pmatrix}
\;=\;
\frac{1}{2}\;
(\one\,+\, T^* T)
\;,
$$
and
$$
(\widehat{\Psi}_0)^*\widehat{ T}\,\widehat{\Psi}_0
\;=\;
\frac{1}{2}\;
\begin{pmatrix}
0 & \one \\
\one & 0 \\
\one & 0 \\
0 & \one
\end{pmatrix}^*
\begin{pmatrix}
0 & \one \\
a & b \\
\one & 0 \\
c & d
\end{pmatrix}
\;=\;
\frac{1}{2}\;
(\one\,+\,
 T)
\;.
$$
Replacing shows the claim.
\hfill $\Box$

\vspace{.2cm}

\begin{proposi} 
\label{prop-derivV(T)calc} 
Let $t\mapsto T_t=\binom{a_t\;b_t}{c_t\;d_t}$ be a differentiable path in $\UM(\Kk,J)$. Then
$$
V(T_t)^*\partial_tV(T_t)
\;=\;
\begin{pmatrix}
\one & 0 \\
-d_t^{-1}c_t & d_t^{-1}
\end{pmatrix}^*
\bigl(T_t^*J\partial_t T_t\bigr)
\begin{pmatrix}
\one & 0 \\
-d_t^{-1}c_t & d_t^{-1}
\end{pmatrix}
\;.
$$
\end{proposi}

\noindent {\bf Proof.} First of all,
$$
V(T_t)^*\partial_tV(T_t)
\;=\;
\widehat{\pi}(\widehat{T}_t\widehat{\Psi}_0)^*
\,\partial_t\; \widehat{\pi}(\widehat{T}_t\widehat{\Psi}_0)
\;=\;
\widehat{T}_t\cdot\widehat{\pi}(\widehat{\Psi}_0)^*
\,\partial_t \;\widehat{T}_t\cdot\widehat{\pi}(\widehat{\Psi}_0)
\;.
$$
Now one can apply Proposition~\ref{prop-derivV(T)calc} in the Krein space $(\widehat{\Kk},\widehat{J})$:
$$
V(T_t)^*\partial_tV(T_t)
\;=\;
A^*\;
\widehat{\Psi}_0^*
\widehat{T}_t^*\widehat{J}\,\partial_t\, \widehat{T}_t
\widehat{\Psi}_0
\;A
\;,
$$
where the second identity in \eqref{eq-Psi0rep} was also used and $A$ is given by
$$
A\;=\;
\left[
\begin{pmatrix}
0 & 0 \\ 0 & c_t
\end{pmatrix}
\widehat{\pi}(\widehat{\Psi}_0)
+
\begin{pmatrix}
\one & 0 \\ 0 & d_t
\end{pmatrix}
\right]^{-1}
\;=\;
\begin{pmatrix}
\one & 0 \\
-d_t^{-1}c_t & d_t^{-1}
\end{pmatrix}
\;.
$$
Now the result follows from consecutively using the identities
$$
\widehat{T}_t^*\widehat{J}\;\partial_t\, \widehat{T}_t
\;=\;
0\;\widehat{\oplus}\;T_t^*J\,\partial_t\, T_t
\;,
\qquad
\widehat{\Psi}_0^*\;0\;\widehat{\oplus}\;T\;\widehat{\Psi}_0
\;=\;
T\;,
$$
in the above.
\hfill $\Box$

\vspace{.2cm}

As a further preparation for applications below,  let us calculate the derivative of $V(\overline{z}\,T)$ w.r.t. the phase $z=e^{\imath t}\in\SM^1$ and thus set
\begin{equation}
\label{eq-speedop}
Q_z(T)\;=\;\frac{1}{\imath}\; V(e^{-\imath t}T)^*\,\partial_t\,V(e^{-\imath t}T)
\;.
\end{equation}
This is a self-adjoint operator on $\Kk$ and can hence also be interpreted as a quadratic form which by the following result is non-degenerate.

\begin{proposi}
\label{prop-Vderive}
Let $ T=\binom{a\;b}{c\;d}$ be a $ J$-unitary. Then
\begin{eqnarray*}
Q_z(T)
& = &
\left(
\begin{array}{cc}
(a^*)^{-1} & 0 \\
0 & d^{-1} 
\end{array}
\right)^*
\left(
\begin{array}{cc}
-\one & -z\,b \\
-\overline{z}\,b^* & \one
\end{array}
\right)
\left(
\begin{array}{cc}
(a^*)^{-1} & 0 \\
0 & d^{-1} 
\end{array}
\right)
\\
& = &
\left(
\begin{array}{cc}
-(a^*a)^{-1} & -za^{-1}bd^{-1} \\
-(za^{-1}bd^{-1})^* & (dd^*)^{-1} 
\end{array}
\right)
\;.
\end{eqnarray*}
Moreover, $0\not\in\sigma(Q_z(T))$.
\end{proposi}

\noindent {\bf Proof.} 
Deriving \eqref{eq-VZdef} and using \eqref{eq-VTinv} for $\overline{z}\,T$ shows
$$
Q_z(T)
\;=\;
\left(
\begin{array}{cc}
{z}\,a^{-1} & -a^{-1}b \\
(d^*)^{-1}b^* & \overline{z}\,(d^*)^{-1} 
\end{array}
\right)
\,\left(
\begin{array}{cc}
-\overline{z}\,(a^*)^{-1} & 0 \\
0 & z\, d^{-1}
\end{array}
\right)
\;,
$$
from what the identity can readily be deduced.  Furthermore, for the block operator in the middle of the expression for $Q_z(T)$, a vector $\binom{v}{w}$ in the kernel has to verify $v+{z}bw=0$ and $\overline{z}b^*v-w=0$, so that $(\one+bb^*)v=0$ and hence $v=w=0$. Using Weyl sequences one checks that indeed there can be no spectrum at all close to $0$, a property that is stable under conjugation with an invertible operator.
\hfill $\Box$

\section{$\SM^1$-Fredholm operators}
\label{sec-S1Fred}

\subsection{Definition and basic properties}
\label{sec-S1Fredbasic}

The definition and basic properties of Fredholm operators are recalled in Appendix~\ref{app-Fredholm}.

\begin{defini}
\label{def-S1Fredholm} 
Let $T\in\UM(\Kk,J)$ and $z\in\SM^1$. Then $T$ is called $z$-Fredholm if and only if $T-z\,\one$ is a Fredholm operator, and $T$ is called $\SM^1$-Fredholm if and only if $T-z\,\one$ is a Fredholm operator for all $z\in\SM^1$. The set of $z$-Fredholm and $\SM^1$-Fredholm operators are denoted by $\FM\UM(\Kk,J,z)$ and $\FM\UM(\Kk,J)$ respectively. 
\end{defini}

Let us collect a few first facts about such operators.

\begin{proposi}
\label{prop-zFredholmproperties} 
Let $z\in\SM^1$ and $T\in\FM\UM(\Kk,J,z)$. Then the following holds.

\vspace{.1cm}

\noindent {\rm (i)} If $S\in\UM(\Kk,J)$ is such that $S-T$ is compact, then $S\in\FM\UM(\Kk,J,z)$.

\vspace{.1cm}

\noindent {\rm (ii)} One has $\Ind(T-z\,\one)=0$.

\vspace{.1cm}

\noindent {\rm (iii)} There exists an $r>0$ such that $T-\zeta\,\one$ is Fredholm for all $\zeta\in\CM$ with $|\zeta-z|<r$.

\vspace{.1cm}

\noindent {\rm (iv)} $S\in\UM(\Kk,J)$ is $z$-Fredholm if and only if $S-z\,\one$ is essentially bounded from below.

\vspace{.1cm}

\noindent {\rm (v)}  $\FM\UM(\Kk,J,z)$ is open in $\UM(\Kk,J)$.

\end{proposi}

\noindent {\bf Proof.} (i) follows directly from the compact stability of the Fredholm property.
(ii) From $(T-z\,\one)^*=\overline{z}\,JT^{-1}(T-z\,\one)J$ follows that $\dim((T-z\,\one)^*)=\dim(T-z\,\one)$, which already implies the result. (iii) By Proposition~\ref{prop-Fredcrit} the hypothesis means that $(T-z\,\one)^*(T-z\,\one)$ is bound below by a positive constant except on a subspace of finite dimension. Now
\begin{equation}
\label{eq-S1id}
(T-z\,\one)(T-z\,\one)^*
\;=\;
(TJ)(T-z\,\one)^*(T-z\,\one)(TJ)^*
\;,
\end{equation}
implies that also $(T-z\,\one)^*$ is essentially bounded from below. Therefore again Proposition~\ref{prop-Fredcrit} concludes the proof. (iv) The identity \eqref{eq-S1id} for $S$ instead of $T$ implies that $(S-z\,\one)^*$ is essentially bounded from below if and only if $S-z\,\one$ is essentially bounded from below. Therefore Corollary~\ref{coro-Fredcrit} concludes the proof. (v) This follows from the fact that the property of being essentially bounded from below, see (iv), is stable under small perturbations.
\hfill $\Box$

\begin{proposi}
\label{prop-S1Fredholmproperties} 
Let $T\in\FM\UM(\Kk,J)$. Then the following holds.

\vspace{.1cm}

\noindent {\rm (i)} If $S\in\UM(\Kk,J)$ is such that $S-T$ is compact, then $S\in\FM\UM(\Kk,J)$.

\vspace{.1cm}

\noindent {\rm (ii)} One has $\Ind(T-z\,\one)=0$ for all $z\in\SM^1$.

\vspace{.1cm}

\noindent {\rm (iii)} There exists an $h>0$ such that $T-\zeta\,\one$ is Fredholm for all $z\in\CM$ with $e^{-h}<|z|<e^h$.

\vspace{.1cm}

\noindent {\rm (iv)} $S\in\UM(\Kk,J)$ is $z$-Fredholm if and only if $S-z\,\one$ is essentially bounded from below $\forall\;z\in\SM^1$.

\vspace{.1cm}

\noindent {\rm (v)} $\FM\UM(\Kk,J)$ is open in $\UM(\Kk,J)$.

\vspace{.1cm}

\noindent {\rm (vi)} $T\in\FM\UM(\Kk,J)$ is equivalent to $T^*\in\FM\UM(\Kk,J)$ as well as $T^{-1}\in\FM\UM(\Kk,J)$.

\vspace{.1cm}

\noindent {\rm (vii)} For $T\in\FM\UM(\Kk,J)$, $\sigma(T)\cap\SM^1\subset\sigma_p(T)$ and each eigenvalue has finite multiplicity.

\end{proposi}

\noindent {\bf Proof.} 
(i)-(v) are obvious from the above.
(vi) This follows immediately from the identities $T^*-z\one=(T-\overline{z}\one)^*$ and $T^{-1}-z\one=-T^{-1}z(T-\overline{z})$ and the facts that the adjoint of a Fredholm operator is Fredholm, and the product of a Fredholm operator with an invertible operator is Fredholm.
(vii) Combining item (ii) with Proposition~\ref{prop-Fredspec} shows that there is neither continuous nor residual spectrum on $\SM^1$. (That there is no residual spectrum on $\SM^1$ also follows from Proposition~\ref{prop-specana}(vi).)  The second claim is part of the definition of $T-z\,\one$ being a Fredholm operator. 
\hfill $\Box$


\vspace{.2cm}

Let us point out that the product $ST$ of $S,T\in\FM\UM(\Kk,J)$ is in general not $\SM^1$-Fredholm. Indeed, the product $TT^{-1}=\one$ is a counterexample. It is reasonable to expect that there is not too much spectrum on $\SM^1$ for a $\SM^1$-Fredholm operator. In fact, this is not true as shows the example in Section~\ref{sec-exampleperturbation}. On the other hand, there is always little spectrum of $V(\overline{z}\,T)$ near $1$ as shown in the following result. 

\begin{theo}
\label{theo-S1Fredholm} 
Let $T\in\UM(\Kk,J)$. Then $T$ is $z$-Fredholm if and only if $1$ is not in the essential spectrum of $V(\overline{z}\,T)$, or equivalently $\sigma_\ess(\Re e\,V(\overline{z}\,T))<1$. Therefore,
$$
T\in\FM\UM(\Kk,J)
\qquad
\Longleftrightarrow
\qquad
\max_{z\in\SM^1}\;\sigma_\ess(\Re e\,V(\overline{z}\,T))\;<\;1
\;.
$$
\end{theo}

\noindent {\bf Proof.} The fact that $1$ is not in the essential spectrum of $V(\overline{z}\,T)$ is equivalent to $V(\overline{z}\,T)-\one=\widehat{\pi}(\widehat{\Psi}_0)^*\widehat{\pi}(\widehat{\overline{z}\,T}\,\widehat{\Psi}_0)-\one$ being a Fredholm operator. By Theorem~\ref{theo-Fredholm} this is in turn equivalent to the fact that the subspaces $\Ran(\widehat{\Psi}_0)$ and $\Ran(\widehat{\overline{z}\,T}\,\widehat{\Psi}_0)$ form a Fredholm pair of subspaces in $\widehat{\Kk}\cong\Hh^{\otimes 4}$. Reordering the first three components $1,2,3$ in  $\Hh^{\otimes 4}$ to $3,1,2$, one obtains equivalence with $\Ran(\binom{\one}{\one})$ and $\Ran(\binom{\one}{\overline{z}\,T})$ being a Fredholm pair of subspaces in $\Hh^{\otimes 4}$. As a frame of the orthogonal complement of $\Ran(\binom{\one}{\one})$ is $\frac{1}{\sqrt{2}}\binom{\one}{-\one}$ and a frame for $\Ran(\binom{\one}{\overline{z}\,T})$ is $\binom{\one}{\overline{z}\,T}(\one+T^*T)^{-\frac{1}{2}}$, Theorem~\ref{theo-Fredholmpair} shows equivalence with $(\one+T^*T)^{-\frac{1}{2}}(\one-z T^*)$ being a Fredholm operator on $\Hh^{\otimes 2}= \Kk$. As multiplication by an invertible operator does not change the Fredholm property, one concludes that $1\not\in\sigma_\ess(V(\overline{z}\,T))$ is equivalent to $T-z\one$ being a Fredholm operator.
\hfill $\Box$

\vspace{.2cm}

The gap in the essential spectrum of $V(T)$ is studied in more detail in Section~\ref{sec-gapV}.

\subsection{Intersection index for paths of $z$-Fredholm operators}
\label{sec-S1FredCZ}

Let $\Gamma=(T_t)_{t\in[t_0,t_1]}$ be a path in $\FM\UM(\Kk,J,z)$. Then $V(\overline{z}\,T_t)_{t\in[t_0,t_1]}$ is a path in the set $\UM_\ess(\Kk)$ of unitaries $U\in\UM(\Kk)$ with $1\not\in\sigma_\ess(U)$. Under the same transversality conditions as in Section~\ref{sec-Index} and Appendix~\ref{sec-SF} it is then possible to define an intersection index $\IN(\Gamma,z)$ by using the spectral flow of $t\in[t_0,t_1]\mapsto V(\overline{z}\,T_t)$. In other words,
$$
\IN(\Gamma,z)
\;=\;
\SF\bigl((V(\overline{z}\,T_t)_{t\in[t_0,t_1]}
\bigr)
\;=\;
\BMa\bigl(\overline{z}\,\Gamma \,\widehat{\Psi}_0,\widehat{\Psi}_0\bigr)
\;,
$$
where $\overline{z}\,\Gamma \,\widehat{\Psi}_0=(\widehat{\overline{z}\,T_t}\,\widehat{\Psi}_0)_{t\in[t_0,t_1]}$ is (after normalization) a path in the $\FM\LM(\widehat{\Kk},\widehat{J},\widehat{\Psi}_0)$. The intersection index $\IN(\Gamma,z)$ counts the number of solutions $T_t\phi_t=z\phi_t$ along the path, but weighted by an orientation given by the inertia. If nothing is known about these inertia, the intersection index is really of little use. But in concrete situations such information may be available, for example by using Proposition~\ref{prop-derivV(T)calc}. Examples are the solutions of a Hamiltonian system (Section~\ref{sec-HamSys}), concrete paths (Sections~\ref{sec-application}, \ref{sec-nontrivialCZ} and \ref{sec-nontrivialex}) as well as the more structural result given in Section~\ref{sec-FredholmIndex}.

\subsection{The essential gap of $V(T)$}
\label{sec-gapV}

The aim of the following results is to make the gap in the spectrum and essential spectrum of $V(T)$ more quantitative and to provide tools to check whether a given $J$-unitary is $\SM^1$-Fredholm. This section is not crucial for the understanding of the sequel. For an arbitrary operator $A$ on some Hilbert space, let us set
\begin{eqnarray}
g(A)
& = &
\min\;\sigma
\bigl(
\bigl(\one+ A^* A\bigr)^{-\frac{1}{2}}
(\one- A)^*(\one- A)
\bigl(\one+ A^* A\bigr)^{-\frac{1}{2}}
\bigr)
\label{eq-gaprecall}
\\
& = &
\sup\,
\left\{g\geq 0\,\left|\,
g\,\bigl(\one+ A^* A\bigr)
<
( A-\one)^*( A-\one)\right.\right\}
\;,
\label{eq-gaprecall2}
\end{eqnarray}
and similarly for $g_\ess(A)$ where in \eqref{eq-gaprecall2} one asks the operator inequality to hold only on a subspace of finite codimension. This quantity satisfies
$$
\frac{1}{1+\|T\|^2}\;\frac{1}{\|\one-T\|^2}\;\leq\;g(T)\;
\leq\;\frac{1}{\|\one-T\|^2}
\;,
$$
and is therefore closely connected to the notion of pseudospectrum (near $1$) \cite{Dav}. If $A$ is normal, then 
by spectral calculus
$$
g(A)
\;=\;
\min_{z\in\sigma(A)}
\;\frac{|1-z|^2}{1+|z|^2}
\;,
\qquad
g_\ess(A)
\;=\;
\min_{z\in\sigma_\ess(A)}
\;\frac{|1-z|^2}{1+|z|^2}
\;.
$$
If $A=V$ is unitary, then 
\begin{equation}
\label{eq-gVT}
g(V)
\;=\;
1\,-\,\max\;\sigma(\Re e(V))
\;=\;
\mbox{\rm dist}(1,\sigma(\Re e(V)))
\;,
\qquad
g_\ess(V)
\;=\;
\mbox{\rm dist}(1,\sigma_\ess(\Re e(V)))
\;.
\end{equation}
This applies, in particular, to $V(T)$ so that $g(V(T))$ provides a good way to measure the gap at $1$. Using this notation, one has for a $J$-unitary $T$ that
$$
T \mbox{ is }\SM^1\mbox{-Fredholm }
\qquad
\Longleftrightarrow
\qquad
\min_{z\in\SM^1}\;g_\ess(V(\overline{z}\,T))
\;>\;0
\;.
$$
Next let us show that $g(V(T))$ and $g_\ess(V(T))$ can also be calculated more directly from $T$, namely $g(T)$ and $g_\ess(T)$ as defined in \eqref{eq-gaprecall}.

\begin{proposi}
\label{prop-gequivalence}
For any $J$-unitary $T$,
$$
g(V(T)) \;=\;g(T)
\;,
\qquad
g_\ess(V(T)) \;=\;g_\ess(T)
\;,
$$
and
$$
g( T)\;=\;g( T^*)\;=\;g( T^{-1})
\;,
\qquad
g_\ess( T)\;=\;g_\ess( T^*)\;=\;g_\ess( T^{-1})
\;.
$$
\end{proposi}

\noindent {\bf Proof.} One can focus on $g$ because $g_\ess$ is dealt with in the same way. Let us begin with an obvious reformulation of \eqref{eq-gVT}:
$$
g(V(T))
\;=\;
1\,-\,\max\;\sigma(\Re e(V(T)))
\;.
$$
Using Proposition~\ref{prop-anglesagain} one now finds:
$$
g(V(T)) 
\;=\;
2\,-\,\max\;\sigma
\bigl(
(\one+ T)
\bigl(\one+ T^* T\bigr)^{-1}
(\one+ T)^*
\bigr)
\;.
$$
If one sets $A=(\one+ T)(\one+ T^* T)^{-\frac{1}{2}}$, then on the r.h.s. intervenes the spectrum $\sigma(AA^*)$, which except for the point $0$ is equal to the spectrum $\sigma(A^*A)$. Therefore
\begin{eqnarray*}
g(V(T)) 
& = &
2\,-\,\max\;\sigma
\bigl(
\bigl(\one+ T^* T\bigr)^{-\frac{1}{2}}
(\one+ T)^*(\one+ T)
\bigl(\one+ T^* T\bigr)^{-\frac{1}{2}}
\bigr)
\\
& = &
\min\;\sigma
\bigl(
\bigl(\one+ T^* T\bigr)^{-\frac{1}{2}}
(\one- T)^*(\one- T)
\bigl(\one+ T^* T\bigr)^{-\frac{1}{2}}
\bigr)
\;.
\end{eqnarray*}
Hence comparing again with the definition \eqref{eq-gaprecall} the first claim follows. For the others, let us conjugate the inequality in \eqref{eq-gaprecall2} with $J$ and then $T$:
\begin{eqnarray*}
g\,
\bigl(\one+ T^* T\bigr)
\;\leq\;
( T-\one)^*( T-\one)
\;\;\;
& \Longleftrightarrow &
\;\;\;
g\,
\bigl(\one+ T^{-1}( T^{-1})^*\bigr)
\;\leq\;
( T^{-1}-\one)( T^{-1}-\one)^*
\\
& \Longleftrightarrow &
\;\;\;
g\,
\bigl(\one+ T T^*\bigr)
\;\leq\;
( T-\one)\,( T-\one)^*
\;.
\end{eqnarray*}
From this the last claims all follow.
\hfill $\Box$

\vspace{.2cm}

\begin{coro}
\label{coro-S1charac}
For any $J$-unitary $T\in\UM(\Kk,J)$, one has
$$
T\in\FM\UM(\Kk,J)
\qquad
\Longleftrightarrow
\qquad
\inf_{z\in\SM^1}\;g_\ess(\overline{z}\,T)\,>\,0
\;.
$$
\end{coro}

\vspace{.2cm}

Due to the importance of the gaps $g(T)$ and $g_\ess(T)$, let conclude this section with a stability analysis for them. Actually, such an analysis is even possible for $g(A)$ and $g_\ess(A)$ and a general operator $A$ on a Hilbert space, but for $J$-unitaries the estimates can be considerably improved.

\begin{proposi}
\label{prop-S1stability}
Let $T\in\FM\UM(\Kk,J)$ and $K\in\BM(\Kk)$ be a $J$-anti-selfadjoint operator, namely satisfying $JK=-K^* J$. Then

\vspace{.1cm}

\noindent {\rm (i)} If $K$ is compact, then $Te^K\in \FM\UM(\Kk,J)$ and $g_\ess(Te^K)=g_\ess(T)$.

\vspace{.1cm}

\noindent {\rm (ii)} If $g(T)>6\,\|T\|\,\|K\|$, then $Te^K\in \FM\UM(\Kk,J)$ with $g(Te^K)\geq g(T)-6\,\|T\|\,\|K\|$.

\vspace{.1cm}

\noindent {\rm (iii)} If $g_\ess(T)>6\,\|T\|\,\|K\|$, then $Te^K\in \FM\UM(\Kk,J)$ with $g_\ess(Te^K)\geq g_\ess(T)-6\,\|T\|\,\|K\|$.
\end{proposi}

\noindent {\bf Proof.}  The claim (i) follows immediately from general principles, but also from the argument below.  Let us use the notations
$$
T
\;=\;
\begin{pmatrix}
a & b \\ c & d
\end{pmatrix}
\;,
\qquad
e^K
\;=\;
\one\,+\,
\begin{pmatrix}
\alpha & \beta \\ \gamma & \delta
\end{pmatrix}
\;,
$$
so that
$$
Te^K
\;=\;
\begin{pmatrix}
a & b \\ c & d
\end{pmatrix}
\,+\,
\begin{pmatrix}
a\alpha+b\gamma & a\beta+b\delta \\ c\alpha+d\gamma & c\beta+d\delta
\end{pmatrix}
\;=\;
\begin{pmatrix}
a+a\alpha+b\gamma & b+a\beta+b\delta \\ c+ c\alpha+d\gamma & d+c\beta+d\delta
\end{pmatrix}
\;.
$$
Because $Te^K\in\UM(\Kk,J)$, it follows that the diagonal entries $a'=a+a\alpha+b\gamma$ and $d'=d+c\beta+d\delta$ satisfy $(a')^*a'\geq \one$ and $(d')^*d'\geq \one$ and are therefore, in particular, invertible. It follows that
\begin{eqnarray}
V(Te^K)
& = &
\begin{pmatrix}
(a^{-1})^*\bigl((\one+\alpha+a^{-1}b\gamma)^{-1}\bigr)^* & (b+a\alpha+b\delta)(\one+d^{-1}c\beta+\delta)^{-1} d^{-1}
\\
-(\one+d^{-1}c\beta+\delta)^{-1}d^{-1}(c+c\alpha+d\gamma)
&
(\one+d^{-1}c\beta+\delta)^{-1}d^{-1}
\end{pmatrix}
\nonumber
\\
& = &
V(T)\,+\,
\begin{pmatrix}
-((a')^{-1})^*(\alpha+a^{-1}b\gamma)^* & (a\alpha-a-(a^*)^{-1})(d')^{-1}
\\
(d')^{-1}\bigl((c\beta+d\delta)d^{-1}c-c\alpha-d\gamma\bigr)
&
-(d^{-1}c\beta+\delta)(d')^{-1}
\end{pmatrix}
\;,
\label{eq-expandV}
\end{eqnarray}
where for the upper right entry in the second equation the identity $bd^{-1}c=a-(a^*)^{-1}$ was used. From this one can also readily write out a formula for $\Re e(V(Te^K))-\Re e(V(T))$.

\vspace{.1cm}

Now let us suppose that $K$ is compact. Then also its entries $\alpha$, $\beta$, $\gamma$ and $\delta$ are compact. Thus \eqref{eq-expandV} implies immediately that $V(Te^K)-V(T)$ is compact. But now Weyl's theorem implies that the essential spectra are the same.

\vspace{.1cm}

Now let us bound $V(Te^K)-V(T)$. Thus one needs to control the norms of the entries of \eqref{eq-expandV}. Clearly the four entries of $T$ and $K$ are bounded by $\|T\|$ and $\|K\|$ respectively. As $\|(a')^{-1}\|\leq 1$ and  $\|(a')^{-1}\|\leq 1$, and by Proposition~\ref{prop-basicpropUKL} also $\|a^{-1}b\|<1$ and $\|d^{-1}c\|<1$, it follows that
$$
\|((a')^{-1})^*(\alpha+a^{-1}b\gamma)^*\|\;\leq\;2\,\|K\|\;,
\qquad
\|(d')^{-1}\bigl((c\beta+d\delta)d^{-1}c-c\alpha-d\gamma\bigr)\|
\;\leq\;
4\,\|T\|\,\|K\|
\;,
$$
and similarly
$$
\|(a\alpha-a-(a^*)^{-1})(d')^{-1}\|\;\leq\;3\,\|T\|\,\|K\|
\;,
\qquad
\|(d^{-1}c\beta+\delta)(d')^{-1}\|\;\leq\;2\,\|K\|
\;.
$$
As $1\leq\|T\|$, the claimed bound (ii) now follows from Lemma~\ref{lem-blockpos} below which holds for arbitrary block operators and is stated without proof (see, {\it e.g.} \cite{AI}). For (iii) one, moreover, needs to appeal to a standard argument with singular Weyl sequences.
\hfill $\Box$

\vspace{.2cm}

\begin{lemma}
\label{lem-blockpos}
For $T\in\BM(\Kk)$, one has
$$
 T\;=\;\begin{pmatrix} a & b \\ c & d \end{pmatrix}\;>\; 0
\qquad
\Longleftrightarrow
\qquad
a>0\;,\;\;d>0\;,\;\;c=b^*\;,\;\;\;\|a^{-\frac{1}{2}}bd^{-\frac{1}{2}}\|<1
\;.
$$
\end{lemma}

\section{Essentially $\SM^1$-gapped $J$-unitaries}
\label{sec-S1gapped}

\subsection{Definition and basic properties}
\label{sec-S1gappedbasic}

Let us introduce the notation $\SM^1_h=\{z\in\CM\,|\,e^{-h}\leq|z|\leq e^{h}\}$ for a closed annulus of width $h\geq 0$.

\begin{defini}
\label{def-S1gapped}
A $J$-unitary $T$ is said to be essentially $\SM^1$-gapped if and only if $\sigma_\ess(T)\cap\SM^1=\emptyset$, and strictly $\SM^1$-gapped if $\sigma(T)\cap\SM^1=\emptyset$. The set of all essentially $\SM^1$-gapped $J$-unitaries is denoted by $\GM(\Kk,J)$. Moreover, $T$ is said to have an essential $\SM^1$-gap $h(T)>0$ if the closed annulus $\SM^1_{h(T)}$ contains no essential spectrum of $T$ and there is no spectrum on the boundary of the annulus. 
\end{defini}

Note that if $n=\dim(\Hh)<\infty$, all $J$-unitaries are essentially $\SM^1$-gapped. An alternative terminology would be to call essentially $\SM^1$-gaped $J$-unitaries essentially hyperbolic, and strictly $\SM^1$-gapped $J$-unitaries hyperbolic, but this will not be used here as we prefer to stick with the notion linked to spectral theory.

\begin{proposi}
\label{prop-S1gapprop}
Any $\SM^1$-gapped $J$-unitary is also $\SM^1$-Fredholm, that is $\GM(\Kk,J)\subset\FM\UM(\Kk,J)$.  The set of $\SM^1$-gapped $J$-unitaries $\GM(\Kk,J)$ is open in the norm topology.
\end{proposi}

\noindent {\bf Proof.} The first claim results from general principles ({\it e.g.} \cite{EE}), the second one follows directly from Theorem I.3.1 of \cite{GK}.
\hfill $\Box$

\vspace{.2cm}

A natural question about essentially $\SM^1$-gapped $J$-unitaries is whether this property is stable under compact perturbations, namely whether a Weyl type stability theorem holds as for selfadjoint and unitary operators. A positive result in this respect is the following. 

\begin{proposi}
\label{prop-S1gapstab}
Let $T\in\GM(\Kk,J)$ and suppose that there is a path in the resolvent set from $0$ to $\SM^1$. Then, if $S\in\UM(\Kk,J)$ is such that $S-T$ is a compact operator, then $S\in\GM(\Kk,J)$.
\end{proposi}

\noindent {\bf Proof.} As $S$ is bounded, $0$ is always in its resolvent set. Therefore the claim follows immediately from analytic Fredholm theory as reviewed in Appendix~\ref{sec-analFred}.
\hfill $\Box$

\vspace{.2cm}

Thus remains open the situation when there is no path in the resolvent set from $0$ to $\SM^1$. Again by analytic Fredholm theory, see Appendix~\ref{sec-analFred}, it is possible that the component of the resolvent set containing $\SM^1$ is filled with point spectrum after a compact perturbation. That precisely this can indeed happen is shown by example in Section~\ref{sec-exampleperturbation}.

\subsection{The signature of essentially $\SM^1$-gapped operators}
\label{sec-FredholmSigDif}

Recall that, given $T\in\GM(\Kk,J)$ and $h>0$ sufficiently small, $\Ee_=$ denotes the finite dimensional subspace consisting of the span of generalized eigenspaces for eigenvalues of $T$ lying in the annulus $\SM^1_h$. For this collection of eigenvalues and generalized eigenspace $\Ee_=$, all the notations and results of Section~\ref{sec-eigenspaceGunitary} apply. If $h$ is chosen sufficiently small, one can assure that the spectrum of $T$ in $\SM^1_h$ lies only on the unit circle (which does not necessarily mean that $T|_{\Ee_=}$ is unitary because there may be non-diagonal Jordan blocks). However, this will not be assumed because we will be interested in paths of essentially $\SM^1$-gapped $J$-unitary operators along which eigenvalues may leave or enter the unit circle, and also $\SM^1_h$.

\begin{defini}
\label{def-S1FredholmSig} 
The signature $\Sig(T)$ of $T\in\GM(\Kk,J)$ is the signature of $\Ee_=$, namely 
$$
\Sig(T)\;=\;\nu_+(\Ee_=)-\nu_-(\Ee_=)\;.
$$
\end{defini}

By Proposition~\ref{prop-signaturessum}, the signature can also be expressed in terms of the inertia $\nu_\pm(\lambda)$ of the eigenvalues $\lambda\in\SM^1_h$:
$$
\Sig(T)
\;=\;
\sum_{\lambda\in\SM^1_h\cap\sigma(T)}\nu_+(\lambda)
\;-\;
\sum_{\lambda\in\SM^1_h\cap\sigma(T)}\nu_-(\lambda)
\;.
$$
The following example shows that in infinite dimension any value of the signature is atteined.

\vspace{.2cm}

\noindent {\bf Example} Let $\Kk=(\Hh\oplus \CM^N)\oplus \Hh$ where $\Hh$ is an infinite dimensional Hilbert space. Let the fundamental symmetry be given by $J=(\one\oplus\one)\oplus (-\one)$. Then for $\lambda\in\CM$ with $|\lambda|<1$ and $\eta\in\RM$, the operator $T=(\lambda\,\one)\oplus (e^{\imath \eta}\one)\oplus (\overline{\lambda}^{-1}\one)$ is essentially $\SM^1$-gapped and satisfies $\Sig(T)=N$.
\hfill $\diamond$

\vspace{.2cm}

\begin{theo}
\label{theo-sigdiffhomo}
The signature is well-defined {\rm (}independent of the choice of $h${\rm )} and a homotopy invariant on $\GM(\Kk,J)$. 
\end{theo}

\noindent {\bf Proof.} Changing $h$ in a continuous manner (such that there is only discrete spectrum of $T$ in $\SM^1_h$, of course) may lead to more or less eigenvalues of $T$ in $\SM^1_h$. However, these eigenvalues always come in pairs $(\lambda,\overline{\lambda}^{-1})$ off the unit circle with $\dim(\Ee_\lambda)=\dim(\Ee_{\overline{\lambda}^{-1}})$ by Corollary~\ref{coro-dimension}. Moreover, by Propostion~\ref{prop-decomp}, the subspace $\Ee_\lambda+\Ee_{\overline{\lambda}^{-1}}$ is non-degenerate and has inertia $(\dim(\Ee_\lambda),\dim(\Ee_\lambda),0)$. Hence this subspace and hence this eigenvalue pair do not contribute to the the signature $\Sig(T)$ anyway. Next let us consider a path $T_t$ in $\GM(\Kk,J)$. Again eigenvalues may leave the annulus $\SM^1_h$ (now of fixed size), but the above argument transposes directly to this case. Concerning bifurcations of eigenvalues on the unit circle it was already shown in Proposition~\ref{prop-stabilitysignature} that the inertia are constant (which is a local statement) and therefore also their contribution to the signature is constant.
\hfill $\Box$

\vspace{.2cm}

\begin{proposi}
\label{prop-signaturefinite}
If $\dim(\Hh)<\infty$, then $\Sig(T)=0$ for all $T\in\GM(\Kk,J)=\FM\UM(\Kk,J)=\UM(\Kk,J)$.  
\end{proposi}

\noindent {\bf Proof.} Because one can choose $h$ such that it contains the whole (only discrete) spectrum of $T$, it follows that $\Ee_==\Kk$ and therefore $\Sig(T)=\Sig(\Kk)=\dim(\Hh)-\dim(\Hh)=0$.
\hfill $\Box$

\vspace{.2cm}

Actually, it is not an easy endeaver to produce examples with non-trivial signature. This is the main object of Section~\ref{sec-nontrivialex}. Let us first provide an alternative way to calculate $\Sig(T)$.

\subsection{The spectral flow of essentially $\SM^1$-gapped operators}
\label{sec-FredholmIndex}

It follows from the explicit formula \eqref{eq-VZdef} for $V(e^{-\imath t}T)$ that the path
\begin{equation}
\label{eq-S1path}
\Gamma_T\;=\;
(V(e^{-\imath t}T))_{t\in[0,2\pi)}
\;,
\end{equation}
in $\UM(\Kk)$ is real analytic. Therefore all results of analytic perturbation theory \cite{Kat}  apply and assure that all discrete eigenvalues depend real analytically on $t$ (if adequate branches are chosen at level crossings). Moreover, if $T\in\FM\UM(\Kk,J)$, then $1$ is not in the essential spectrum of $V(e^{-\imath t}T)$ for all $t\in[0,2\pi)$. Therefore it is possible to define an associated intersection index as in Section~\ref{sec-S1FredCZ} provided that $1$ is not a constant eigenvalue of $\Gamma_T$. If $1$ is a constant eigenvalue, then Proposition~\ref{prop-TtoU2} implies that each point on $\SM^1$ is an eigenvalue of $T$. As already pointed out, this may happen as shows the example in Section~\ref {sec-exampleperturbation}. On the other hand, for essentially $\SM^1$-gapped $J$-unitaries this cannot happen, namely one has the following:

\begin{proposi}
\label{prop-S1gappedtransverse}
Let $T\in\FM\UM(\Kk,J)$. The path $\Gamma_T$ defined in {\rm \eqref{eq-S1path}} is transversal in the sense that there is no constant eigenvalue equal to $1$ if and only if $T\in\GM(\Kk,J)$.  
\end{proposi}

Therefore $\SF(\Gamma_T)$ is well-defined for $T\in\GM(\Kk,J)$. Of course, the spectral flow is just a particular case of the Bott-Maslov index and the intersection index defined in Sections~\ref{sec-Index} and \ref{sec-S1FredCZ}:
$$
\SF(\Gamma_T)
\;=\;
\IN\bigl((e^{-\imath t}T)_{t\in[0,2\pi)},1\bigr)
\;=\;
\BMa\bigl((\widehat{e^{-\imath t}T}\,\widehat{\Psi}_0)_{t\in[0,2\pi)},\widehat{\Psi}_0\bigr)
\;.
$$
Therefore, it counts the number of eigenvalues of $T$ on the unit circle, but weighted by an orientation.  Let us first note that the homotopy invariance of the spectral flow (in the sense of Section~\ref{sec-Index} immediately implies the following.

\begin{proposi}
\label{prop-homotopyinvarianceindex} If $s\mapsto T_s$ is a continuous path in $\GM(\Kk,J)$, then $s\mapsto\SF(\Gamma_{T_s})$ is constant, namely $\SF(\Gamma_T)$ is a homotopy invariant.
\end{proposi}

Furthermore, let us provide an explicit proof of the fact that $\SF(\Gamma_T)$ is trivial in finite dimensions, even though this also follows from Proposition~\ref{prop-signaturefinite} combined with Theorem~\ref{theo-sigSFeq} below.

\begin{proposi}
\label{prop-finitedimindex}
If $n=\dim(\Hh)<\infty$, then $\SF(\Gamma_T)=0$ for all $T\in\GM(\Kk,J)$.
\end{proposi}

\noindent {\bf Proof.} The unitaries $V(\overline{z}\,T)\in \mbox{\rm U}(2n)$ are  traceclass. Therefore, the spectral flow by $1$ can be calculated as the total winding number:
$$
\SF(\Gamma_T)
\;=\;
\int^{2\pi}_0\frac{dt}{2\pi\imath}\;\Tr\left(V(e^{-\imath t}T)^*\partial_tV(e^{-\imath t}T)\right)
\;.
$$
Now by Proposition~\ref{prop-TtoU2} one finds
\begin{eqnarray*}
\Tr\left(V(e^{-\imath t}T)^*\partial_tV(e^{-\imath t}T)\right)
& = &
\imath\,\Tr
\left(\,
\left(
\begin{array}{cc}
e^{-\imath t}a^{-1} & -\,c^*(d^{-1})^* \\
(d^{-1})^*b^*  & e^{\imath t}\,(d^{-1})^* 
\end{array}
\right)
\left(
\begin{array}{cc}
- e^{\imath t}(a^*)^{-1} & 0 \\
0 & e^{-\imath t}\,d^{-1} 
\end{array}
\right)
\,\right)
\\
& = &
\imath\,
\Tr\left(-(a^*a)^{-1}+(dd^*)^{-1}\right)
\end{eqnarray*}
But as $a^*a=\one+c^*c$ and $dd^*=\one+cc^*$, the unitary invariance of the trace implies that the last expression vanishes so that $\SF(\Gamma_T)=0$.
\hfill $\Box$

\vspace{.2cm}

In general, the operator $V(e^{-\imath t}T)^*\partial_tV(e^{-\imath t}T)$ is not trace class so that the above calculation cannot be carried through.  It may not come as a surprise that this orientation is given by the inertia of the eigenvalue. Actually, this follows from an explicit formula proved next.

\begin{proposi}
\label{prop-speedcalc}
Let $T\in\FM\UM(\Kk,J)$ have a simple unit eigenvalue $e^{\imath\varphi}$ with eigenvector $\phi$. Let  $e^{\imath \theta_t}$ be a discrete non-degenerate eigenvalue of $V(e^{-\imath t}T)$ depending analytically on $t$ and satisfying $\theta_0=\varphi$. Then 
$$
\partial_t\,\theta_t\big|_{t=0}
\;=\;
-\,\phi^*J\,\phi
\;.
$$
\end{proposi}

\noindent {\bf Proof.} By considering $e^{-\imath\varphi}T$ instead of $T$, one sees that it is sufficient to consider the case $\varphi=0$. Thus one has $T\phi=\phi$ as well as $JT^*J\phi=\phi$ and $V(T)\phi=\phi$. The phase speed can now be calculated using the quadratic from $Q_1(T)$ defined in {\rm \eqref{eq-speedop}}:
$$
\partial_t\,\theta_t\big|_{t=0}
\;=\;
\phi^*Q_1(T)\,\phi
\;=\;
-\,\imath\;\phi^*\partial_t V(e^{-\imath t}T)\big|_{t=0}\,\phi
\;=\;
\phi^*
\begin{pmatrix}
-(a^*)^{-1} & 0 \\
0 & d^{-1}
\end{pmatrix}
\phi
\;.
$$
Now one can proceed as follows:
\begin{eqnarray*}
\partial_t\,\theta_t\big|_{t=0}
& = &
\phi^*
\begin{pmatrix}
-(a^*)^{-1} & 0 \\
0 & 0
\end{pmatrix}
JT^*J\,\phi
\;+\;
\phi^*JTJ\,
\begin{pmatrix}
0 & 0 \\
0 & d^{-1}
\end{pmatrix}
\phi
\\
& = &
\phi^*
\begin{pmatrix}
-(a^*)^{-1} & 0 \\
0 & 0
\end{pmatrix}
\begin{pmatrix}
a^* & -c^* \\
-b^* & d^*
\end{pmatrix}
\phi
\;+\;
\phi^*
\begin{pmatrix}
a & -b \\
-c & d
\end{pmatrix}
\begin{pmatrix}
0 & 0 \\
0 & d^{-1}
\end{pmatrix}
\phi
\;\,=\;\,
-\,\phi^*J\,\phi
\;,
\end{eqnarray*}
where in the last equality the relation $(a^*)^{-1}c^*=bd^{-1}$ was used which is equivalent to the relation $c^*d=a^*b$, see Propostion~\ref{prop-basicpropUKL}. 
\hfill $\Box$

\vspace{.2cm}

Based on the previous proposition, one can now prove that the spectral flow and the signature are equal, up to a sign.

\begin{theo}
\label{theo-sigSFeq}
For all $T\in\GM(\Kk,J)$, one has $\Sig(T)=-\SF(\Gamma_T)$.
\end{theo}

\noindent {\bf Proof.} Because both $\Sig(T)$ and $\SF(\Gamma_T)$ are homotopy invariants by Theorem~\ref{theo-sigdiffhomo} and Proposition~\ref{prop-homotopyinvarianceindex}, it is possible to deform $T$ in such a manner that all eigenvalues on $\SM^1$ are simple. But for such $T$ the orientation of the spectral flow of the eigenvalues of $V(e^{-\imath t}T)$ through $1$ is due to Proposition~\ref{prop-speedcalc} given by  the inertia of the corresponding eigenvalue of $T$ multiplied by $-1$. Summing over all inertia is thus equal to minus the spectral flow.
\hfill $\Box$

\subsection{Block diagonalization of $J$-unitaries}
\label{sec-S1diag}

Let us begin with the construction of a number of  objects naturally associated to a $\SM^1$-gapped $J$-unitary $T$ with a given (and fixed) gap $h(T)$. First of all,  the spectrum can be split in three subsets:
$$
\Delta_>\;=\;\{z\in\sigma(T)\,|\,|z|\geq e^{h(T)}\}
\;,
\qquad
\Delta_<\;=\;\{z\in\sigma(T)\,|\,|z|\leq e^{-h(T)}\}
\;,
$$
and $\Delta_==\sigma(\Delta)\setminus (\Delta_+\cup\Delta_-)\subset\SM^1_{h(T)}$. Of course, this splitting depends on the choice of $h(T)$, but this will be irrelevant for our purposes below. Associated to the splitting are Riesz projections denoted by $P_>=P_{\Delta_>}$, $P_<=P_{\Delta_<}$ and $P_==P_{\Delta_=}$, as well as $T$-invariant subspaces by $\Ee_>=\Ran(P_>)$, $\Ee_<=\Ran(P_<)$ and $\Ee_==\Ran(P_=)$ and $T^*$-invariant subspaces $\Ff_>=\Ker(P_>)$, $\Ff_<=\Ker(P_<)$ and $\Ff_==\Ker(P_=)$. If necessary, also the notation $\Ee_<(T)=\Ee_<$, {\it etc.},  will be used. Let us also use $P_\geq=P_>+P_=$ and $P_\leq=P_<+P_=$ as well as $\Ee_\geq$ and $\Ee_\leq$. Frames for $\Ee_<$, $\Ee_=$, $\Ee_>$ and so on will be denoted by $\Phi_<$, $\Phi_=$, $\Phi_\geq$. Then $\Phi_\leq=(\Phi_<,\Phi_=)$ is also a frame for $\Ee_\leq$. Of course, for a strictly $\SM^1$-gapped operator $T$, one has $P_==0$ and $\Phi_==0$. From general principles (see Appendix~\ref{app-Riesz}),
\begin{equation}
\label{eq-Rieszprop1}
\one\;=\;
P_<+P_=+P_>\;=\;P_<+P_\geq
\;,
\qquad
0\;=\;P_<P_\geq\;=\;P_\geq P_<\;=\;P_<P_=\;=\;P_<P_>
\;,
\end{equation}
as well as related identities. Every $v\in\Ee_<$ in the range of the idempotent $P_<$ satisfies $P_<v=v$, and similarly for vectors in $\Ee_\geq$. This implies
\begin{equation}
\label{eq-Rieszprop2}
\Ee_<\;=\;\Ff_\geq
\;,
\qquad
\Ff_<\;=\;\Ee_\geq
\;,
\end{equation}
as well as again related identities. Moreover, it follows from Proposition~\ref{prop-RieszProperties} that
\begin{equation}
\label{eq-Rieszprop3}
P_<^*\;=\; J\,P_>\, J
\;,
\qquad
P_=^*\;=\; J\,P_=\, J\;.
\end{equation}
This implies
\begin{equation}
\label{eq-Rieszprop4}
\Ee_<\;=\;J\,\Ff_>^\perp\;,
\qquad
\Ee_>\;=\;J\,\Ff_<^\perp\;,
\qquad
\Ff_=^\perp\;=\;J\,\Ee_=\;.
\end{equation}
Finally, Proposition~\ref{prop-eigenspacesJortho} shows that $\Ee_<\,\widehat{\perp}\,\Ee_\leq$ and $\Ee_\geq\,\widehat{\perp}\,\Ee_>$. In particular, $\Ee_<$ and $\Ee_>$ are isotropic (and Lagrangian for a strictly $\SM^1$-gapped $J$-unitary). Thus
\begin{equation}
\label{eq-Rieszprop5}
\Phi_<^*\,J\,\Phi_\geq\;=\;0
\;,
\qquad
\Phi_\leq^*\,J\,\Phi_>\;=\;0
\;.
\end{equation}
Another fact that will be of importance later on is that $\Ee_=$ is non-degenerate by Proposition~\ref{prop-decomp}. As a finite-dimensional space $\Ee_=$ is also closed and therefore Proposition~\ref{prop-signaturecalc} combined with $\Kk=(\Ee_<+\Ee_>)\widehat{+}\Ee_=$ implies that
\begin{equation}
\label{eq-Rieszprop6}
(\Ee_=)^{\widehat{\perp}}
\;=\;
\Ee_<\,+\,\Ee_>
\;.
\end{equation}
Next it follows from Proposition~\ref{prop-dimensionadd}(ii) that  $(\Ee_=\widehat{+}\,\Ee_<)^{\widehat{\perp}}= (\Ee_=)^{\widehat{\perp}}\cap(\Ee_<)^{\widehat{\perp}}$, so that the equality \eqref{eq-Rieszprop6} shows $(\Ee_=\widehat{+}\,\Ee_<)^{\widehat{\perp}}=(\Ee_<+\Ee_>)\cap(\Ee_<)^{\widehat{\perp}}=\Ee_<$. From this and  a similar argument one deduces
$$
(\Ee_<)^{\widehat{\perp}}
=\Ee_=\,\widehat{+}\,\Ee_<\;,
\qquad
(\Ee_>)^{\widehat{\perp}}
\;=\;\Ee_=\,\widehat{+}\,\Ee_>
\;.
$$
Finally let us note that there are also connections between the spectral subspaces of $T$ and $T^*$, denoted by $\Ee_<(T)$, $\Ee_<(T^*)$, {\it etc.}, for now. Indeed,
$$
\Ee_<(T^*)
\;=\;
\Ran(P^*_<)\;=\;\Ker(P_<)^\perp
\;=\;
\Ff_<(T)^\perp
\;=\;
J\,\Ee_>(T)\;,
\qquad
\Ee_>(T^*)\;=\;J\,\Ee_<(T)\;.
$$

\vspace{.2cm}

\begin{proposi}
\label{prop-S1Fredholmprop}
Let $T\in\GM(\Kk,J)$.  Then $\Ee_<$ and $\Ee_>$ form a Fredholm pair. 
\end{proposi}

\noindent {\bf Proof.} As $\Ee_<$, $\Ee_>$ and $\Ee_<+\Ee_>$ are all ranges of Riesz projections, they are closed (see Proposition~\ref{prop-RieszProj} in Appendix~\ref{app-angles}). Moreover, $\Ee_<\cap\Ee_>=\{0\}$ and $\dim((\Ee_<+\Ee_>)^\perp)=\dim(\Ee_=)<\infty$ so that all conditions of Definition~\ref{def-Fredpair} in Appendix~\ref{app-angles} are satisfied.
\hfill $\Box$

\vspace{.2cm}

Associated to the Fredholm pair $\Ee_<$ and $\Ee_>$ is always a Fredholm index, which in the present situation is $\Ind(\Ee_<,\Ee_>)=-\,\mbox{codim}(\Ee_<+\Ee_>)$. This index is of little interest in the present situation though. For example, it does depend on the choice of the gap $h(T)$ because this may modify the appearing codimension. The signature defined in Section~\ref{sec-FredholmSigDif} below is a more interesting and topological index. The main consequence of the Fredholm property relevant in the present context is the following.

\begin{proposi}
\label{prop-projS1Fred}
Let $T\in\GM(\Kk,J)$. Then
$$
P_<\;=\;\Phi_<\,\bigl(\Phi_>^*\, J\,\Phi_<\bigr)^{-1}\,\Phi_>^*\, J
\;,
\qquad
P_>\;=\;\Phi_<\,\bigl(\Phi_<^*\, J\,\Phi_>\bigr)^{-1}\,\Phi_<^*\, J
\;,
$$
and
$$
P_=\;=\;\Phi_=\,\bigl(\Phi_=^*\, J\,\Phi_=\bigr)^{-1}\,\Phi_=^*\, J
\;.
$$
\end{proposi}

\noindent {\bf Proof.} The oblique projection $P_<$ has by definition range $\Ee_<$ and kernel $\Ff_<$.  But $\Ff_<=\Ee_\geq$ by \eqref{eq-Rieszprop2} and $\Ee_\geq$ and $\Ee_>$ only differ by a finite dimensional subspace, it follows that $\Ee_<$ and $\Ff_<$ form a Fredholm pair. Moreover, $\Ee_<\cap\Ff_<=\{0\}$ and $\Ee_<+\Ff_<=\Kk$ by \eqref{eq-Rieszprop1}. Thus one can appeal to Proposition~\ref{prop-projectionconstruct} to write out $P_<$. Now $\Phi_<$ is a frame for $\Ee_<$ by definition. Because $\Ff_<^\perp=J\Ee_>$ by \eqref{eq-Rieszprop4}, a frame for $\Ff_<^\perp$ is given by $J\Phi_>$. Thus  Proposition~\ref{prop-projectionconstruct} allows to deduce the formula for $P_<$. The one for $P_>$ follows in a similar manner. For the last formula, let us note that $\Ee_=$ is non-degenerate by Proposition~\ref{prop-decomp}. Therefore the matrix $\Phi_=^*J\Phi_=$ is invertible. Hence the formula is well-defined and obviously an idempotent with range $\Ee_=$. As $\Ff^\perp_==J\Ee_=$ by \eqref{eq-Rieszprop4}, the kernel of the r.h.s. is indeed $\Ff_=$ as desired. 
\hfill $\Box$

\vspace{.2cm}

Finally let us come to a block diagonalization of $T\in\GM(\Kk,J)$.  For that purpose, let us set
\begin{equation}
\label{eq-roughdiag}
{ M}\;=\;(\Phi_<,\Phi_=,\Phi_>)
\;.
\end{equation}
One can check explicitly using \eqref{eq-Rieszprop5} that 
$$
M^*\,J\,M
\;=\;
\begin{pmatrix} 0 & 0 & \Phi_>^* J\Phi_< \\
0 & \Phi_=^*\, J\,\Phi_= & 0 \\
\Phi_<^* J\Phi_> & 0 & 0
\end{pmatrix}
\;=\;
\begin{pmatrix} 0 & 0 & \Phi_>^* J\Phi_< \\
0 & \Phi_=^*\, J\,\Phi_= & 0 \\
(\Phi_>^* J\Phi_<)^* & 0 & 0
\end{pmatrix}
\;.
$$
Thus the inverse of $M$ is given by
$$
 M^{-1}
\;=\;
\begin{pmatrix} 0 & 0 & (\Phi_>^* J\Phi_<)^{-1} \\
0 & (\Phi_=^*\, J\,\Phi_=)^{-1} & 0 \\
(\Phi_<^* J\Phi_>)^{-1} & 0 & 0
\end{pmatrix}
\, M^*\, J
\;,
$$
where the matrix is in the grading of \eqref{eq-roughdiag}. The invertibility of the three operators follows as in Proposition~\ref{prop-projS1Fred}.  Now the spaces $\Ee_<$, $\Ee_=$ and $\Ee_>$ are invariant for $T$. Therefore the span of $ T\Phi_<$ lies in $\Ee_<$ and thus \eqref{eq-Rieszprop5} implies $\Phi_\leq^* J T\Phi_<=0$. Similarly $\Phi_\geq^* J T\Phi_>=0$. Therefore one readily checks that
$$
 M^{-1}\, T\, M
\;=\;
\begin{pmatrix}
(\Phi_>^* J\Phi_<)^{-1}\,\Phi_>^*\, J\, T\,\Phi_<
& 0 & 0 \\
0 & (\Phi_=^* J\Phi_=)^{-1}\,\Phi_=^*\, J\, T\,\Phi_= & 0
\\
0 & 0 &
(\Phi_<^* J\Phi_>)^{-1}\,\Phi_<^*\, J\, T\,\Phi_>
\end{pmatrix}
\;.
$$
But as $ T\Phi_<=\Phi_<\Phi_<^* T\Phi_<$ by the invariance property and because $\Phi_<\Phi_<^*$ is the orthogonal projection on $\Ee_<$, and similarly for $ T\Phi_>$ and $T\Phi_=$, it follows that
\begin{equation}
\label{eq-roughdiag2}
 M^{-1}\, T\, M
\;=\;
\begin{pmatrix}
\Phi_<^*\, T\,\Phi_<
& 0 & 0 \\
0 & \Phi_=^*\, T\,\Phi_= & 0
\\
0 & 0 &
\Phi_>^*\, T\,\Phi_>
\end{pmatrix}
\;.
\end{equation}
The three entries are the restrictions of $ T$ to the invariant subspaces $\Ee_<$, $\Ee_=$ and $\Ee_>$. Hence their spectra lie inside, on or respectively outside the unit disc. One can now further choose the frames such that the diagonal blocks become Jordan blocks, in particular, the finite dimensional matrix $\Phi_=^*T\Phi_=$. It is also possible to  use formula \eqref{eq-roughdiag2} to construct $\SM^1$-gapped $J$-unitary operators from a disjoint decomposition
$\Kk=\Ee_<+\Ee_=+\Ee_>$ into two isotropic subspaces $\Ee_<$ and $\Ee_>$ and a non-degenerate subspace $\Ee_=$.

\section{Examples}
\label{sec-example}

The aim of this section is to illustrate the concepts and results of the paper by some examples. The formulas are often a bit easier if one works with another fundamental symmetry $G$ obtained from $J$ by the Cayley transform $C$:
$$
G\;=\;C^*JC\;=\;
\left(
\begin{array}{cc}
0 & -\imath\,\one \\
\imath\,\one & 0
\end{array}
\right)
\;,
\qquad
C\;=\;
\frac{1}{\sqrt{2}}\,
\left(
\begin{array}{cc}
\one & -\imath\,\one \\
\one & \imath\,\one
\end{array}
\right)
\;.
$$
One then has $G$-Lagrangian subspaces, $G$-unitaries, $\FM\UM(\Kk,G)$, {\it etc.}, and the passage from one to the other is just the Cayley transform. Note that $G$ is the standard symplectic form, multiplied by a complex unit in order to make it self-adjoint and involutive. Symplectic matrices and transfer operators (see Section~\ref{sec-transfer} below) are usually given in a form that is $G$-unitary rather than $J$-unitary, so that, pending on the scientific background, the reader may be more familiar with it.  Another reason why to prefer one or the other representation in concrete situations is the following fact (which can be deduced from the examples in Section~\ref{sec-2times2}): $J$-unitary unitaries may be diagonal, but not $G$-unitaries; on the other hand, $G$-unitaries with spectrum off the unit circle can be diagonal, but not $J$-unitaries. Hence one can diagonalize unitaries within the group of $J$-unitaries, and hyperbolic operators within the group of $G$-unitaries.

\subsection{Positivity in linear Hamiltonian systems}
\label{sec-HamSys}

In this section, let us consider a Hamiltonian system of the form
\begin{equation}
\label{eq-HamSys}
J\,\partial_t\,T^E_t
\;=\;
\imath\,(H_t+E\,P_t)\,T^E_t
\;,
\qquad
T^E_0\in\FM\UM(\Kk,J,1)
\;.
\end{equation}
Here $E\in\RM$ is a spectral parameter, and $H_t=H_t^*$ and $P_t\geq 0$ are Lipshitz continuous functions in $\BM(\Kk)$. As one readily checks that $\partial_t \bigl((T^E_t)^*JT^E_t\bigr)=0$ and as the initial condition $T^E_0$ is $J$-unitary, it follows that the solution $T^E_t$ of \eqref{eq-HamSys} is also $J$-unitary. Because the initial condition is, moreover,  a $1$-Fredholm operator and the set of these operators is open, the solution $T^E_t$ is also $1$-Fredholm, at least for sufficiently small $t$. For fixed (small) $t$, it is also possible to vary $E$ in compact intervals such that still $T^E_t\in\FM\UM(\Kk,J,1)$. Therefore there is a two-parameter family $T^E_t$ in $\FM\UM(\Kk,J,1)$ associated to \eqref{eq-HamSys} and one can study the intersection index defined in Section~\ref{sec-S1FredCZ} via the spectral flow of the unitaries $V(T^E_t)$. The transversality in $t$ and $E$ follows from the following positivity results, which are also crucial in the context of Sturm-Liouville type oscillation theory associated to the Hamiltonian system. In the finite-dimensional case, (i) and (ii) are due to Lidskii \cite{Lid}, and (iii) to Bott \cite{Bot}. Similar monotonicity results hold for the unitaries $\pi_\Psi(T^E_t\Phi)$ whenever $\Phi$ and $\Psi$ form a Fredholm pair of Lagrangian subspaces.

\begin{theo} 
\label{theo-osci} 
The solution of the Hamiltonian system {\rm \eqref{eq-HamSys}} satisfies the following.

\vspace{.1cm}

\noindent {\rm (i)} If $H_t+EP_t>0$, then $\frac{1}{\imath}\;V(T^E_t)^*\partial_t V(T^E_t)>0$.

\vspace{.1cm}

\noindent {\rm (ii)} If $H_t+EP_t<0$, then $\frac{1}{\imath}\,V(T^E_t)^*\partial_t V(T^E_t)<0$.

\vspace{.1cm}

\noindent {\rm (iii)} If $\partial_ET^E_0=0$, then $\frac{1}{\imath}\,V(T^E_t)^*\partial_E V(T^E_t)\geq 0$.

\end{theo}

\noindent {\bf Proof.} Recall that $V(T^E_t)=\widehat{\pi}(\widehat{\Psi}_0)^*\,\widehat{\pi}(\widehat{T^E_t}\,\widehat{\Psi}_0)=\widehat{\pi}(\widehat{\Psi}_0)^*\,\widehat{T^E_t}\cdot\widehat{\pi}(\widehat{\Psi}_0)$ so that
\begin{eqnarray*}
\frac{1}{\imath}\;V(T^E_t)^*\partial_t V(T^E_t)
& = &
\frac{1}{\imath}\;
\bigl(\widehat{T^E_t}\cdot\widehat{\pi}(\widehat{\Psi}_0)\bigr)^*
\partial_t \bigl(\widehat{T^E_t}\cdot\widehat{\pi}(\widehat{\Psi}_0)\bigr)^*
\\
& = &
\frac{1}{\imath}\;
A^*\,\binom{\widehat{\pi}(\widehat{\Psi}_0)}{\one}^*
(\widehat{T^E_t})^*\widehat{J}\,\partial_t\widehat{T^E_t}
\binom{\widehat{\pi}(\widehat{\Psi}_0)}{\one}\,A
\\
& = &
\frac{2}{\imath}\;
A^*\,\widehat{\Psi}_0^*
\bigl(0\;\widehat{\oplus}\;({T^E_t})^*{J}\,\partial_t{T^E_t}\bigr)
\widehat{\Psi}_0\,A
\\
& = &
\frac{2}{\imath}\;
A^*\,({T^E_t})^*{J}\,\partial_t{T^E_t}
\,A
\;,
\end{eqnarray*}
where in the second equality Proposition~\ref{prop-derivcalc} was used and $A$ is some invertible operator. Now using \eqref{eq-HamSys},
$$
\frac{1}{\imath}\;V(T^E_t)^*\partial_t V(T^E_t)
\;=\;
2\;
A^*\,({T^E_t})^*(H_t+EP_t)\,{T^E_t}\,A
\;,
$$
which implies (i) and (ii). Similarly,
$$
\frac{1}{\imath}\;V(T^E_t)^*\partial_E V(T^E_t)
\;= \;
\frac{2}{\imath}\;
A^*\,({T^E_t})^*{J}\,\partial_E{T^E_t}\,A
\;.
$$
Again from \eqref{eq-HamSys} one finds for $\epsilon>0$
$$
\partial_s\,\left((T^E_s)^*\,J\,T^{E+\epsilon}_s\right)
\;=\;
\imath\,\epsilon\;(T^E_s)^*\,P_s\,T^{E+\epsilon}_s
\;.
$$
As $(T_t^E)^*J T_t^{E}=J=(T^E_0)^*JT_0^{E}$, one thus has
\begin{eqnarray*}
\frac{1}{\imath}\;(T^E_t)^*\,J\,\partial_E T_t^E
& = & 
\lim_{\epsilon\to 0}\;
(\imath\,\epsilon)^{-1}\,
\left((T_t^E)^*\,J\,T_t^{E+\epsilon}\,-\,
(T_t^E)^*\,J\,T_t^{E}
\right)
\\
& = & 
\lim_{\epsilon\to 0}\;
(\imath\,\epsilon)^{-1}\,
\left((T_t^E)^*\,J\,T_t^{E+\epsilon}\,-\,(T_0^E)^*\,J\,T_0^{E+\epsilon}\,+\,
(T_0^E)^*\,J\,T_0^{E+\epsilon}\,-\,
(T_0^E)^*\,J\,T_0^{E}
\right)
\\
& = &
\int^t_0ds\;(T_s^E)^*\,P_s\,T_s^E
\;+\;
(T^E_0)^*\,J\,\partial_E T_0^E
\;.
\end{eqnarray*}
Because $P_s$ is non-negative and the last summand vanishes by hypothesis, this implies (iii).
\hfill $\Box$

\subsection{The calculation of bound states by a Bott-Maslov index}
\label{sec-application}

Let $H=H_0+K$ be a self-adjoint bounded operator on $\Hh$ given as a sum of a self-adjoint operator $H_0$  called the free Hamiltonian and a compact operator $K$ called the perturbation. In typical applications, $H_0$ has a band spectrum which is purely absolutely continuous.  However, all that is needed for the below is an interval $\Delta\subset\RM$, possibly unbounded, in the resolvent set $\rho(H_0)$ of $H_0$. It follows from Weyl's stability theorem for essential spectra that $H$ has only discrete spectrum in $\Delta$. The new eigenvalues and eigenvectors of $H$ are called bound states. The aim of this section is to reformulate the calculation of these bound states as an intersection problem for a path of Lagrangian Fredholm pairs. With some more technical effort, one can also deal with unbounded-selfadjoint operators.

\vspace{.2cm}

For that purpose, let us use the frames
$$
\Phi^E\;=\;\binom{\one}{H-E\,\one}\;(\one+(H-E\,\one)^2)^{-\frac{1}{2}}
\;,
\qquad
\Phi_0^E\;=\;\binom{\one}{H_0-E\,\one}\;(\one+(H_0-E\,\one)^2)^{-\frac{1}{2}}
\;,
$$
as well as $\Psi=\binom{\one}{0}$. These three frames are $G$-Lagrangian on $\Kk$. 
%

\begin{proposi}
\label{prop-scatsetup}  Let $E\in\Delta\subset\rho(H_0)\cap\RM$. Then the Lagrangian subspaces associated to $\Phi^E_0$ and $\Psi$ form a Fredholm pair and their intersection is empty for $E\in\Delta$. Furthermore, also the Lagrangian subspaces associated to  $\Phi^E$ and $\Psi$ form a Fredholm pair and the dimension of the intersection is equal to the multiplicity of $E$ as bound state of $H$.
\end{proposi}

\noindent {\bf Proof.}  In order to check the Fredholm pair property, let us use the characterization of Theorem~\ref{theo-Fredholm}(ii), forumated for $G$ instead of $J$. As $\Psi^*G\Phi^E_0= (E\,\one-H_0)(\one+(H_0-E\,\one)^2)^{-\frac{1}{2}}$ is self-adjoint, it is Fredholm if and only if $0$ is not in its essential spectrum. But this is given because $E$ is not even in the spectrum. Next let us suppose that $\Phi^E_0$ and $\Psi$ have a non-trivial intersection, namely that there are non-vanishing vectors $v,w\in\Hh$ such that $\Phi_0^Ev=\Psi w$. Then the upper equation of this identity implies $(\one+(H_0-E\,\one)^2)^{-\frac{1}{2}}v=w$ and the lower one $(H_0-E\,\one)w=0$. This would mean that $E$ is an eigenvalue of $H_0$, in contradiction to the hypothesis. For the pair $\Phi^E$ and $\Psi$ one proceeds in the same way. Alternatively, one can use the fact that $\Phi^E$ is obtained from $\Phi_0^E$ by compact perturbation.
\hfill $\Box$

\vspace{.2cm}

Due to this proposition, one can calculate the eigenvalues of $H$ in $\Delta$ (namely the bound states) via the spectral flow of the unitaries
$$
E\in\Delta\;\mapsto\;V^E\;=\;\pi(C\Psi)^*\pi(C\Phi^E)\in\UM(\Hh)
\;.
$$
By Theorem~\ref{theo-Fredholm}, this path is actually in $\UM_\ess(\Hh)$. Of course, $\pi(C\Psi)=\one$ and by the explicit expression of $\Phi^E$:
\begin{equation}
\label{eq-VofE}
V^E
\;=\;
\bigl(\one-\imath(H-E\,\one)\bigr)
\bigl(\one+\imath(H-E\,\one)\bigr)^{-1}
\;=\;
(G^E+\imath\,\one)(G^E-\imath\,\one)^{-1}
\;,
\end{equation}
where $G^E=(E\,\one-H)^{-1}$ is the resolvent whenever it exists. Precisely at the bound states this inverse does not exist, but nevertheless $V^E$ is well-defined by the first formula.

\begin{proposi}
\label{prop-monotonicity}  Let $\Delta\subset\rho(H_0)$. Then $E\in\Delta\mapsto V^E$ is a real analytic path in $\UM_\ess(\Hh)$ satisfying
$$
\frac{1}{\imath}\;(V^E)^*\,\partial_EV^E\;>\;0
\;,
\qquad
\lim_{|E|\to\infty}\; V^E\;=\;-\,\one
\;.
$$
In particular, all eigenvalues of $V^E$ rotate in the positive sense as a function of $E$ and cross $1$ transversally in the positive orientation.  These latter eigenvalue crossings correspond exactly at the eigenvalues of $H$, with equal multiplicities.
\end{proposi}

\noindent {\bf Proof.} It only remains to verify the positivity of
\begin{eqnarray*}
\frac{1}{\imath}\; (V^E)^*\partial_E V^E
& = &
\frac{1}{\imath}\; (V^E)^* 
\Bigl[\imath\,
\bigl(\one+\imath(H-E\,\one)\bigr)^{-1}
-
V^E\,(-\imath)\,
\bigl(\one+\imath(H-E\,\one)\bigr)^{-1}
\Bigr]
\\
& = &
2\;\Bigl(\bigl(\one+\imath(H-E\,\one)\bigr)^{-1}\Bigr)^*
\,\bigl(\one+\imath(H-E\,\one)\bigr)^{-1}
\\
& = &
2\;\bigl(\one+(H-E\,\one)^2\bigr)^{-1}
\;,
\end{eqnarray*}
which is obvious.
\hfill $\Box$

\subsection{The unitary $V(T)$ for $2\times 2$ matrices}
\label{sec-2times2}

This elementary section provides the normal forms of $2\times 2$ $J$-unitaries $T$, their $G$-unitary Cayley transform $C^*TC$, and the associated unitaries $V(e^{-\imath t}T)$

\vspace{.2cm}

\noindent {\bf Example 1} Let us consider, for $\varphi,\eta\in\RM$,
$$
T
\;=\;
e^{\imath\varphi}
\begin{pmatrix}
\cosh(\eta) & \sinh(\eta) \\
\sinh(\eta) & \cosh(\eta)
\end{pmatrix}
\;
\qquad
C^*TC\;=\;\begin{pmatrix} e^{\eta+\imath\varphi} & 0 \\ 0 & e^{-\eta+\imath\varphi}
\end{pmatrix}
\;.
$$
One then finds
$$
V(e^{-\imath t}T)
\;=\;
\begin{pmatrix}
\frac{e^{\imath(\varphi-t)}}{\cosh(\eta)} & \tanh(\eta) \\
-\tanh(\eta) & \frac{e^{-\imath(\varphi-t)}}{\cosh(\eta)}
\end{pmatrix}
\;.
$$
This matrix satisfies
$$
\Tr(V(e^{-\imath t}T))
\;=\;
\frac{2\,\cos(\varphi-t)}{\cosh(\eta)}\;,
\qquad
\det(V(e^{-\imath t}T))
\;=\;1\;.
$$
Hence the eigenvalues of $ V(e^{-\imath t}T)$ are
$$
e^{\imath\theta_{\pm,t}}
\;=\;
\frac{\cos(\varphi-t)}{\cosh(\eta)}
\;\pm\;\imath\,
\sqrt{1-\frac{\cos^2(\varphi-t)}{\cosh^2(\eta)}}
\;.
$$
In particular,
$$
\max_{t\in[0,2\pi]}\;\Re e(e^{\imath\theta_{\pm,t}})
\;=\;
\frac{1}{\cosh(\eta)}
\;,
$$
which is thus strictly less than $1$. Hence there is a gap to the value $1$ on the circle.
\hfill $\diamond$

\vspace{.2cm}

\noindent {\bf Example 2} Now let us consider
$$
T
\;=\;
e^{\imath\varphi}
\begin{pmatrix}
e^{-\imath\eta} & 0 \\
0 & e^{\imath\eta}
\end{pmatrix}
\;,
\qquad
C^*TC\;=\;e^{\imath\varphi}\begin{pmatrix} \cos(\eta) & -\sin(\eta) \\ \sin(\eta) & \cos(\eta)
\end{pmatrix}
\;.
$$
Then
$$
V(e^{-\imath t}T)
\;=\;
\begin{pmatrix}
e^{\imath(\eta-t-\varphi)} & 0 \\  0
 & e^{-\imath(\eta-t+\varphi)}
\end{pmatrix}
\;.
$$
Hence the spectrum can immediately be read off.
\hfill $\diamond$

\vspace{.2cm}

\noindent {\bf Example 3} Let us consider the Jordan block 
$$
T
\;=\;
e^{\imath\varphi}
\begin{pmatrix}
1-\imath a & \imath a \\
-\imath a & 1+\imath a
\end{pmatrix}
\;,
\qquad
C^*TC\;=\;e^{\imath\varphi}\begin{pmatrix} 1 & -2a \\ 0 & 1
\end{pmatrix}
\;,
$$
where $a\in\RM$. Then one finds
$$
V(e^{-\imath t}T)
\;=\;
\begin{pmatrix}
\frac{e^{\imath(\varphi-t)}}{1+\imath a} & \frac{\imath a}{1+\imath a} \\  \frac{\imath a}{1+\imath a}
 & \frac{e^{-\imath(\varphi-t)}}{1+\imath a}
\end{pmatrix}
\;.
$$
This matrix satisfies
$$
\Tr(V(e^{-\imath t}T))
\;=\;
\frac{2\,\cos(\varphi-t)}{1+\imath a}\;,
\qquad
\det(V(e^{-\imath t}T))
\;=\;\frac{1-\imath a}{1+\imath a}\;.
$$
Hence the eigenvalues are
$$
e^{\imath\theta_{\pm,t}}
\;=\;
\frac{1}{1+\imath a}
\left(
\cos(\varphi-t)
\;\pm\;\imath\,
\sqrt{a^2+\sin^2(\varphi-t)}
\right)
\;.
$$
In particular, for $t=\varphi$ and say $a>0$, one has $e^{\imath\theta_{+,\varphi}}=1$ and $\Re e(e^{\imath\theta_{-,\varphi}})=1-a^2$. Hence there is a gap to the value $1$ on the circle. Furthermore,
$$
\Im m(e^{\imath\theta_{+,t}})
\;=\;
\frac{1}{2}\;\frac{a+\frac{1}{a}}{1+a^2}\;(t-\varphi)^2\;+\;
+\Oo((t-\varphi)^3)
\;,
$$
that is, the eigenvalue is touching $1$ only from above. Therefore, there is no spectral flow of $t\mapsto 
V(e^{-\imath t}T)$ through $1$. 
\hfill $\diamond$

\vspace{.2cm}

\noindent {\bf Example 4} Finally, let us study the family
$$
T^E
\;=\;
\begin{pmatrix}
E-\imath & E \\
E & E+\imath
\end{pmatrix}
\;,
\qquad
C^*T^EC\;=\;\begin{pmatrix} 2\,E & -1 \\ 1 & 0
\end{pmatrix}
\;.
$$
Then
$$
V(e^{-\imath t}T)
\;=\;
\begin{pmatrix}
\frac{e^{-\imath t}}{E+\imath} & \frac{E}{E+\imath} \\
\frac{-\,E}{E+\imath} & \frac{e^{\imath t}}{E+\imath}
\end{pmatrix}
\;.
$$
Thus
$$
\Tr(V(e^{-\imath t}T))
\;=\;
\frac{2\,\cos(t)}{E+\imath}\;,
\qquad
\det(V(e^{-\imath t}T))
\;=\;\frac{E-\imath}{E+\imath}\;,
$$
so that
$$
e^{\imath\theta_{\pm,t}}
\;=\;
\frac{1}{E+\imath }
\left(
\cos(t)
\;\pm\;\imath\,
\sqrt{E^2+\sin^2(t)}
\right)
\;.
$$
Let us now focus on $E>0$. Then
$$
\Re e(e^{\imath\theta_{+,t}})
\;=\;
\frac{E\,\cos(t)+\sqrt{E^2+\sin^2(t)}}{1+E^2}
\;,
\qquad
\Re e(e^{\imath\theta_{-,t}})
\;\leq\;
0\;.
$$
If $E<1$, then $\cos(t)=E$ has two solutions $t_+>0$ and $t_-<0$ for which then $\theta_{+,t_\pm}=0$. For $E=1$ there is one  solution $t=0$, while for $E>1$ there is no solution and hence $\Re e(e^{\imath\theta_{+,t}})<1$ uniformly in $t$. This is a bifurcation, but the spectral flow remains invariant.

\subsection{A finite rank perturbation of an $\SM^1$-gapped $G$-unitary}
\label{sec-exampleperturbation}

As motivation to this section, let us recall some results about the operator family $S_t=S+tQ$ on $\ell^2(\ZM)$ where $S$ is the shift and $Q=|0\rangle\langle 1|$. For $t=0$ it is the shift, but $S_1=S={S}_+\oplus{S}_-$ is the sum of a uni-lateral shift and its adjoint, where the direct sum decomposition is $\ell^2(\ZM)=\ell^2(\ZM_+)\oplus\ell^2(\ZM_-)$ with $\ZM_+=\NM=\{1,2,\ldots\}$ and $\ZM_-=\ZM\setminus\ZM_+=\{0,-1,-2,\ldots\}$, and $S_\pm= S|_{\ell^2(\ZM_\pm)}$. Thus the spectrum of $S_1$ is the unit disc filled with point spectrum because $\sum_{n\geq 1}\lambda^n\,|n\rangle$ is an eigenvector with eigenvalue $\lambda\in\CM$ as long as $|\lambda|<1$. As $S_1$ is a finite rank perturbation of $S$, this provides an example for the scenario (ii) in the analytic Fredholm theorem of Appendix~\ref{sec-analFred}. Note that the Fredholm index is $\Ind(S_1)=1-1=0$. On the other hand, the spectrum of $S_t$ for $t\in[0,1)$ is only the unit circle (see {\it e.g.} \cite[Problem 102]{Hal}, or argue as below). On the other hand, the family $S_t$ is norm-continuous in $t$. This example shows that the spectrum as a set is only upper semi-continuous, namely $\limsup_{t\to 1}\sigma(S_t)\subset\sigma(S_1)$, but this inclusion may be strict. 

\vspace{.2cm}

The aim of this section is to construct an example of $T_t$ showing that the same phenomena may happen within the class of $G$-unitary operators (it is impossible in the class of normal operators). Let still $S$ be the two-sided shift on $\ell^2(\ZM)$ and $r\in(0,1)$. Then one starts from
$$
T_0
\;=\;
\begin{pmatrix}
r\,S & 0 \\
0 & r^{-1}\,S
\end{pmatrix}
\;,
$$
acting on the Hilbert space $\ell^2(\ZM)\otimes\CM^2$. The operator $T_0$ is $G$-unitary and its spectrum is 
$$
\sigma(T_0)\;=\;
r\SM^1\cup r^{-1}\SM^1
\;,
$$
so that $T_0$ is clearly $\SM^1$-gapped. Now let $P_+=|1\rangle\langle 1|$ and $P_-=|0\rangle\langle 0|$ be the one-dimensional projections in $\ell^2(\ZM)$ onto the subspaces spanned by $|1\rangle$ and $|0\rangle$ respectively and then set
$$
K
\;=\;
\begin{pmatrix}
0 & P_+ \\
-P_+ & 0
\end{pmatrix}
\;.
$$
One has $GK=-K^*G$ so that $K$ is in the Lie algebra of the $G$-unitaries and thus
$$
e^{tK}
\;=\;
\begin{pmatrix}
\one-P_+ +\cos(t)P_+ & \sin(t)P_+ \\
-\sin(t)P_+ & \one-P_++\cos(t)P_+
\end{pmatrix}
\;,
$$
is $G$-unitary and a compact perturbation of $\one$. Let us set
\begin{equation}
\label{eq-loopdef}
T_t\;=\;T_0\,e^{tK}
\;=\;
\begin{pmatrix}
r S(\one-P_+)+r\cos(t)SP_+ & r\sin(t)SP_+ \\
-r^{-1}\sin(t)SP_+ & r^{-1}S(\one-P_+)+r^{-1}\cos(t)SP_+
\end{pmatrix}
\;.
\end{equation}

\vspace{.2cm}

\noindent {\bf Claim } {\sl $\sigma(T_t)=\sigma(T_0)$ for $t\not =\frac{\pi}{2},\frac{3\pi}{2}$, and 
\begin{equation}
\label{eq-claim}
\sigma(T_{\frac{\pi}{2}})\;=\;
\sigma(T_{\frac{3\pi}{2}})
\;=\;
\{\lambda\in\CM\,|\,r\leq |\lambda|\leq r^{-1}\}
\;.
\end{equation}
Furthermore, in agreement with analytic Fredholm theory one has $\sigma_{\mbox{\rm\tiny dis}}(T_{\frac{\pi}{2}})=\sigma(T_{\frac{\pi}{2}})\setminus \sigma(T_0)$.}

\vspace{.2cm}

This shows that the essentially $\SM^1$-gapped operators are not stable under compact perturbations. Let us begin the proof of the claim by rewriting \eqref{eq-loopdef}. First recall that
$$
S_+^*S_+\;=\;\one-P_+\;,
\qquad
S_+S_+^*\;=\;\one\;,
\qquad
S_-S_-^*\;=\;\one-P_-\;,
\qquad
S_-^*S_-\;=\;\one\;.
$$
Furthermore, with the partial isometry $Q=|0\rangle\langle 1|$ as above, 
$$
Q^*Q\;=\;P_+\;,
\qquad
QQ^*\;=\;P_-\;,
\qquad
Q\;=\;SP_+\;=\;P_-Q
\;.
$$
For the matrix entries of $T_t$ one finds in the grading of $\ell^2(\ZM_+)\oplus\ell^2(\ZM_-)$
$$
SP_+\;=\;
\begin{pmatrix}
0 & 0 \\
Q
& 0
\end{pmatrix}
\;,
\qquad
S(\one-P_+)
\;=\;
\begin{pmatrix}
S_+ & 0 \\
0 & S_-
\end{pmatrix}
\;.
$$
Therefore $T_t$ acting on $\ell^2(\ZM_+)\oplus\ell^2(\ZM_-)\oplus\ell^2(\ZM_+)\oplus\ell^2(\ZM_-)$ becomes 
$$
T_{t}
\;=\;
\begin{pmatrix}
rS_+ & 0 & 0 & 0 \\
r\cos(t)Q & r S_- & r\sin(t)Q & 0 \\
0 & 0 & r^{-1}S_+ & 0 \\
-r^{-1}\sin(t)Q & 0 & r^{-1}\cos(t)Q & r^{-1}S_-
\end{pmatrix}
\;.
$$
Apart from $t=0$, the points $t=\frac{\pi}{2},\frac{3\pi}{2}$ are particularly simple because the $4\times 4$ matrix operator decouples in two $2\times 2$ matrix operators. Let us start with
%
%
$$
T_{\frac{\pi}{2}}
\;=\;
\begin{pmatrix}
rS_+ & 0 & 0 & 0 \\
0 & r S_- & r\,Q & 0 \\
0 & 0 & r^{-1}S_+ & 0 \\
-r^{-1}\,Q & 0 & 0 & r^{-1}S_-
\end{pmatrix}
\;.
$$
Hence it is sufficient to carry out the spectral analysis of the following block operators on $\ell^2(\ZM_+)\oplus\ell^2(\ZM_-)$:
$$
A_+
\;=\;
\begin{pmatrix}
rS_+ & 0  \\
-r^{-1}\,Q  & r^{-1}S_-
\end{pmatrix}
\;,
\qquad
A_-
\;=\;
\begin{pmatrix}
r^{-1}S_+ & 0  \\
r\,Q  & r S_-
\end{pmatrix}
\;.
$$
Indeed, then $\sigma(T_{\frac{\pi}{2}})=\sigma(A_+)\cup\sigma(A_-)$. Note that neither $A_+$ nor $A_-$ are $G$-unitary (for any $G$).  As $S_\pm$ are unilaterl shifts, a singular Weyl sequence argument shows that $r\SM^1\cap r^{-1}\SM^1\subset\sigma(A_\pm)$. Let us focus on $A_-$ and look at the trial state
\begin{equation}
\label{eq-eigenfc}
\psi
\;=\;
\alpha\sum_{n\geq 1}\lambda_+^n|n\rangle\,\oplus\,\beta\;\sum_{n\leq 0}\lambda_-^{|n|}|n\rangle
\;,
\end{equation}
where $|\lambda_\pm|<1$ and $\alpha,\beta\in\CM$. Clearly $\psi\in\ell^2(\ZM_+)\oplus\ell^2(\ZM_-)$. As $Q$ is a perturbation acting non-trivially only on the subspace spanned by $|1\rangle$, any eigenfunction of $A_-$ has to be of this form. Furthermore
$$
A_-\psi
\;=\;
\alpha r^{-1}\lambda_+\sum_{n\geq 1}\lambda_+^n|n\rangle\,\oplus\,\alpha r\lambda_+|0\rangle\,+\,\beta\;r(\lambda_-)^{-1}\sum_{n\leq -1}\lambda_-^{|n|}|n\rangle
\;.
$$
Now for a given $\lambda\in\CM$ with $|\lambda|\in(r,r^{-1})$ let us choose $\lambda_\pm$, $\alpha$ and $\beta$ such that
$$
\lambda\;=\;\frac{\lambda_+}{r}\;=\;\frac{r}{\lambda_-}\;,
\qquad
\alpha\;=\;1\;,
\qquad
\beta\;=\;\frac{r\lambda_+}{\lambda}\;=\;r^2
\;.
$$
Then $A_-\psi=\lambda\psi$. Thus the annulus $\{\lambda\in\CM\,|\,r<|\lambda|<r^{-1}\}$ is in the point spectrum of $A_-$ so that $\sigma(A_-)$ contains the closure of the annulus. Outside of the annulus one could have point spectrum (this is the only spectrum allowed by analytic Fredholm theory because $A_-$ is bounded and invertible), but actually the above argument shows that there are no further eigenvalues outside of the annulus. 

\vspace{.2cm}

One can attempt to apply a similar argument to $A_+$: 
$$
A_+\psi
\;=\;
\alpha r\lambda_+\sum_{n\geq 1}\lambda_+^n|n\rangle\,\oplus\,\alpha(-r^{-1})\lambda_+|0\rangle\,+\,\beta\;(r\lambda_-)^{-1}\sum_{n\leq -1}\lambda_-^{|n|}|n\rangle
\;.
$$
But now $r\lambda_+=(r\lambda_-)^{-1}$ is incompatible with $|\lambda_\pm|<1$ and $r<1$. Hence $A_+$ has no eigenvalue and $\sigma(A_+)=r\SM^1\cup r^{-1}\SM^1$. Combining the above one thus concludes that \eqref{eq-claim} holds for $t=\frac{\pi}{2}$. The case $t=\frac{3\pi}{2}$ is dealt with similarly.

\vspace{.1cm}

Next let us do the spectral analysis of $T_t$ for $t\not = \frac{\pi}{2},\frac{3\pi}{2}$. By analytic Fredholm theory as described in Appendix~\ref{sec-analFred} either $\sigma(T_t)\setminus\sigma(T_0)$ only contains discrete spectrum, or all points in this set are eigenvalues. Thus let us look for eigenvalues of $T_t$. The eigenvectors have to be of the form
$$
\psi
\;=\;
\alpha_\uparrow \,\sum_{n\geq 1}\lambda_{+,\uparrow}^n|n,\uparrow\rangle
\,+\,
\beta_\uparrow\,\sum_{n\leq 0}\lambda_{-,\uparrow}^{|n|}|n,\uparrow\rangle
\,+\,
\alpha_\downarrow \,\sum_{n\geq 1}\lambda_{+,\downarrow}^n|n,\downarrow\rangle
\,+\,
\beta_\downarrow\,\sum_{n\leq 0}\lambda_{-,\downarrow}^{|n|}|n,\downarrow\rangle
\;,
$$
where all $\lambda$'s are of modulus less than $1$ and the coefficients $\alpha$ and $\beta$ are complex numbers, and the states $|n,\uparrow\rangle$ and $|n,\downarrow\rangle$ correspond to the first and second component of $\ell^2(\ZM)\oplus\ell^2(\ZM)$.  Applying $T_t$ now gives
\begin{eqnarray*}
T_t\psi \!
& = & \!
\alpha_\uparrow r\lambda_{+,\uparrow}\sum_{n\geq 1}\lambda_{+,\uparrow}^n|n,\uparrow\rangle
+
\frac{\beta_\uparrow r}{\lambda_{-,\uparrow}}\sum_{n\leq 0}\lambda_{-,\uparrow}^{|n|}|n,\uparrow\rangle
+
\frac{\alpha_\downarrow\lambda_{+,\downarrow}}{r}\sum_{n\geq 1}\lambda_{+,\downarrow}^n|n,\downarrow\rangle
+
\frac{\beta_\downarrow}{ r\lambda_{-,\downarrow}}\sum_{n\leq 0}\lambda_{-,\downarrow}^{|n|}|n,\downarrow\rangle
\\
& & +\;r(\alpha_\uparrow \cos(t)\lambda_{+,\uparrow}
+\alpha_\downarrow \sin(t)\lambda_{+,\downarrow})|0,\uparrow\rangle
\,+\,
\frac{1}{r}(\alpha_\downarrow\cos(t)\lambda_{+,\downarrow}\,-\,\alpha_\uparrow \sin(t)\lambda_{+,\uparrow})
|0,\downarrow\rangle
\;.
\end{eqnarray*}
From this equation one deduces with some care that there cannot be any eigenvectors in the annulus $\{\lambda\in\CM\,|\,r<|\lambda|<r^{-1}\}$ unless $t=\frac{\pi}{2},\frac{3\pi}{2}$. This completes the proof of the claim.

\vspace{.2cm}

Next let us calculate the unitaries $V(\overline{z}\,T_t)$ as given in \eqref{eq-VZdef}. For that purpose, one first has to calculate the Cayley transform of $T$. With $r=e^h$,
$$
C\,T_0C^*
\;=\;
\begin{pmatrix}
\cosh(h) S & \sinh(h)S \\
\sinh(h)S & \cosh(h)S
\end{pmatrix}
\;,
\qquad
Ce^{tK}C^*
\;=\;
\begin{pmatrix}
\one-P_++e^{\imath t}P_+ & 0 \\
0 & \one-P_++e^{-\imath t} P_+
\end{pmatrix}
\;.
$$
From this the $J$-unitary $CT_{t}C^*$ can readily be written out. For \eqref{eq-VZdef}, one needs the inverse of the adjoint of the upper left as well as the lower right corner of $CT_tC^*$, namely
$$
\left(\cosh(h)(\one-P_++e^{-\imath t}P_+)S^*\right)^{-1}
\;=\;
\sech(h)S(\one-P_++e^{\imath t}P_+)
\;,
$$
and 
$$
\left(\cosh(h)S(\one-P_++e^{-\imath t} P_+)\right)^{-1}
\;=\;
\sech(h)(\one-P_++e^{\imath t}P_+)S^*
\;.
$$
Thus
$$
V(\overline{z}\,T_t)
\;=\;
\begin{pmatrix}
\overline{z}\,\sech(h)S(\one-P_++e^{\imath t}P_+)
& \tanh(h) \\
\tanh(h)(P_+-e^{2\imath t}P_+-\one) & z\,\sech(h)(\one-P_++e^{\imath t}P_+)S^*
\end{pmatrix}
\;.
$$
In particular,
$$
V(\overline{z}\,T_0)
\;=\;
\begin{pmatrix}
\overline{z}\,\sech(h)S
& \tanh(h) \\
-\tanh(h) & z\,\sech(h)S^*
\end{pmatrix}
\;.
$$
After Fourier transform the spectral analysis of the $2\times 2$ matrix (see Example 1 above) applies and this shows that $\Re e( V(\overline{z}\,T_0))\leq \sech(h)^{-1}\one$ for all $z\in\SM^1$. Now $V(\overline{z}\,T_{\frac{\pi}{2}})$ is a finite range perturbation of $V(\overline{z}\,T_0)$. Thus the essential spectrum is unchanged by Weyl's theorem. Moreover, as every $z\in\SM^1$ is an eigenvalue of $T$ by the above, it follows that $1$ is always an eigenvalue of  $V(\overline{z}\,T_{\frac{\pi}{2}})$. Note that this is not a contradiction to the the fact $0\not\in\sigma(Q_z(T))$, because the quadratic form $\sigma(Q_z(T))$ has as many negative as positive eigenvalues.

\subsection{A loop in $\FM\UM(\Kk,G)$ with non-trivial intersection index}
\label{sec-nontrivialCZ}

This section completes the analysis of the example in the last section by calculating the intersection number of the loop $\Gamma=(T_t)_{t\in[0,2\pi)}$ where $T_t$ is given by \eqref{eq-loopdef}. As a finite rank perturbation of a $\SM^1$-gapped $G$-unitary $T_0$, this is clearly a path in $\FM\UM(\Kk,G)$. Moreover, the analysis of the proceeding section shows that there are two intersection along the loop, namely at $t=\frac{\pi}{2},\frac{3\pi}{2}$, and both are of multiplicity $1$. Hence $\IN(\Gamma)$ could be $-2$, $0$ or $2$ and it remains to analyze the orientation at the two intersections. For that purpose, let us first note that
$$
\frac{1}{\imath}\;T_t^*G\,\partial_t T_t\
\;=\;
\frac{1}{\imath}\;T_t^*G\,T_t\,K
\;=\;
\frac{1}{\imath}\;G\,K
\;=\;
\begin{pmatrix}
P_+ & 0 \\ 0 & P_+
\end{pmatrix}
\;\geq\; 0
\;.
$$
It follows from Proposition~\ref{prop-derivV(T)calc}, for the fundamental symmetry $G$ instead of $J$,  and the fact that the eigenfunction~\eqref{eq-eigenfc} does not vanish on the range of $P_+$ that the phase speed is positive. Therefore $\IN(\Gamma)=2$.

\subsection{Transfer operators}
\label{sec-transfer}

This section merely presents a class of $G$-unitaries typically associated to discrete Schr\"odinger operators. More specific cases are studied in the following sections. Let $H=H^*\in\BM(\Hh)$ and $A\in\BM(\Hh)$ such that $A^{-1}\in\BM(\Hh)$ exists. Then
\begin{equation}
\label{eq-transfergen}
T\;=\;
\begin{pmatrix}
2\,H A^{-1} & -A^* \\
A^{-1} & 0
\end{pmatrix}
\;,
\end{equation}
is $G$-unitary on $\Kk=\Hh\otimes\Hh$.  Let us look for an eigenvalue $\lambda\in\SM^1$ and eigenvector $\binom{v}{w}$ of $T$:
$$
T
\begin{pmatrix}
v \\
w
\end{pmatrix}
\;=\;
\lambda
\begin{pmatrix}
v \\
w
\end{pmatrix}
\qquad
\Longleftrightarrow
\qquad
\begin{pmatrix}
2\,HA^{-1}v-A^*w \\
A^{-1}v
\end{pmatrix}
\;=\;
\lambda
\begin{pmatrix}
v \\
w
\end{pmatrix}
\;.
$$
The second equation enforces $v=\lambda Aw$. Replacing this into the first one, the eigenvalue equation is equivalent to searching for $w\in\Hh$ with
\begin{equation}
\label{eq-transfereig}
(2\,H-\lambda A-(\lambda A)^*)w\;=\;0\;,
\end{equation}
namely for $w\in\Hh$ in the kernel of the self-adjoint operator $H(\lambda)=2\,H-\lambda A-(\lambda A)^*$. If such a $w$ and $\lambda$ are given, then
$$
T
\begin{pmatrix}
\lambda A w \\ w
\end{pmatrix}
\;=\;
\lambda
\begin{pmatrix}
\lambda A w \\ w
\end{pmatrix}
\;,
$$
and, if the eigenvalue $\lambda$ is simple, its signature is the sign of 
$$
\begin{pmatrix}
\lambda A w \\ w
\end{pmatrix}^*
G
\begin{pmatrix}
\lambda A w \\ w
\end{pmatrix}
\;=\;
-\,2\,\Im m\bigl(\lambda\,w^*Aw\bigr)
\;.
$$
For the rest of the spectrum one deduce from a Weyl sequence argument
$$
0\in\sigma(H(\lambda))
\qquad
\Longleftrightarrow
\qquad
\lambda\in\sigma(T)
\;.
$$
In concrete situations, this also allows to examine whether $T$ is $\SM^1$-gapped. Furthermore, one checks
$$
CTC^*\;=\;
\frac{1}{2}\begin{pmatrix}
(2\,H-\imath\,\one-\imath\,A^*A)A^{-1} & (2\,H-\imath\,\one+\imath\,A^*A)A^{-1} \\
(2\,H+\imath\,\one-\imath\,A^*A)A^{-1} & (2\,H+\imath\,\one+\imath\,A^*A)A^{-1}
\end{pmatrix}
\;.
$$
%
%
%
This, and consequently the formulas for $V(\overline{z}\,T)$ and $Q_z(T)$ considerably simplify if $A$ is unitary. Thus let us assume this from now on so that
$$
CTC^*\;=\;
\begin{pmatrix}
(H-\imath\,\one)A^{-1} & HA^{-1} \\
HA^{-1} & (H+\imath\,\one)A^{-1}
\end{pmatrix}
\;.
$$
Due to the general formulas,  
$$
V(\overline{z}\,T)\;=\;
\begin{pmatrix}
\overline{z}(H+\imath\,\one)^{-1}A^* & H(H+\imath\,\one)^{-1} \\ -A(H+\imath\,\one)^{-1}HA^* & zA(H+\imath\,\one)^{-1}
\end{pmatrix}
\;,
$$
and
$$
Q_z(T)
\;=\;
\begin{pmatrix}
-A (H^2+\one)^{-1}A^* & -zAH(H^2+\one)^{-1} \\ -\overline{z} (H^2+\one)^{-1}HA^* & 
(H^2+\one)^{-1}
\end{pmatrix}
\;.
$$
%
%
The speed of the eigenvalue of $V(\overline{\lambda}\,T)$ crossing $1$ is due to Proposition~\ref{prop-speedcalc} equal to
\begin{eqnarray*}
\frac{1}{\|\lambda Aw\|^2+\|w\|^2}\,
\begin{pmatrix}
\lambda A w \\ w
\end{pmatrix}^*
Q_\lambda(T)
\begin{pmatrix}
\lambda A w \\ w
\end{pmatrix}
\;=\;
\frac{2\,\Im m\bigl(\lambda\,w^*Aw\bigr)}{\|\lambda Aw\|^2+\|w\|^2}\,
\;.
\end{eqnarray*}
%

\subsection{$\SM^1$-gapped $G$-unitaries with discrete spectrum on the unit circle}

For a first concrete example of a transfer operator of the form \eqref{eq-transfergen}, let $A=\one$ and $H=H^*\in\BM(\Hh)$ be such that $\sigma_\ess(H)\cap[-1,1]=\emptyset$  and suppose that $H w=\mu w$ for some $\mu\in(-1,1)$ and a unit vector $w\in\Hh$. For sake of simplicity, let us also suppose that there is only one such eigenvalue.  Then $\lambda$ is an eigenvalue of $T$ if $2\mu=\lambda+\lambda^{-1}$. Actually then both $\lambda$ and $\lambda^{-1}$ are eigenvalues of $T$: 
$$
T
\begin{pmatrix}
\lambda\,w \\
w
\end{pmatrix}
\;=\;\lambda
\begin{pmatrix}
\lambda\,w \\
w
\end{pmatrix}
\;,
\qquad
T
\begin{pmatrix}
\lambda^{-1}w \\
w
\end{pmatrix}
\;=\;\lambda^{-1}
\begin{pmatrix}
\lambda^{-1}w \\
w
\end{pmatrix}
\;.
$$
As, moreover, $\mu\in (-1,1)$, it follows that $\lambda\in\SM^1$. Furthermore, a Weyl sequence argument shows that $T$ is essentially $\SM^1$-gapped. Let us examine the signature of $\lambda=e^{\imath \varphi}$:
$$
\begin{pmatrix}
\lambda\,w \\
w
\end{pmatrix}^*
G
\begin{pmatrix}
\lambda\,w \\
w
\end{pmatrix}
\;=\;
-2\,\sin(\varphi)
\;.
$$ 
Thus the signature is negative for $\lambda$ in the upper arc of $\SM^1$, and positive for $\lambda$ in the lower one. As the signature is the sum over all eigenvalue pairs on the unit circle, one has $\Sig(T)=0$.

\vspace{.2cm}

A concrete example is $2H=E-\widehat{S}-\widehat{S}^*-\cot(\alpha) |0\rangle\langle 0|$ on $\Hh=\ell^2(\NM)$ with $E\not\in[ -1,1]$ and $\widehat{S}$ is the unilateral shift (which was also denoted by $S_+$ in Section~\ref{sec-exampleperturbation}). This models a half-sided discrete Laplacian with adequate boundary condition $\alpha\in[0,\pi]$. 

\subsection{Essentially $\SM^1$-gapped $G$-unitaries with non-vanishing signature}
\label{sec-nontrivialex}

This example is inspired by the two-dimensional Harper model describing a tight-binding electron on a square lattice submitted to a magnetic field (in the Landau gauge). This connection will be made more explicit elsewhere \cite{SV}. Let again $\Hh=\ell^2(\NM)$ and $2H=E-\widehat{S}-\widehat{S}^*$ where $\widehat{S}$ is the unilateral shift. Furthermore, the unitary is $A=e^{\imath\theta X}$ where $X$ is the position operator on $\ell^2(\NM)$ defined by $X|n\rangle=n\,|n\rangle$ and finally  $\theta\in[0,2\pi)$ is a phase (the magnetic flux through the unit cell). The $G$-unitary transfer operator $T$ depending on the two parameters $E\in\RM$ and $\theta$ is then explicitly given by
$$
T\;=\;
\begin{pmatrix}
(E-\widehat{S}-\widehat{S}^*)e^{-\imath\theta X} & -e^{-\imath\theta X}\\
e^{-\imath\theta X} & 0
\end{pmatrix}
\;.
$$
Let us look for eigenvalues $\lambda=e^{\imath\varphi}$ of $T$ by analyzing \eqref{eq-transfereig}, which becomes more explicitly
$$
(E-\widehat{S}-\widehat{S}^*- 2\,\cos(\theta X+\varphi))w\;=\;0
\;.
$$
This is the Schr\"odinger equation $\widehat{h}_{\theta,\varphi}w=Ew$ for the one-dimensional (critical) Harper operator $\widehat{h}_{\theta,\varphi}=\widehat{S}+\widehat{S}^*+ 2\,\cos(\theta X+\varphi)$ restrict to the discrete half-line, that is, acting on $\ell^2(\NM)$. One is thus interested in its bound states (boundary states). For that purpose let us first of all choose and then fix $E\in\RM$ in a spectral gap of the two-sided Harper operator ${h}_{\theta,\varphi}={S}+{S}^*+ 2\,\cos(\theta X+\varphi)$ acting on $\ell^2(\ZM)$, defined with the two-sided shift $S$. Second of all, let us suppose that the phase is rational $\theta=2\pi\frac{q}{p}$. This is actually not a restriction because one can approximate irrational $\theta$ by rational ones and use that $T$ depends continuously on $\theta$ so that the signature does not change. The rationality assumption  implies that the potential of $h_{\theta,\varphi}$ is $p$-periodic. As the half-sided Harper operator $\widehat{h}_{\theta,\varphi}$ is a Jacobi matrix, its bound states can be calculated by transfer matrix methods, as discussed in detail in \cite{ASV}. The transfer matrix over one period is the following real $2\times 2$ $G$-unitary matrix of unit determinant:
$$
\Tt^E_\varphi
\;=\;
\prod_{n=0}^{p-1}
\begin{pmatrix}
E-2\,\cos(\theta n+\varphi) & -1 \\ 1 & 0
\end{pmatrix}
\;.
$$
As $E$ is not in the spectrum of ${h}_{\theta,\varphi}$, the eigenvalues $\kappa$, ${\kappa}^{-1}$ of $\Tt^E_\varphi$ are off the unit circle. This also assures that $T$ is essentially $\SM^1$-gapped. One now has a bound state for $\widehat{h}_{\theta,\varphi}$ if and only if the Dirichlet boundary condition coincides with the contracting direction, namely 
$$
\Tt^E_\varphi\binom{1}{0}
\;=\;\kappa\binom{1}{0}
\;,
$$
where $|\kappa|<1$. For fixed $E$, this is a trigonometric equation for $\varphi$. The (weighted) number of its solutions is actually equal to the Chern number of the Fermi projection of two-dimensional Harper operator on energies below $E$ \cite{ASV}. This number of solutions can thus take any integer value (when one is free to vary $\frac{q}{p}$ as well as $E$, the only free parameters). For each solution $\varphi$, one then knows that $e^{\imath\varphi}$ is an eigenvalue of $T$. A more detailed treatment of this connection can be found in \cite{SV}.

\appendix

\section{Reminders on Fredholm operators}
\label{app-Fredholm}

A bounded operator $T\in\BM(\Hh)$ on a Hilbert space is called a Fredholm operator if and only if $\Ran(T)$ is closed, $\dim(\Ker(T))<\infty$ and $\dim(\Ran(T)^\perp)<\infty$. Its Fredholm index is then defined by $\Ind(T)=\dim(\Ker(T))-\dim(\Ran(T)^\perp)$. Recall also \cite{Hal} that an operator $T\in\BM(\Hh)$ is called bounded from below if there is a constant $g>0$ such that $\|T\phi\|\geq g\|\phi\|$ for all $\phi\in\Hh$. Then one can prove the following: $T$ is invertible (as operator, that is, there exists a bounded inverse) if and only if $T$ is bounded from below and $\Ran(T)$ is dense. Furthermore, $T$ is said to be essentially bounded from below if there exists a constant $g>0$ such that $T^*T\geq g\,\one$ except on a finite dimensional subspace. The following shows that this is connected to Fredholm properties.

\begin{proposi}
\label{prop-Fredcrit}
All of the following statements are equivalent:

\vspace{.1cm}

\noindent {\rm (i)}  There exists $g>0$ such that $T^*T\geq g\,\one$ except on a finite dimensional subspace

\vspace{.1cm}

\noindent {\rm (ii)} $0\not\in\sigma_\ess(T^*T)$ 

\vspace{.1cm}

\noindent {\rm (iii)} $T^*T$ is a Fredholm operator

\vspace{.1cm}

\noindent {\rm (iv)} There is no singular Weyl sequence $(\phi_n)_{n\geq 1}$ of pairwise orthogonal unit vectors with $T\phi_n\to0$

\vspace{.1cm}

\noindent {\rm (v)} $T$ is a left semi-Fredholm operator, namely $\Ran(T)$ is closed and $\dim(\Ker(T))<\infty$

\vspace{.1cm}

\noindent {\rm (vi)} There exists $S\in\BM(\Hh)$ such that $ST-\one$ is compact {\rm (}$S$ is called a left pseudo-inverse{\rm )}

\vspace{.1cm}

\noindent {\rm (vii)} There exists $S\in\BM(\Hh)$ such that $T^*S-\one$ is compact 

\vspace{.1cm}

\noindent {\rm (viii)} $T^*$ is a right semi-Fredholm operator, namely $\Ran(T^*)$ is closed and $\dim(\Ran(T^*)^\perp)<\infty$

\end{proposi}

\noindent {\bf Proof.} The equivalence of (i) to (iv) can be found in any text book. Equivalence with (v) is \cite[Corollary~I.4.7]{EE} and with (vi) \cite[Theorem~I.3.13]{EE} (actually, it rather follows from the proof therein). Finally equivalence of (vi) with (vii) is \cite[Proposition~I.3.9]{EE} and with (viii) again the proof of \cite[Theorem~I.3.13]{EE}, combined with \cite[Theorem~I.3.7]{EE}
\hfill $\Box$

\vspace{.2cm}

\begin{coro}
\label{coro-Fredcrit}
$T$ is a Fredholm operator if and only if $T$ and $T^*$ are essentially bounded from below.
\end{coro}

\begin{proposi}
\label{prop-Fredspec}
For a Fredholm operator $T$ with $\Ind(T)=0$, one has $0\not\in\sigma_c(T)\cup\sigma_r(T)$.
\end{proposi}

\noindent {\bf Proof.} For a Fredholm operator with vanishing index there exists a compact operator $K$ such that $S=T+K$ is invertible, namely $0\not\in\sigma(S)$. Suppose that $0\in\sigma_c(T)$. Then there exists a singular Weyl sequence $(\phi_n)_{n\geq 1}$ of pairwise orthogonal unit vectors with $T\phi_n\to 0$. But as $K\phi_n\to 0$, one also has $S\phi_n\to 0$ and therefore $0\in\sigma(S)$, which is a contradiction. Next recall that $0\in\sigma_r(T)$ if and only if $T$ has trivial kernel and $\Ran(T)$ is not dense, namely if $\dim(\Ker(T))=0$ and $\dim(\Ker(T^*))>0$. But these two properties are not reconcilable with $\Ind(T)=0$ so that $0\not\in\sigma_r(T)$.   
\hfill $\Box$

\section{Frames, angle spectrum and Fredholm pairs}
\label{app-angles}

This appendix resembles a few definitions and known results about subspaces of a separable Hilbert space $\Hh$.

\begin{defini}
\label{def-frame}
Let $\Ee$ be a closed subspace of a Hilbert space $\Hh$ of dimension $k\in\NM\cup\{\infty\}$. Denote $\ell^2=\ell^2(\{1,\ldots,k\})$. A frame for $\Ee$ is an operator $\Phi:\ell^2\to \Hh$ with $\Ee=\Ran(\Phi)$ and $\Phi^*\Phi=\one$. Then $\Phi^\perp$ denotes a frame for $\Ee^\perp$.
\end{defini}

If $\Phi$ is a frame for $\Ee$, then $\Phi\Phi^*$ is obviously the orthogonal projection in $\Hh$ on $\Ee$. Note that frames are also partial isometries. Clearly frames are not unique. The following definition naturally generalizes finite-dimensional notions. In fact, if $\phi$ and $\psi$ are two unit vectors, then the angle $\theta$ between them is given by $\cos(\theta)^2=|\phi^*\psi|^2$.

\begin{defini}
\label{def-anglespec}
Let $\Ee$ and $\Ff$ be two closed subspaces of a Hilbert space $\Hh$ with frames ${\Phi}$ and $\Psi$ respectively. Suppose $\dim(\Ee)\geq\dim(\Ff)$. Then the angle spectrum $\sigma(\Ee,\Ff)\subset[0,\frac{\pi}{2}]$ between $\Ee$ and $\Ff$ is given by the those $\theta\in [0,\frac{\pi}{2}]$ the cosine square of which are in the spectrum of $(\Psi)^*\Phi\Phi^*\Psi$:
$$
\cos^2(\sigma(\Ee,\Ff)\bigr)\;=\;\sigma((\Psi)^*\Phi\Phi^*\Psi)
\;.
$$
The angle spectrum can be decomposed into point, singular and absolutely continuous spectrum as well as discrete and essential spectrum by using these notions for the self-adjoint operator $(\Psi)^*\Phi\Phi^*\Psi$.
\end{defini}

Because for any operator $T$ one has $\sigma(T^*T)\cup\{0\}=\sigma(TT^*)\cup\{0\}$, it follows that $\sigma(\Ee,\Ff)\cup\{\frac{\pi}{2}\}=\sigma(\Ff,\Ee)\cup\{\frac{\pi}{2}\}$ in case that $\dim(\Ee)=\dim(\Ff)$. It is also possible to calculate the angle spectrum from the associated orthogonal projections. For example, if $\Ee+\Ff=\Hh$, then $\sin(\sigma(\Ee,\Ff))=\sigma(\Phi\Phi^*-\Psi\Psi^*)\cap\RM_\geq$. 

\vspace{.2cm}

It is possible that the angle spectrum $\sigma(\Ee,\Ff)$ between two closed subspaces is essential at $0$ so that there are infinitely many almost common directions. Then, even if $0$ is not in the point spectrum of $\sigma(\Ee,\Ff)$ so that the intersection of the subspaces is empty as a set, one cannot speak of transversal subspaces. The following notion due to Kato \cite{Kat} is to be interpreted as a transversality condition on two subspaces  as well as their orthogonal complements up to finite dimensional intersections.  

\begin{defini}
\label{def-Fredpair}
Two closed subspaces $\Ee$ and $\Ff$ of a Hilbert space $\Hh$ are said to form a Fredholm pair if the dimension of the intersection $\Ee\cap\Ff$ and the codimension of the sum $\Ee+\Ff$ are finite and, moreover, $\Ee+\Ff$ is a closed subspace. Their index is then defined by
$$
\mbox{\rm ind}(\Ee,\Ff)
\;=\;
\dim(\Ee\cap\Ff)-\mbox{\rm codim}(\Ee+\Ff)
\;.
$$
\end{defini}

Note that the closedness of $\Ee+\Ff$ has to be assumed, in general. It does not follow from the closedness of $\Ee$ and $\Ff$ unless one of the latter is finite dimensional.
Note also that if, say, $\Ff$ is finite dimensional, then $\Ee$ has to span all but finitely many directions of $\Hh$ if $\Ee$ and $\Ff$ form a Fredholm pair.  One is, however, mainly interested in the case where both $\Ee$ and $\Ff$ are infinite dimensional. The above definition coincides with \cite{Fur}, but in \cite{ASS} $\Ee$ and $\Ff$ (or the associated orthogonal projections) are said to form a Fredholm pair if the dimension of the intersection $\Ee\cap\Ff^\perp$ and the codimension of the sum $\Ee+\Ff^\perp$ are finite and $\Ee+\Ff^\perp$ is closed. In view of the connections to the angles between the subspaces presented next, we stick to the above choice.

\begin{theo}
\label{theo-Fredholmpair} 
Let $\Ee$ and $\Ff$ be two closed subspaces and ${\Phi}$ and ${\Psi}$ be frames for $\Ee$ and $\Ff$. Then the following are equivalent:

\vspace{.1cm}

\noindent {\rm (i)} $\Ee$ and $\Ff$ form a Fredholm pair

\vspace{.1cm}

\noindent {\rm (ii)} $\Phi^* \Psi^\perp$ is a Fredholm operator

\vspace{.1cm}

\noindent {\rm (iii)}  The essential angle spectra $\sigma_\ess(\Ee,\Ff)$ and $\sigma_\ess(\Ee^\perp,\Ff^\perp)$ do not contain $0$


\end{theo}

\noindent {\bf Proof.}  Let $P=\Phi(\Phi)^*$ be the orthogonal projection on $\Ee$ and $Q= \Psi^\perp(\Psi^\perp)^*$ the orthogonal projection on $\Ff^\perp$. It is proved in  \cite{ASS,Fur} that (i) is equivalent to the fact that $QP$ seen as an operator from $P\Hh$ to $Q\Hh$ is a Fredholm operator. But $\Phi:\ell^2\to P\Hh$ and  $(\Psi^\perp)^*: Q\Hh\to\ell^2$ are isomorphisms so that $QP$ is Fredholm if and only if $(\Psi^\perp)^* QP\Phi=(\Psi^\perp)^* \Phi$ is Fredholm.  For the equivalence (ii)$\Leftrightarrow$(iii) let us recall that an operator $T$ is Fredholm if and only if $T^*T$ and $TT^*$ are Fredholm, thus if and only if $0\not\in\sigma_\ess(T^*T)$ and $0\not\in\sigma_\ess(TT^*)$. This is applied to $T=\Phi^* \Psi^\perp$. Hence (ii) is equivalent to 
$$
0\not\in\sigma_\ess\bigl((\Psi^\perp)^*\Phi\Phi^* \Psi^\perp\bigr)
\,=\,1-\sigma_\ess\bigl(\Psi^\perp)^*\Phi^\perp(\Phi^\perp)^* \Psi^\perp\bigr)
\qquad
\Longleftrightarrow
\qquad
0\not\in\sigma_\ess\bigl(\Ee^\perp,\Ff^\perp\bigr)
\;,
$$
and 
$$
0\not\in\sigma_\ess\bigl(\Phi^*\Psi^\perp(\Psi^\perp)^* \Phi\bigr)
\,=\,1-\sigma_\ess\bigl(\Phi^*\Psi\Psi^* \Phi \bigr)
\qquad
\Longleftrightarrow
\qquad
0\not\in\sigma_\ess\bigl(\Ff,\Ee)
\;.
$$
But $0\not\in\sigma_\ess\bigl(\Ff,\Ee)$ is equivalent to $0\not\in\sigma_\ess\bigl(\Ee,\Ff)$ because these two angle spectra differ at most by the point $\frac{\pi}{2}$.
\hfill $\Box$

\vspace{.2cm}

Next let us recall that if $\Hh=\Ee+\Ff$ is the sum of two closed subspaces with trivial intersection $\Ee\cap\Ff=\{0\}$, then there is an associated oblique projection (idempotent) defined by $P\phi=\psi$ if $\phi=\psi+\psi'$ with $\psi\in\Ee$ and $\psi'\in\Ff$ is the unique decomposition. The range of $P$ is $\Ee$, the kernel $\Ff$. The projection $P$ is orthogonal if and only if $\Ee=\Ff^\perp$.

\begin{proposi}
\label{prop-projectionconstruct}
Let $\Ee$ and $\Ff$ be a Fredholm pair and let $\Phi$ and $\Psi$ be frames for $\Ee$ and $\Ff$. If  $\Hh=\Ee+\Ff$ and $\Ee\cap\Ff=\{0\}$, then
\begin{equation}
\label{eq-obliquegeneral}
P\;=\;\Phi\bigl((\Psi^\perp)^*\Phi\bigr)^{-1}(\Psi^\perp)^*\;,
\end{equation}
is the oblique projection {\rm (}idempotent{\rm )} on $\Hh$ with range $\Ran(P)=\Ee$ and kernel $\Ker(P)=\Ff$. 
\end{proposi}

\noindent {\bf Proof.} By Theorem~\ref{theo-Fredholmpair}, the operator $(\Psi^\perp)^*\Phi$ is Fredholm and therefore $\Ran((\Psi^\perp)^*\Phi)$ is closed so that $\ell^2=\Ran((\Psi^\perp)^*\Phi)\oplus \Ker(\Phi^*\Psi^\perp)$. Let us show that the kernel of $\Phi^*\Psi^\perp$ is trivial. Indeed, if $\Phi^*\Psi^\perp v=0$ for some vector $v$, then $\Psi^\perp v$ is perpendicular to $ \Ee$, thus $\Psi^\perp v\in\Ee^\perp\cap\Ff^\perp=(\Ee+\Ff)^\perp=\{0\}$, the last equality by the hypothesis $\Hh=\Ee+\Ff$, so that $v=0$. Thus $(\Psi^\perp)^*\Phi:\ell^2\to\ell^2$ is surjective. To check the injectivity, suppose that $(\Psi^\perp)^*\Phi v=0$. Then $\Phi v\in\Ee\cap\Ff$ so that $v=0$ by the hypothesis $\Ee\cap\Ff=\{0\}$. Hence $(\Psi^\perp)^*\Phi$ is bijective and bounded and therefore invertible by the inverse mapping theorem.
\hfill $\Box$

\section{Riesz projections}
\label{app-Riesz}

The following proposition resembles a few facts about Riesz projections associated to a bounded operator $T$ on a Hilbert space. Proofs can be found, {\it e.g.}, in \cite{Kat}.

\begin{proposi}
\label{prop-RieszProj}
Let $\Delta\subset\sigma(T)$ be a separated spectral subset, namely a closed subset which has trivial intersection with the closure of $\sigma(T)\setminus\Delta$. Associated to $\Delta$ let  $\Gamma$ be a curve in $\CM\setminus\sigma(T)$ with winding number $1$ around each point of $\Delta$ and $0$ around all points of $\sigma(T)\setminus\Delta$. The Riesz projection of $T$ on $\Delta$ is defined as
\begin{equation}
\label{eq-RieszProj}
P_\Delta
\;=\;
\oint_\Gamma \frac{dz}{2\pi\imath}\;
(z- T)^{-1}
\;.
\end{equation}
Range and kernel of $P_\Delta$ are denoted by $\Ee_\Delta=\Ran(P_\Delta)$ and $\Ff_\Delta=\Ker(P_\Delta)$. If $\Delta=\{\lambda\}$ is an isolated point in $\sigma(T)$, let us also use the notation $P_\lambda=P_\Delta$, $\Ee_\lambda=\Ee_\Delta$ and so on. The following properties hold.

\vspace{.1cm}

\noindent {\rm (i)} $P_\Delta$ is idempotent, namely an oblique projection and $\Ee_\Delta$ and $\Ff_\Delta$ are closed subspaces. Moreover, 

$P_\Delta$ is independent of the choice of $\Gamma$. 

\vspace{.1cm}

\noindent {\rm (ii)} If $T$ is invertible and $(\Gamma)^{-1}$ denotes the path of inversed complex points, one has
\begin{equation}
\label{eq-RieszProjInv}
P_\Delta
\;=\;
\oint_{(\Gamma)^{-1}} \frac{dz}{2\pi\imath}\;
(z- T^{-1})^{-1}
\;.
\end{equation}

\vspace{.1cm}

\noindent {\rm (iii)} If $\Delta$ and $\Delta'$ are disjoint separated spectral subsets, then $P_\Delta P_{\Delta'}=0$ and $P_{\Delta\cup\Delta'}=P_\Delta +P_{\Delta'}$.

\vspace{.1cm}

\noindent {\rm (iv)} 
For is a disjoint decomposition $\sigma(T)=\bigcup_{l=1}^L\Delta_l$ in separated spectral subsets, $\sum_{l=1}^L P_{\Delta_l}=\one$.  

\vspace{.1cm}

\noindent {\rm (v)} $\Ee_\Delta$ is invariant for $T$ and $\Ff_\Delta$ is invariant for $T^*$. Moreover, $\dim(\Ee_\Delta)=\dim(\Ff_\Delta^\perp)$. 

\vspace{.1cm}

\noindent {\rm (vi)} If $\Phi_\Delta$ and $\Psi_\Delta$ are frames for $\Ee_\Delta$ and $\Ff_\Delta^\perp$ and $\Psi_\Delta^*\Phi_\Delta$ is invertible, then 
$$
P_\Delta
\;=\;
\Phi_\Delta\,\bigl(\Psi_\Delta^*\Phi_\Delta)^{-1}\Psi_\Delta^*
\;.
$$

\vspace{.1cm}

\noindent {\rm (vii)} The orthogonal projections on $\Ee_\Delta$ and $\Ff_\Delta$ are $P_\Delta P_\Delta^*=\Phi_\Delta\Phi_\Delta^*$ and $P_\Delta^* P_\Delta=\Psi_\Delta\Psi_\Delta^*$. 

\vspace{.1cm}

\noindent {\rm (viii)}  If $\dim(\Ee_\lambda)<\infty$, then $\Ee_\lambda$ is the span of the generalized eigenvectors of $T$ to $\lambda$.

\vspace{.1cm}

\noindent {\rm (ix)}  Let $f$ be an analytic function on the convex closure of $\sigma(T)$. Suppose that 
$$
f(\sigma(T)\cap\Delta)\;\cap\; f(\sigma(T)\setminus\Delta)\;=\;\emptyset\;.
$$ 

Denote by $Q_{f(\Delta)}$ the Riesz projection of $f(T)$ on $f(\Delta)$, which is a separated spectral subset 

for $f(T)$. Then $Q_{f(\Delta)}=P_{\Delta}$.

\end{proposi}
\section{Analytic Fredholm theory}
\label{sec-analFred}

This appendix is there to illustrate the difficulty linked to the stability of the essential spectrum of non-self-adjoint operators. The following criterion is extracted from the arguments in Section XIII.4 of  \cite{RS}. Examples that case (ii) below actually does appear are given in Section~\ref{sec-exampleperturbation}.

\begin{theo}
\label{theo-analFredholm}
Let $ T$ and $S$ be two bounded operators on some separable Hilbert space such that $ T-S$ is compact. Let $C\subset\CM$ be one connected component of $\CM\setminus\sigma( T)$. Then one of the following two claims holds true:

\vspace{.1cm}

\noindent {\rm (i)} $C$ contains a point in the resolvent set of $S$.

\vspace{.1cm}

\noindent {\rm (ii)} All points of $C$ are eigenvalues of $S$.

\vspace{.1cm}

\noindent In the case {\rm (i)}, the spectrum of $S$ in $C$ is discrete.

\end{theo}

\noindent {\bf Proof.}  Let $K= T-S$. Then $z\in C\mapsto K( T-z)^{-1}$ is analytic and compact-valued. For $z\in C$ one has $S-z= (1-K( T-z)^{-1})( T-z)$, so the inverse $(S-z)^{-1}$ exists if and only if $(1-K( T-z)^{-1})^{-1}$ exists.  Now one clearly has the dichotomy that either $C$ contains some point in the resolvent set of $S$ or it contains none. The first case corresponds to (i). Then there exists some $z_0\in C$ such that the inverse $(1-K( T-z_0)^{-1})^{-1}$ exists. But by the analytic Fredholm theorem \cite[Theorem VI.16]{RS}, the inverse $(1-K( T-z)^{-1})^{-1}$ then exists for all $z\in C$ except for a discrete set of points. Hence also the spectrum of $S$ lying in $C$ only consists of a discrete set of points. In the second possibility, where no point of $C$ lies in the resolvent set of $S$, the operator $1-K( T-z)^{-1}$ is not invertible for any $z\in C$. By the Fredholm alternative this implies that for each $z\in C$ there is a vector $v_z$ lying in the kernel of $1-K( T-z)^{-1}$. Setting $w_z=( T-z)^{-1}v_z$, one then has $( T-z)w_z=K w_z$, that is $Sw_z=z w_z$. Therefore in the second possibility all points $z\in C$ are eigenvalues of $S$.
\hfill $\Box$

\section{Spectral flow of paths of essentially gapped unitaries}
\label{sec-SF}

For the convenience of the reader, some folklore facts about spectral flow are recollected in this appendix. Let $\UM_\ess(\Hh)$ be the set of those unitary operators $u$ on a Hilbert space $\Hh$ with the property that $1\not\in\sigma_\ess(u)$. Such unitaries will be called essentially gapped.  Associated to every (continuous) path $\gamma=(u_t)_{t\in[t_0,t_1]}$ in $\UM_\ess(\Hh)$ one can now define the spectral flow $\SF(\gamma)$ as the number of eigenvalue crossings through $1$, weighted by the orientation of the crossing. For that purpose, let us first suppose that the number of crossings  $\{t\in [t_0,t_1)\;|\;1\in\sigma(u_t)  \}$ is finite and does not contain the initial point $t_0$. It is explained below that these transversality and boundary conditions can be considerably relaxed in a straightforward manner.  At a crossing $u_t$ with $1\in\sigma(u_t)$, let $e^{\imath\theta_{1,s}},\ldots,e^{\imath\theta_{l,s}}$ be those eigenvalues of the unitary $u_s$ which are all equal to $1$ at $s=t$. Let us choose them to be continuous in $s$ and call the $\theta_{k,s}\in(-\pi,\pi]$ also the eigenphases of $u_s$. Choose $\epsilon,\delta>0$ such that $\theta_{k,s}\in[-\delta,\delta]$ for $k=1,\ldots,l$ and $s\in [t-\epsilon,t+\epsilon]$ and that there are no other eigenphases in $[-\delta,\delta]$ for $s\neq t$ and finally $\theta_{k,s}\neq 0$ for those parameters. Let $n_-$ and $n_+$ be the number of those of the $l$ eigenphases less than $0$ respectively before and after the intersection, and similarly let $p_-$ and $p_+$ be the number of eigenphases larger than $0$ before and after the crossing. Then the signature of $u_t$ is defined by
\begin{equation}
\label{eq-signature}
\mbox{\rm sgn}(u_t)
\;=\;
\frac{1}{2}\;(p_+-n_+-p_-+n_-)
\;=\;
l-n_+-p_-
\;.
\end{equation}
Note that $-l\leq\mbox{\rm sgn}(\gamma_t)\leq l$ and that $\mbox{\rm sgn}(\gamma_t)$ is the effective number of eigenvalues that have crossed $1$ in the counter-clock sense. Furthermore the signature is stable under perturbations of the path in the following sense: if a crossing is resolved by a perturbation into a series of crossings with lower degeneracy, then the sum of their signatures is equal to $\mbox{\rm sgn}(u_t)$. Finally let us remark that, if the phases are differentiable and $\partial_t \theta_{k,t}\neq 0$ for $k=1,\ldots,l$, then $\mbox{\rm sgn}(u_t)$ is equal to the sum of the $l$ signs $\mbox{\rm sgn}(\partial_t \theta_{k,t})$, $k=1,\ldots,l$. Yet another equivalent way to calculate  $\mbox{\rm sgn}(u_t)$ is as the signature of $\frac{1}{\imath}(u_t)^*\partial_tu_t$ seen as quadratic from on the eigenspace of $u_t$ to the eigenvalue $1$ (again under the hypothesis that the form is non-degenerate). Now the spectral flow of the path $\gamma$  is defined by
\begin{equation}
\label{eq-SFdef}
\SF(\gamma)
\;=\;
\sum_{t}
\;\mbox{\rm sgn}(u_t)
\;,
\end{equation}
where the sum is over the finite number of points at which $1\in\sigma(u_t)$.
If the initial point $t_0$ is on the singular cycle, but say the speeds of the eigenvalues passing through $1$ are non-vanishing, the index can still be defined. In order to conserve an obvious concatenation property, only the initial point $t_0$ in \eqref{eq-SFdef} is included and not the final point $t_1$ (an alternative would be to give each a weight $\frac{1}{2}$).  Furthermore, if a path $\gamma$ is such that $1\in\sigma(u_t)$ for an interval of parameters $t$, but it is clearly distinguishable how many eigenvalues pass through $1$ in the process, then the index can be defined as well. A formal, but obvious definition in these cases is not written out. It is clear from the definition that $\SF(\gamma)$ is a homotopy invariant under homotopies keeping the end points of $\gamma$ fixed.

\vspace{.2cm}

\noindent {\bf Acknowledgements:} The author profited from several discussions with Stephane Merigon and received financial support by the DFG. This work has a follow up \cite{SV} which analyzes homotopy invariants for essentially gapped operators with a supplementary symmetry and also deals with unbounded $J$-isometries.


\end{document}